\documentclass{IEEEtaes}

\usepackage{amsmath,amsfonts}
\usepackage{algorithmic}
\usepackage{algorithm}
\usepackage{array}
\usepackage[caption=false,font=normalsize,labelfont=sf,textfont=sf]{subfig}
\usepackage[utf8]{inputenc}
\usepackage{textcomp}
\usepackage{stfloats}
\usepackage{url}
\usepackage{verbatim}
\usepackage{graphicx}
\usepackage{cite}
\usepackage{tabularx}
\usepackage{makecell}

\usepackage{bm}
\usepackage{amsthm,amssymb}
\usepackage{mathrsfs}
\usepackage{soul} % 导入 soul 包
\usepackage{color, xcolor} % 颜色包，color 必须导入，xcolor 建议导入

\usepackage{multirow}

 \usepackage {CJKutf8}

\renewcommand{\thesection}{\Roman{section}}

\jvol{XX}
\jnum{XX}
\jmonth{XXXXX}
\pubyear{2025}
\begin{document}

\title{Learning-Based Multi-Stage Strategy for a Fixed-Wing Aircraft to Evade a Missile Detected at a Short Distance}

\author{Zhiguan Niu}
\author{Xiaochao Zhou}
\author{Hao Xiong}
% \author{Hantao Jiang}
% \author{Yixin Zhang}
% \author{Bernd R. Noack}

\affil{Harbin Institute of Technology Shenzhen, China}

% \receiveddate{This work was partially supported by
% % the Research Foundation for Advanced Talents, Harbin Institute of Technology Shenzhen (Grant No. CA11409019),
% the Shenzhen Science and Technology Program (Grant No. RCBS20210609103819024),
% the Natural Science Foundation of Guangdong Province China (Grant No. 2023A1515011010),
% and the Guangdong Basic and Applied Basic Research Foundation for Young Scientists (Grant No. 2021A1515110021).}

\authoraddress{
% The next few paragraphs should contain the authors' current affiliations, including current address and e-mail. 
Zhiguan Niu, Xiaochao Zhou, and Hao Xiong 
are with the School of Intelligence Science and Engineering, Harbin Institute of Technology Shenzhen, Shenzhen, China. \\
Corresponding author: Hao Xiong (e-mail: xionghao@hit.edu.cn). 
% Second B. Author, Jr., was with Rice University, Houston, TX 77005 USA. He is now with the Department of Physics, Colorado State University, Fort Collins, CO 80523 USA (e-mail: \href{mailto:author@lamar.colostate.edu}{author@lamar.colostate.edu}). 
% Third C. Author is with the Electrical Engineering Department, University of Colorado, Boulder, CO 80309 USA, on leave from the National Research Institute for Metals, Tsukuba 305-0047, Japan 
% (e-mail: \href{mailto:author@nrim.go.jp}{author@nrim.go.jp}).
}

% \corresp{{\itshape (Corresponding author: Hao Xiong)}.}

\markboth{Niu ET AL.}{LEARNING MULTI-STAGE EVADE}
\maketitle

% \author{
% Yiming Ou, Hao Xiong, Hantao Jiang, and Bernd R. Noack
        % <-this % stops a space
% \thanks{Corresponding author: Hao Xiong (e-mail: xionghao@hit.edu.cn).}
% \thanks{Yiming Ou, Hao Xiong, Hantao Jiang, and Bernd R. Noack are with the School of Mechanical Engineering and Automation, Harbin Institute of Technology Shenzhen, Shenzhen, China.}
% }


\IEEEpubid{0000--0000/00\$00.00~\copyright~2025 IEEE}

\maketitle

\begin{abstract}
\textcolor{black}{
Missiles pose a major threat to aircraft in modern air combat. 
Advances in technology make them increasingly difficult to detect until they are close to the target and highly resistant to jamming.
The evasion maneuver is the last line of defense for an aircraft. However, conventional rule-based evasion strategies are limited by computational demands and aerodynamic constraints, and existing learning-based approaches remain unconvincing for manned aircraft against modern missiles. To enhance aircraft survivability, this study investigates missile evasion inspired by the pursuit–evasion game between a gazelle and a cheetah and proposes a multi-stage reinforcement learning-based evasion strategy. 
% The strategy learns a large-azimuth turning policy to initiate evasion, a small-azimuth escape policy to maintain distance, and a short-range aggressive maneuver policy for final avoidance.
The strategy learns a large azimuth policy to turn to evade, a small azimuth policy to keep moving away, and a short distance policy to perform agile aggressive maneuvers to avoid.
One of the three policies is activated at each stage based on distance and azimuth. 
To evaluate performance, a high-fidelity simulation environment modeling an F-16 aircraft and missile under various conditions is used to compare the proposed approach with baseline strategies. 
Experimental results show that the proposed method achieves superior performance, enabling the F-16 aircraft to successfully avoid missiles with a probability of 80.89 percent for velocities ranging from 800 m/s to 1400 m/s, maximum overloads from 40 g to 50 g, detection distances from 5000 m to 15000 m, and random azimuths. 
When the missile is detected beyond 8000 m, the success ratio increases to 85.06 percent.
}
\end{abstract}


% \IEEEpubidadjcol

\section{Introduction}

\IEEEpubidadjcol

\textcolor{black}{
Missiles pose a major threat to aircraft in modern air combat \cite{Yan2024AGame,Wang2023ParameterMechanism,Tian2025AircraftField-of-View}. 
Missiles typically have much higher speed and superior maneuverability than their targets \cite{Cook2025MissileLearning}. 
Advances in technology enable missiles to remain undetected until they are close to the target. 
Due to their agility and the limited reaction and detection time available, missiles are increasingly difficult for pilots to evade \cite{Zhang2025AdaptiveGame,Gong2024Closed-FormSystem,Peng2022State-Following-Kernel-BasedTarget}. 
During an evasion maneuver, pilots have only seconds to react while interpreting warning data and visually locating the threat \cite{Yan2024AGame}. Consequently, manual avoidance imposes a heavy cognitive workload and requires significant estimation, making real-time missile avoidance increasingly challenging for human pilots.
}

\textcolor{black}{
However, conventional missile avoidance techniques \cite{Tian2023MissileAircraft,Evdokimenkov2021UnmannedAttack,Tian2023Estimation-basedAircraft,Du2025Three-dimensionalInformation}, such as programmed maneuvers and differential games, perform poorly \cite{Yan2024AGame} when an aircraft encounters a missile at close range. These methods are time-consuming \cite{Yan2024AGame} and often fail to accurately incorporate aircraft aerodynamics, which are critical for close-range avoidance \cite{Ou2024DynamicLearning}. 
Recent advances in reinforcement learning (RL) \cite{Clarke2020DeepAircraft,Qu2022SpacecraftLearning,Jiang2023SafelyRecovery} provide a promising approach to enhance evasion maneuvers and improve pilot survivability. 
An RL-based evasion policy can learn to respond rapidly and precisely to an incoming missile, optimizing maneuvers according to the aircraft’s state and performance.}

\textcolor{black}{
Scholar have achieved amazing process is air combat \cite{Chai2023ACombat,Wang2025DynamicCombat,Pope2023HierarchicalTrials} and pursuit-evasion game of the aircraft \cite{Ma2025HierarchicalMethod,Bertram2021AnGame,gao2023intelligentVehicles} based on RL in recent years.
}
\textcolor{black}{
Also, scholar have also applied RL to address the missile evasion problem for aircraft.
Narne et al. developed a real-time cooperative guidance strategy for an aircraft evading an incoming missile. The aircraft executes evasive maneuvers while launching a defensive missile to intercept or divert the attacker \cite{Li2021Real-timeLearning}.
Liu et al. proposed a prediction-information-based twin delayed deep deterministic policy gradient decision-making algorithm for aircraft, enabling autonomous evasion against interceptor attacks in an adversarial environment \cite{Liu2024PITD3-basedAircraft}.
Chen et al. proposed a Monte Carlo–based Deep Q-Network algorithm that integrates the Monte Carlo reinforcement learning framework with the DQN approach to address the evasion problem for high-speed aircraft \cite{Chen2023TheLearning}.
Scukins et al. proposed a decision-support tool to assist pilots in Beyond Visual Range air combat by addressing the complexity of multiple missile threats \cite{Scukins2024DeepEvasion}.
Chen et al. developed a deep reinforcement learning–based evasion strategy for unpowered aircraft, achieving coordinated control of angle of attack, bank angle, and body morphing \cite{Chen2022TheLearning}.
Ozbek proposed a reinforcement learning approach for generating real-time missile evasion maneuvers using the twin delayed deep deterministic policy gradient algorithm with a two-term reward function, and validated it using the Harfang simulator \cite{Ozbek2023MissileLearning}.
Li et al. developed a deep reinforcement learning–based maneuver evasion strategy to enhance an aircraft’s autonomous capability to penetrate defended airspace \cite{Li2022ResearchLearning}.
Cook et al. focused on evading surface-to-air missiles and proposed using deep reinforcement learning to train a guidance policy that assumes control of the aircraft upon missile detection \cite{Cook2025MissileLearning}.
Yuan et al. investigated a hierarchical, goal-guided learning method enabling an aircraft to evade multiple missiles by detecting threats at long range \cite{Yuan2023HierarchicalLearning}.
Yan et al. introduced a maneuvering strategy that combines line-of-sight (LOS) angle rate correction with RL to enable high-speed aircraft to evade pursuers \cite{Yan2024AGame}.
Zhang et al. introduced a risk-sensitive PPO algorithm and a training framework incorporating multi-head attention and dual-population adversarial training to improve adaptability to various unknown missile types \cite{Zhang2025AdaptiveGame}.
}
\textcolor{black}{
However, the above-mentioned studies have one or more limitations in the implementation details, reward function design, aircraft aerodynamics model, applicability to general scenarios, integration of the aircraft characteristics into the architecture and training of a strategy, statistical validation, or dependence on additional survival technologies (e.g., decoy).
}

\textcolor{black}{
This study is inspired by our previous research \cite{Ou2024DynamicLearning}, which explored a method for aircraft based on RL to respond to dynamic obstacles within short distances.
This study observes that 1) a missile can remain silent until it is at a short distance from the target and 2) the movement of a missile is insensitive to evasion maneuvers at a large distance but is sensitive to evasion maneuvers within a certain short distance. 
Thus, this study starts from addressing a missile at a short distance and then extends to addressing a missile at a farther distance.
Based on the above work, this study proposes a multi-stage RL-based evasion strategy for an aircraft evading a missile at a short distance.
The strategy is activated only upon detection of an approaching missile. 
In particular, this study focuses on a general mainstream air combat scenario of an aircraft encountering a missile. 
In such as a scenario, a missile can have different velocities and maneuver capabilities that are considerably higher or better than those of the aircraft and come from a different direction. The aircraft can have a different velocity.
A general scenario is challenging for the training of an effective RL-based evasion strategy and has not been comprehensively solved and validated in the above-mentioned references.
Moreover, a multi-stage architecture and corresponding reward functions are designed based on the observations of multi-stage characteristic. 
Previous related studies have not utilized the characteristic to improve the architecture but usually apply a learning-based policy to address all possible states, 
making the training of a policy challenging.
}

\textcolor{black}{
The major contributions of this paper are as follows.}
\begin{itemize}
	\item \textcolor{black}{
    A multi-stage RL-based strategy without requiring modification of an easy-to-use vanilla RL algorithm
    is developed for aircraft to address an approaching missile with different conditions at a short distance.
    % Compared with traditional geometric methods, the proposed strategy consider the aerodynamic characteristics and dynamic constraints of the aircraft, enabling it to take more realistic actions and extend its survival time.
    Different from existing reinforcement learning-based evasion strategies,
    this study addresses a general scenario of an aircraft evading a missile through integrating insights of 
    the multi-stage scenario into strategy architecture with corresponding reward functions, and includes detailed validation.
    }
        
    % and design a reward function that encourages highly maneuverable evasive actions.
    % Our strategy not only compels obstacle to reach in maximum overload, but also prolongs the survival time of the aircraft.
    % The strategy addresses not only non-cooperative collision avoidance with small sensing range, but also altitude, course and airspeed recover.

    \item \textcolor{black}{
    Training and comparison experiments have been conducted in a close-to-real simulation environment of an F-16 aircraft based on the JSBSim simulator and a numerical missile with conditions set according to realistic scenarios, accounting for the aerodynamics of the aircraft and the different maneuvering capabilities, velocities, incoming directions, and detected distance of the missile.
    Experimental results suggest that the multi-stage evasion strategy empowered by machine intelligence can considerably enhance the last line of defense for aircraft in modern air combat.    
    }
    
    % \item The time delay is introduced into the obstacle's tracking strategy. Experimental results show that the when the delay is set to 0.03 seconds, the aircraft can reach a survival rate of 60$\%$.
    
\end{itemize}

\textcolor{black}{
The remaining sections of this paper are structured as follows. 
Section \ref{sec:preliminaries} demonstrates the preliminaries of this study, including the problem statement and the notation for the scenario of an aircraft encountering a missile.
Section \ref{sec:method} proposes a multi-stage RL-based evasion strategy for an aircraft to address an incoming missile at different distances.
Section \ref{sec:experiments} presents the details of the implementation and training of the multi-stage RL-based strategy and conducts comparative experiments based on the JSBSim simulator to demonstrate the effectiveness of the proposed multi-stage RL-based strategy. 
Finally, Section \ref{sec:conclusion} summarizes the paper and suggests potential directions for future research.
}

%%%%%%%%%%%%%%%%%%%%%%%

\section{Preliminaries \label{sec:preliminaries}}

\textcolor{black}{
In this section, the problem statement and notations of the scenario of an aircraft evading a missile are introduced.}

\subsection{Problem statement}

\textcolor{black}{
This study addresses a general scenario of an aircraft evading a missile, as shown in Fig. \ref{fig: notations}, to investigate and improve the survivability of the aircraft in an adversarial environment. 
In such a scenario, the aircraft can have a different initial velocity.
The aircraft may encounter a missile incoming from different directions and observe the missile at an uncertain short distance.
The missile can have a different velocity that is considerably larger than the velocity of the aircraft.
The maneuvering capability of the missile, represented by maximum overload, is considerably larger than that of the aircraft.
The missile tracks the aircraft according to a mainstream navigation law such as Proportional Navigation (PN) or Augmented Proportional Navigation (APN).
}
\textcolor{black}{
It is assumed that the aircraft can observe the velocity and attitude of itself.
Also, the aircraft can observe or estimate the position and velocity of a detected missile with respect to the aircraft with advanced devices.}

\subsection{Notations}

% To present the state of a scenario, 

% The origin of the world frame locates below the initial position of the aircraft and on the ground.
% The heading of the aircraft is defined from the north in a clockwise order.

\begin{figure}[!htb]
    \centering
    \includegraphics[scale=0.45]{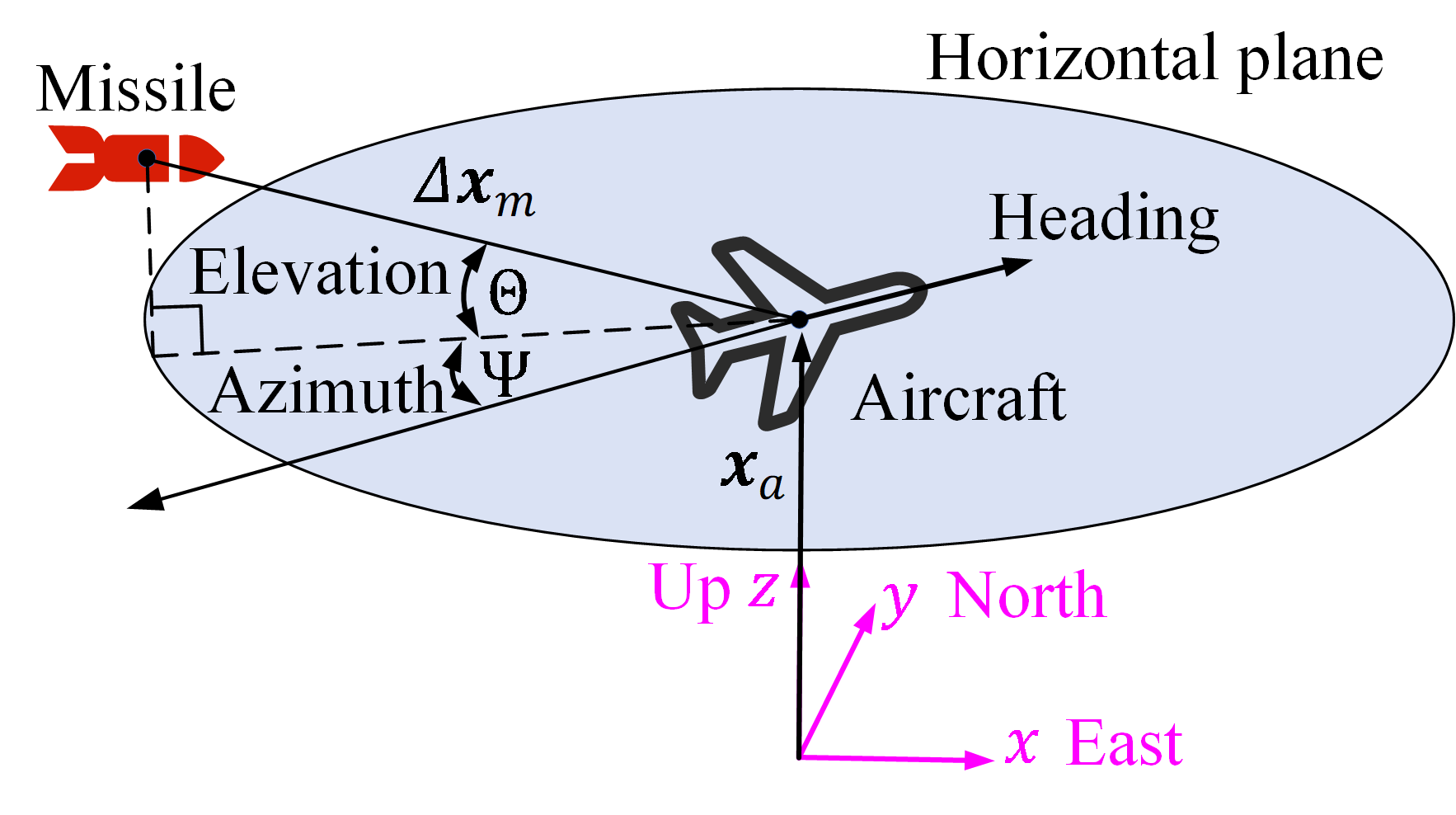}
    \caption{\textcolor{black}{Notations of the scenario of an aircraft evading a missile.}}
    \label{fig: notations}
\end{figure}

\textcolor{black}{
The notations of the scenario of an aircraft encountering a missile are illustrated in Fig. \ref{fig: notations}.
A world frame is defined to specify the state of the aircraft and the missile.
This study defines the world frame based on the east-north-up (ENU) coordinate system.
The positive directions of the x-axis, y-axis, and z-axis correspond to east, north, and up, respectively.
The origin of the world frame is located below the initial position of the aircraft and at sea level.
The position of the aircraft is defined as 
$\bm{x}_{a} = [x_a, y_a, z_a]$.
The velocity of the aircraft is defined as 
$\dot{\bm{x}}_{a} = [\dot{x}_a, \dot{y}_a, \dot{z}_a]$.
The attitude of the aircraft is defined as 
$[\theta, \phi, \psi]$.
The position of the missile with respect to the aircraft is defined as 
$\Delta\bm{x}_m = [\Delta x_m, \Delta y_m, \Delta z_m]$.
The velocity of the missile with respect to the aircraft is defined as 
$\Delta \dot{\bm{x}}_{m} = [\Delta \dot{x}_m, \Delta \dot{y}_m, \Delta \dot{z}_m]$.
The azimuth of the missile with respect to the aircraft is defined as 
$\Psi$ based on the opposite direction of the heading of the aircraft.
The azimuth is positive if the missile is on the right side and vice versa.
The elevation of the missile with respect to the aircraft is defined as 
$\Theta$.
The elevation is positive if the missile is above the horizontal plane defined by the aircraft and vice versa.
}

% The following is the defined state space:
% \[
% \left[ v_x, v_y, v_z, \theta, \phi, \psi, \Delta x, \Delta y, \Delta z, \Delta v_x, \Delta v_y, \Delta v_z \right]
% \]
% where $ v_x, v_y, v_z$ is aircraft's velocity components in x, y and z axis, $\theta, \phi, \psi$ is roll, pitch, heading of the aircraft, $\Delta y, \Delta z, \Delta v_x, \Delta v_y, \Delta v_z $ is the relative position and relative velocity between the aircraft and the obstacle.

% \subsubsection{Observation Space}

% \subsubsection{Action Space}

% The action space of the aircraft is defined by four continuous control variables:

% \[
% \delta_{Elevator}, \delta_{Aileron}, \delta_{Rudder}, \delta_{Throttle}
% \]

% In short-range air combat scenarios, fixed-wing aircraft face persistent threats from high-speed aggressive obstacles. 
% Traditional obstacle avoidance strategies is hard to ensure both survivability and maneuverability in complex dynamic environments.
% In this paper, a dynamic obstacle avoidance problem of the aircraft with a limited sensing range is addressed.
% The situation tend to discuss a high-threat, short-range air combat scenarios, with a high-speed aggressive obstacle chasing the aircraft.

%%%%%%%%%%%%%%%%%%%%%%%%%%%%%%%%%%%%%%%%%%%%%%%%
% \subsection{6-DOF Aiarcraft Model} \label{sec:VO}

% \subsubsection{PN and APN}

% The principle of 3DPN and the principle of APN.

%%%%%%%%%%%%%%%%%%%%%%%%%%%%%%%%%%%%%%%%%%

\section{Method \label{sec:method}}

\textcolor{black}{
In this section, a multi-stage RL-based strategy is developed for an aircraft to address a missile at a short distance.
The multi-stage RL-based strategy is tailored for the scenario of an aircraft encountering a missile at a short distance from the perspectives of strategy architecture, reward functions, and training for generalization.}

\textcolor{black}{
The multi-stage RL-based strategy has a multi-stage architecture, including a short distance stage, a relatively long distance and small azimuth stage, and a relatively long distance and large azimuth stage.
The short distance stage aims to guide the aircraft to avoid the missile.
The relatively long distance and small azimuth stage and relatively long distance and large azimuth stage aim to provide better conditions for the aircraft to avoid the missile in the short distance stage and postpone the start of the short distance stage to wear out the effective time of the missile.}

\begin{figure}[!htb]
    \centering
    \includegraphics[scale=0.5]{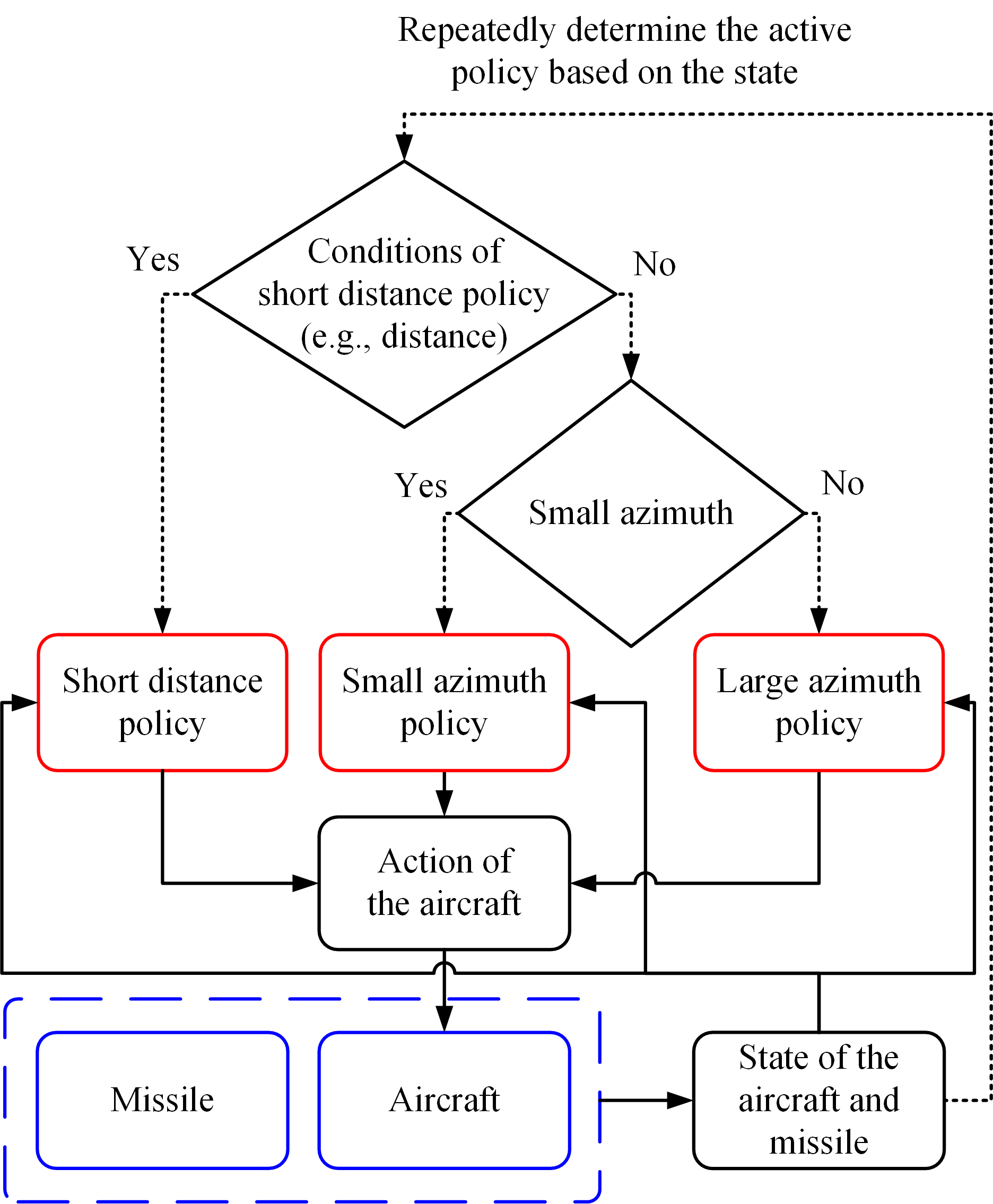}
    \caption{\textcolor{black}{Flow diagram of the multi-stage RL-based evasion strategy.}}
    \label{fig: flow diagram}
\end{figure}

\textcolor{black}{
The flow diagram of the multi-stage RL-based strategy is shown in Fig. \ref{fig: flow diagram}.}
\textcolor{black}{
To achieve the multi-stage RL-based strategy, one should train a short distance policy for the short distance stage (detailed in Section \ref{sec:method short distance policy}), and then analyze the performance of the short distance policy with different initial conditions (detailed in Section \ref{sec:method policy swithing}).
Based on the performance, one should determine the switching conditions that are beneficial for the short distance policy.
According to the switching conditions, one should train a small azimuth policy (detailed in Section \ref{sec:method small azimuth policy}) and a large azimuth policy (detailed in Section \ref{sec:method large azimuth policy}) for scenarios that do not satisfy the switching conditions.}
\textcolor{black}{
Eventually, the multi-stage RL-based strategy can be achieved by combining the three policies.
The multi-stage RL-based strategy determines the stage based on the initial conditions of a scenario and switches to the activated policy according to Fig. \ref{fig: flow diagram}.}

% is  decomposed into two policies: a short-distance evasion strategy and a long-distance moving away strategy.

% (1) Short-distance strategy: A traditional circling strategy is first trained, and then an improved version is developed by adding the line-of-sight (LOS) rate into the reward function.

% (2) Long-distance strategy: A moving-away policy is trained, enabling the aircraft to move away from the obstacle, and build favorable flight states for the short-distance phase.

% (3) Policy switching: A distance threshold for switching between the two strategies is determined, based on the escape success ratio in different initial conditions.

% \hl{how to determine threshold?}

% (4)Integrate the escape and circling strategy using behavior cloning.

% The obstacle evasion strategy is divided into two parts bases on the distance between the obstacle and the aircraft. This design is based on the following two considerations:

% (1) The energy consumption is extremely high during high-overload circling maneuvers, and remaining in a circling state continuously would quickly deplete its energy;

% (2) A fixed-wing aircraft cannot sustain extreme overloads (e.g., 8-9 g) for a long time without stalling or sustaining damage. \cite{}

% Therefore, dividing the strategy into two parts helps improve overall survival probability. With the following method, we tested the performance of the short-range obstacle evasion strategy.

\subsection{Short distance policy} \label{sec:method short distance policy}

% The circling maneuver is a classical aircraft evasion strategy.
\textcolor{black}{
The short distance policy is trained to increase the LOS of the missile, inspired by the PN law, and thus survive the aircraft.
The short distance policy tends to guide the aircraft to perform large overload maneuvers when the missile is close the the aircraft, and thus force the missile to miss the aircraft due to the limit of the maximum overload.
 The make the short distance policy more trainable, this paper proposes to train a pretrain policy that make the aircraft to perform large overload maneuvers (e.g., a steep turn policy), according to curriculum learning \cite{Xiao2021FlyingSim2Real}. Then, based on the pretrain policy, it has a better chance of successfully training the short distance policy.}

% main purpose is to force the obstacle to exceed its load limit, thereby achieving successful evasion.

\subsubsection{Steep turn policy}

\textcolor{black}{
Steep turn is a common evasion maneuver \cite{Yang2018EvasiveCombat}.
One can obtain a steep turn policy as the pretrain policy for the short distance policy.
% Inspired by the conventional circling maneuver, the short distance RL-based evasion strategy designs a reward function, enabling the aircraft to learn the circling-like maneuvers and circling direction.
% During training, a traditional circling strategy is first developed, then used as a pre-trained model to accelerate the learning of the final short-distance strategy.
% For the circling strategy, 
The reward function of the steep turn policy can be expressed as
\begin{equation}
    \label{eq: steep turn reward}
        r_{\rm{turn}}  = \omega_{\rm{turn}}^{\rm{roll}} r_{\rm{turn}}^{\rm{roll}} + \omega_{\rm{turn}}^{\rm{pitch}} r_{\rm{turn}}^{\rm{pitch}}
\end{equation}
where $r_{\rm{turn}}^{\rm{roll}}$ and $r_{\rm{turn}}^{\rm{pitch}}$ are the roll reward and pitch reward, respectively. 
$\omega_{\rm{turn}}^{\rm{roll}}$ and $\omega_{\rm{turn}}^{\rm{pitch}}$ are the corresponding weights of roll reward and pitch reward, respectively.}

\textcolor{black}{
The roll reward $r_{\rm{turn}}^{\rm{roll}}$ guides the aircraft to perform a large roll angle to make a steep turn and can be expressed as
% The roll angle was guided to approach a target angle to encourage aircraft to generate higher normal load factors, close to extreme overload.
\begin{equation}
    \label{eq: steep turn roll reward}
        r_{\rm{turn}}^{\rm{roll}}  =  e^{-\displaystyle\frac{|\phi - {\rm{sgn}}((\Delta\bm{x}_m \times \dot{\bm{x}}_{a}) \cdot \bm{u}_{\rm{up}}) \cdot \phi^*|}{\tau_{\rm{turn}}^{\rm{roll}}}}
\end{equation}
where 
$\phi$ denotes the roll of the aircraft and 
$\phi^*$ denotes the target roll. 
$\tau_{\rm{turn}}^{\rm{roll}}$ is an adjustable parameter.} 
\textcolor{black}{
% The circling direction coefficient 
${\rm{sgn}}((\Delta\bm{x}_m \times \dot{\bm{x}}_{a}) \cdot \bm{u}_{\rm{up}})$ determines a left or right turn based on the position of the missile with respect to the aircraft, represented by $\Delta\bm{x}_m$, and the velocity of the aircraft, represented by $\dot{\bm{x}}_{a}$.
% \begin{equation}
%     \label{eq:circling direction coefficient}
%     % d = \text{sign}\left( (\dot{\bm{x}}_{a} \times \Delta\bm{x}_m) \cdot \bm{u}_{\rm{up}} \right)
%     d = 
%     \begin{cases} 
%     1 & \text{if } (\dot{\bm{x}}_{a} \times \Delta\bm{x}_m) \cdot \bm{u}_{\rm{up}} > 0, \\
%     -1 & \text{else}.
%     \end{cases}
% \end{equation}
% $\dot{\bm{x}}_{a}$ is the velocity of the aircraft. $\Delta\bm{x}_m$ represents the aircraft-to-obstacle vector from the position of the missile to the position of the aircraft. 
$\bm{u}_{\rm{up}}$ is the upward unit vector.}

\textcolor{black}{
The pitch reward $r_{\rm{turn}}^{\rm{pitch}}$ tends to maintain the altitude of the aircraft and can be expressed as
% The pitch angle is encouraged to remain close to zero, ensuring the aircraft maintains stable flight during circling.
\begin{equation}
    \label{eq: steep turn ptich reward}
        r_{\rm{turn}}^{\rm{pitch}}  = e^{-\displaystyle\frac{|\theta - \theta^*|}{\tau_{\rm{turn}}^{\rm{pitch}}}}
\end{equation}
where 
% $\theta$ denotes the pitch of the aircraft
$\theta^*$ denotes the target pitch for altitude keeping. 
$\tau_{\rm{turn}}^{\rm{pitch}}$ is an adjustable parameter.} 
    
    % \begin{equation}
    % \label{eq: circling reward}
    %     \text{r}_{circ}  =  \omega^s_{\phi}e^{-\displaystyle\frac{|\phi - d \cdot \phi^*|}{\tau_{1}}} + \omega^s_{\theta}e^{-\displaystyle\frac{|\theta|}{\tau_{2}}}
    % \end{equation}

% where $\phi, \theta$ denotes the roll and pitch angle of the aircraft, and $\phi^*$ denotes the desired roll angle, $\tau_{1}$, $\tau_{2}$ are scaling parameters controlling sensitivity, 

% TODO: why target roll = 85?   %放在实验部分说

% if the obstacle initially appeared on the left side of the aircraft, the aircraft tends to counterclockwise circling.

% otherwise, if the obstacle appeared on the right side, the aircraft tends to clockwise circling.

\subsubsection{Short distance policy}

% In this article, the obstacle is modeled as a high-speed aggressive pursuer. 
% Thus, a simple circling maneuver is no longer enough: the aircraft cannot force the obstacle to exceed its overload limit, let alone achieve complete evasion. 
% Considering that the obstacle is guided by 3-dimension proportional navigation (3DPN), where its acceleration is directly proportional to the line-of-sight (LOS) rate, 
% the LOS rate is introduced into the reward function.

\textcolor{black}{
A short distance policy is trained based on the steep turn policy 
and the reward function with an additional LOS term.
The reward function can be expressed as
\begin{equation}
    \label{eq: short distance reward}
       r_{\rm{short}}  = \omega_{\rm{turn}}^{\rm{roll}} r_{\rm{turn}}^{\rm{roll}} + \omega_{\rm{turn}}^{\rm{pitch}} r_{\rm{turn}}^{\rm{pitch}} + \omega_{\rm{short}}^{\rm{LOS}} r_{\rm{short}}^{\rm{LOS}}
\end{equation}
where $r_{\rm{short}}^{\rm{LOS}}$ is the LOS reward and $\omega_{\rm{short}}^{\rm{LOS}}$ is the corresponding weight of the LOS reward.
This study proposes to learn a short distance policy that tends to avoid a missile by horizontal aggress maneuvers. In this way, the short distance policy is still applicable even if the altitude of the aircraft is not high.
Thus, the LOS reward is designed based on the assumption that the pitch reward can guide the aircraft to maneuver in a horizontal plane roughly, and then $\Delta\bm{x}_m$ is approximately horizontal. 
$r_{\rm{short}}^{\rm{LOS}}$ can be expressed as
\begin{equation}
    \label{eq: short distance LOS reward}
        r_{\rm{short}}^{\rm{LOS}}  = \tanh \left(\frac{{\rm{sgn}}((\Delta\bm{x}_m \times \dot{\bm{x}}_{a}) \cdot \bm{u}_{\rm{up}}) \cdot \dot{\lambda}^{s}}{\tau_{\rm{short}}^{\rm{LOS}}}\right)
\end{equation}
${\rm{sgn}}((\Delta\bm{x}_m \times \dot{\bm{x}}_{a}) \cdot \bm{u}_{\rm{up}})$ determines if the missile is on the left or right side of the aircraft, and has been included in (\ref{eq: steep turn roll reward}) also.
For instance, if a missile is on the right side of the aircraft, ${\rm{sgn}}((\Delta\bm{x}_m \times \dot{\bm{x}}_{a}) \cdot \bm{u}_{\rm{up}})$ is positive.
$\tau_{\rm{short}}^{\rm{LOS}}$ is an adjustable parameter.
$\dot{\lambda}^{s}$ is a signed LOS rate defined as 
\begin{equation}
    \dot{\lambda}^{s} = {\rm{sgn}}((\bm{u}_{m}(t-1) \times \bm{u}_{m}(t)) \cdot \bm{u}_{\rm{up}}) \cdot \frac{||\bm{u}_{m}(t) - \bm{u}_{m}(t-1)||}{\Delta t}
\end{equation}
where $\bm{u}_{m} = \frac{\Delta\bm{x}_m}{||\Delta\bm{x}_m||}$ is the unit vector of the position of the missile with respect to the aircraft.
For instance, if the aircraft make a missile on the right side and moving right further, ${\rm{sgn}}((\bm{u}_{m}(t-1) \times \bm{u}_{m}(t)) \cdot \bm{u}_{\rm{up}})$ is positive and thus the signed LOS rate is positive.  
% The unit vector of the position of the aircraft with respect to the missile is defined as $\bm{u}_{\rm{up}}_a = \frac{\Delta\bm{x}_m}{||\Delta\bm{x}_m||}$.
% The sign is defined as
% \begin{equation}
%     {\rm{sgn}}_{\dot{\lambda}^{s}} = {\rm{sgn}}((\bm{u}_{\rm{up}}_{a, t-1} \times \bm{u}_{\rm{up}}_{a,t}) \cdot u)
% \end{equation}
% where $\dot{\lambda}$ the line-of-sight (LOS) rate.
It should be noted that this study proposes the signed LOS rate $\dot{\lambda}^{s}$ to guide the short distance policy to learn to increase the LOS of the missile for the benefit of the aircraft, rather than maximizing the magnitude of the LOS rate according to \cite{Yan2024AGame,Li2022HierarchicalAvoidance}.
}

% By encouraging the aircraft to maneuver in a way that increases the obstacle’s overload,  the pursuer is more likely to miss the target, thereby improving the survivability of the aircraft.

% \begin{equation} 
% \label{eq:roll reward}
% \begin{aligned}    
% r_{improve} = &  \omega^s_{\phi}e^{-\displaystyle\frac{|\phi - d \cdot \phi^*|}{\tau_{1}}} 
%               + \omega^s_{\theta}e^{-\displaystyle\frac{|\theta|}{\tau_{2}}} \\
%              & + \omega^s_{\lambda}\tanh\left(-\frac{\dot{\lambda}}{\tau_{3}}\right)
% \end{aligned}     
% \end{equation}

% and the new reward function is designed as follow:

% Since the raw LOS rate $\dot{\lambda}{raw}$ changes rapidly, an exponential moving average (EMA) smoothing filter is introduced to obtain a stable LOS rate. The smoothed LOS rate $\dot{\lambda}{t}$ is defined as:

% \begin{equation}
%     \label{eq: circling reward}
%         \dot{\widetilde\lambda}_t = \alpha \dot{\lambda}_{raw,t} + (1-\alpha) \dot{\widetilde\lambda}_{t-1}
%     \end{equation}
% where $\alpha \in (0,1)$ is the smoothing coefficient controlling the degree of temporal smoothing.

\subsection{Policy switching} \label{sec:method policy swithing}

% \hl{tests show the importance of distance and speed}

% \hl{distance analysis shows the importance of roll}

% \hl{small angle away policy: 1) small angle, 2) roll, 3) pitch, 4) speed}

% \hl{large angle away policy: 1) turn, 2) pitch, 3) zero roll at 8 km, 4) speed}

% To effectively cope with aggressive dynamic obstacles, a two-stage strategy with switching is required, since a single circling maneuver is not sufficient in all scenarios.

% Experimental results show that the aircraft achieves the highest escape success ratio when circling begins at a special distance $d_c$. 
% This indicates that there exists an optimal switching point, rather than circling from the beginning.

\textcolor{black}{
With different initial conditions, it is hard for the short distance policy to have the same performance in success ratio.
The initial distance, the initial velocity and roll of the aircraft, the initial velocity of the missile, the initial azimuth of the missile etc. can make a big difference.
This study proposes to investigate the performance of the short distance policy with different initial conditions, including but not limited to the above-mentioned conditions.
The initial conditions that can lead to the best performance of the short distance policy should be the target of the small azimuth policy and large azimuth policy, as well as the conditions for switching from the small azimuth policy or large azimuth policy to the short distance policy.}

% The following factors are all considered:
% (1) \textbf{Distance} is the dominant factor. Starting circling at around a certain distance yields the highest success ratio, and this distance is thus chosen as the switching threshold.
% (2) \textbf{Aircraft speed} also plays a significant role. The highest initial speed results in a more successful result.
% (3) \textbf{Obstacle’s speed and initial angle} were tested but found to have limited influence on the success ratio.

% So in long distance stage, the aircraft is encouraged to accelerate and separate from the obstacle, building a velocity advantage for the next phase;
% when the distance threshold is reached, the aircraft switches to the short distance strategy;
% In the short-distance stage, the aircraft performs high-overload circling maneuvers to evade the obstacle.
% The short strategy can avoid obstacles under two conditions:

% (1)there must be sufficient distance between the aircraft and the obstacle to allow for evasion maneuvers.

% (2) aircraft's radar must be capable of detecting obstacles at this threshold.(add reference)

\subsection{Small azimuth policy}  \label{sec:method small azimuth policy}

% In the long-distance stage and at small angle, 
% the main purpose of the aircraft is to moving away from the obstacle, maintain the stability of the roll and pitch angles, and accelerate, with acceleration aimed at improving the success ratio of short-distance strategies.

\textcolor{black}{
If the conditions of the short distance policy are not satisfied and the missile incomes from a small azimuth direction, the small azimuth policy will be activated and guide the aircraft to move away to wear out the effective time of the missile as possible and achieve beneficial conditions for the short distance policy. 
According to this study, the objectives of the small azimuth policy include 
(1) ensuring a small azimuth of the missile,
(2) maintaining the altitude and attitude of the aircraft,
(3) making the aircraft move in the opposite direction from the missile, 
and (4) accelerating the aircraft.}

% \hl{At this stage, the primary goals are:
% % (1) maintaining the angle between the vector from the missile to the aircraft within a safe range: Keep the angle between –30° and 30°.
% % (2) Maintain attitude stability: Constrain pitch and roll angles to ensure a steady flight.
% % (3) Reduce the relative angle between the aircraft and the obstacle.
% % (4) Accelerate: Increase speed to ensure a higher success ratio when transitioning to the short-distance phase.
% }

\textcolor{black}{
The reward function of the small azimuth policy can be expressed as
\begin{equation}\label{eq: small azimuth reward}
    \begin{aligned}
    r_{\rm{small}} =& r_{\rm{small}}^{\rm{roll}} + r_{\rm{small}}^{\rm{pitch}} + r_{\rm{small}}^{\rm{azimuth}} + r_{\rm{small}}^{\rm{vel}} +  \\
                & c_{\rm{small}}^{\rm{roll}} + c_{\rm{small}}^{\rm{pitch}} + c_{\rm{small}}^{\rm{azimuth}} + c_{\rm{small}}^{\rm{vel}}
    \end{aligned} 
\end{equation}
where $r_{\rm{small}}^{\rm{roll}}$ and $c_{\rm{small}}^{\rm{roll}}$ are the roll reward and the roll constraint, respectively.
$r_{\rm{small}}^{\rm{roll}}$ and $c_{\rm{small}}^{\rm{roll}}$ can be expressed as    
\begin{equation}
    r_{\rm{small}}^{\rm{roll}} =e^{-\displaystyle \frac{|\phi|}{\tau_{\rm{small}}^{\rm{roll}}}}
\end{equation}
\begin{equation}
    c_{\rm{small}}^{\rm{roll}} =
    \begin{cases}
        P_{\rm{small}}^{\rm{roll}}, & \text{if } |\phi| > \phi_{\rm{small}}^{\rm{max}} \\
        0,   & \text{otherwise}
    \end{cases}
\end{equation}
where 
% $\phi$ denotes the roll of the aircraft, 
$\tau_{\rm{small}}^{\rm{roll}}$ is an adjustable parameter.
$\phi_{\rm{small}}^{\rm{max}}$ denotes the upper boundary of the magnitude of roll set for the small azimuth policy. 
$P_{\rm{small}}^{\rm{roll}}$ is a negative constant as punishment.}

\textcolor{black}{
$r_{\rm{small}}^{\rm{pitch}}$ and $c_{\rm{small}}^{\rm{pitch}}$ are the pitch reward and the pitch constraint, respectively.
$r_{\rm{small}}^{\rm{pitch}}$ and $c_{\rm{small}}^{\rm{pitch}}$ can be expressed as
\begin{equation}
    r_{\rm{small}}^{\rm{pitch}} = e^{-\displaystyle \frac{|\theta - \theta^*|}{\tau_{\rm{small}}^{\rm{pitch}}}}
\end{equation}
\begin{equation}
    c_{\rm{small}}^{\rm{pitch}} =
    \begin{cases}
        P_{\rm{small}}^{\rm{pitch}}, & \text{if } |\theta| > \theta_{\rm{small}}^{\rm{max}} \\
        0,   & \text{otherwise}
    \end{cases}
\end{equation}
where 
% $\theta$ denotes the pitch angle of the aircraft, 
$\tau_{\rm{small}}^{\rm{pitch}}$ is an adjustable parameter.
$\theta_{\rm{small}}^{\rm{max}}$ denotes the upper boundary of the magnitude of pitch set for the small azimuth policy. 
$P_{\rm{small}}^{\rm{pitch}}$ is a negative constant as punishment.}

\textcolor{black}{
$r_{\rm{small}}^{\rm{azimuth}}$ and $c_{\rm{small}}^{\rm{azimuth}}$ are the azimuth reward and the azimuth constraint, respectively.
$r_{\rm{small}}^{\rm{azimuth}}$ and $c_{\rm{small}}^{\rm{azimuth}}$ can be expressed as
\begin{equation}
    r_{\rm{small}}^{\rm{azimuth}} = e^{-\displaystyle\frac{|\Psi|}{\tau_{\rm{small}}^{\rm{azimuth}}}}
\end{equation}
\begin{equation}
    c_{\rm{small}}^{\rm{azimuth}} =
    \begin{cases}
        P_{\rm{small}}^{\rm{azimuth}}, & \text{if } |\Psi| > \Psi_{\rm{small}}^{\rm{max}} \\
        0,   & \text{otherwise}
    \end{cases}
\end{equation}
where $\Psi$ denotes the azimuth of the missile.  
$\tau_{\rm{small}}^{\rm{azimuth}}$ is an adjustable parameter.
$\Psi_{\rm{small}}^{\rm{max}}$ represents the upper boundary of the magnitude of azimuth set for the small azimuth policy. 
$P_{\rm{small}}^{\rm{azimuth}}$ denotes a negative constant as punishment.}

\textcolor{black}{
$r_{\rm{small}}^{\rm{vel}}$ and $c_{\rm{small}}^{\rm{vel}}$ are the velocity reward and the velocity constraint, respectively.
$r_{\rm{small}}^{\rm{vel}}$ and $c_{\rm{small}}^{\rm{vel}}$ can be expressed as
\begin{equation}
    r_{\rm{small}}^{\rm{vel}} = \tanh(\frac{||\dot{\bm{x}}_{a}||-\dot{x}_{a}^{\rm{ref}}}{\tau_{\rm{small}}^{\rm{vel}}})
\end{equation}
\begin{equation}
    c_{\rm{small}}^{\rm{vel}} =
    \begin{cases}
        P_{\rm{small}}^{\rm{vel}}, & \text{if} ||\dot{\bm{x}}_{a}|| > \dot{x}_{a}^{\rm{max}} \text{ or } ||\dot{\bm{x}}_{a}|| < \dot{x}_{a}^{\rm{min}} \\
        0,   & \text{otherwise}
    \end{cases}
\end{equation}
where $\dot{\bm{x}}_{a}$ denotes the velocity of the aircraft. 
$\dot{x}_{a}^{\rm{ref}}$ denotes a reference velocity used to adjust the effective range of tanh. 
$\dot{x}_{a}^{\rm{max}}$ and $\dot{x}_{a}^{\rm{min}}$ denote the upper and lower boundaries of the aircraft, respectively. 
$P_{\rm{small}}^{\rm{vel}}$ is a negative constant as punishment.}

\subsection{Large azimuth policy}  \label{sec:method large azimuth policy}

% In the long-distance stage, the main purpose of the aircraft is to move away from the obstacle while accelerating.
% adjusting its flight state to create favorable conditions for the short-distance phase.

\textcolor{black}{
If the conditions of the short distance policy are not satisfied and the missile incomes from a large azimuth direction, the large azimuth policy will be activated and guide the aircraft to turn to the opposite direction from the missile for moving away and achieve beneficial conditions for the short distance policy.
If the large azimuth policy makes the aircraft achieve a small azimuth and the conditions of the short distance policy are not satisfied, the large azimuth policy will switch to the small azimuth policy.
If the conditions of the short distance policy are satisfied, the large azimuth policy will switch to the short distance policy.}

% Experiments show that aircraft will achieve a higher success ratio if the aircraft has a higher speed and a smaller roll angle. Therefore, in the long-distance stage, the aircraft is encouraged to reduce its roll angle to zero, and accelerate to build an advantage in the short-distance stage.

\textcolor{black}{
According to this study, the objectives of the large azimuth policy include 
(1) reducing the azimuth of the missile, 
(2) maintaining the altitude of the aircraft, 
(3) reducing the roll of the aircraft to around 0 deg quickly if the distance condition of the short distance policy is about to be satisfied, 
and (4) accelerating the aircraft.}

% \hl{When the direction of the missile from the aircraft in a horizontal plane is large, the large azimuth policy focuses on the following goals:
% 1) Reduce the direction, thereby delaying potential collision.
% (2) Maintain attitude stability: Constrain pitch deviations to get a  steady flight.
% (3) Once the distance between the aircraft and the obstacle less than 8500 m, maintaining a near-zero roll angle becomes more important, ensuring the high success ratio of the short-distance evasion strategy.
% (4) Acceleration: Accelerate to improve the probability of successful escape during the short-distance stage.}

\textcolor{black}{
The reward function of the large azimuth policy can be expressed as
\begin{equation} 
    \label{eq: large azimuth reward fucntion}
    \begin{aligned}
    r_{\rm{large}} = & r_{\rm{large}}^{\rm{roll}} + r_{\rm{large}}^{\rm{pitch}} + r_{\rm{large}}^{\rm{vel}} +  \\
                & c_{\rm{large}}^{\rm{roll}} + c_{\rm{large}}^{\rm{pitch}} + c_{\rm{large}}^{\rm{vel}}
    \end{aligned} 
\end{equation}
% (1) Roll reward: Guides the aircraft to align its velocity direction with the line-of-sight (LOS) vector from the obstacle to the aircraft, thereby extending the time to potential collision.
% (1) Roll reward: 
% Guides the aircraft to move away from obstacles in the correct direction, thereby extending the time to potential collision.
where $r_{\rm{large}}^{\rm{roll}}$ and $c_{\rm{large}}^{\rm{roll}}$ are the roll reward and the roll constraint, respectively.
$r_{\rm{large}}^{\rm{roll}}$ can be expressed as 
\begin{equation}
r_{\rm{large}}^{\rm{roll}} =
\begin{cases}
    r_{\rm{large}}^{\rm{far}}, & || \Delta\bm{x}_m || > d_c \\
    r_{\rm{large}}^{\rm{close}}, & || \Delta\bm{x}_m ||  \le d_c
\end{cases}
\label{eq: large azimuth roll reward}
\end{equation}
where $d_c$ is a critical distance. 
If the distance between the aircraft and the missile is larger than the critical distance (i.e., $ || \Delta\bm{x}_m || > d_c$), the roll reward is
\begin{equation}
    r_{\rm{large}}^{\rm{far}} =
    e^{\displaystyle - \frac{|\phi - {\rm{sgn}}((\Delta\bm{x}_m \times \dot{\bm{x}}_{a}) \cdot \bm{u}_{\rm{up}}) \cdot \phi^*|}{\tau_{\rm{large}}^{\rm{far}}}}
    \label{eq: large azimuth roll far reward}
\end{equation}
where 
% $\phi$ denotes the roll angle of the aircraft.
$\phi^*$ denotes the target roll.
$\tau_{\rm{large}}^{\rm{far}}$ is an adjustable parameter.
If the distance between the aircraft and the missile is smaller than the critical distance (i.e., $ || \Delta\bm{x}_m || \le d_c$), the roll reward is given by
\begin{equation}
    r_{\rm{large}}^{\rm{close}} =
    \begin{cases}
        P_{\rm{large}}^{\rm{roll}}, & |\phi| > \phi_{\rm{large}}^{\rm{short}} \\
        R_{\rm{large}}^{\rm{roll}}, & |\phi| \le \phi_{\rm{large}}^{\rm{short}}
    \end{cases}
    \label{eq: large azimuth roll close reward}
\end{equation}
% $\phi$ denotes the roll of the aircraft,
% \hl{$\phi^{lim}$ represents the boundary of roll.}
% $c_{\rm{large}}^{\rm{roll}}$ is 
where $\phi_{\rm{large}}^{\rm{short}}$ represents the upper boundary of the magnitude of roll set for reducing the roll of the aircraft to achieve a beneficial roll condition for the short distance policy.
$P_{\rm{large}}^{\rm{roll}}$ is a negative constant as punishment. 
$R_{\rm{large}}^{\rm{roll}}$ is a positive constant as a reward.}
\textcolor{black}{
The roll constraint $c_{\rm{large}}^{\rm{roll}}$ can be expressed as
\begin{equation}
    c_{\rm{large}}^{\rm{roll}} =
    \begin{cases}
        P_{\rm{large}}^{\rm{roll}}, & \text{if } |\phi| > \phi_{\rm{large}}^{\rm{max}} \\
        0,   & \text{otherwise}
    \end{cases}
\end{equation}
$\phi_{\rm{large}}^{\rm{max}}$ denotes the upper boundary of the magnitude of roll set for the large azimuth policy.}

% (2) Pitch reward: Constrains the pitch angles to maintain stable flight attitude. The reward function encourages the pitch angle to remain close to zero, while constraining the pitch angle in a safe range.

\textcolor{black}{
$r_{\rm{large}}^{\rm{pitch}}$ and $c_{\rm{large}}^{\rm{pitch}}$ are the pitch reward and the pitch constraint, respectively.
$r_{\rm{large}}^{\rm{pitch}}$ and $c_{\rm{large}}^{\rm{pitch}}$ can be expressed as
\begin{equation}
    \label{eq:long distance reward} 
        r_{\rm{large}}^{\rm{pitch}} = e^{-\displaystyle\frac{|\theta - \theta^*|}{\tau_{\rm{large}}^{\rm{pitch}}}}
\end{equation}
\begin{equation}
    c_{\rm{large}}^{\rm{pitch}} =
    \begin{cases}
        P_{\rm{large}}^{\rm{pitch}}, & \text{if } |\theta| > \theta_{\rm{large}}^{\rm{max}} \\
        0,   & \text{otherwise}
    \end{cases}
\end{equation}
where 
% $\theta$ denotes the pitch angle of the aircraft. 
$\theta_{\rm{large}}^{\rm{max}}$ denotes the upper boundary of the magnitude of pitch set for the large azimuth policy.
$\tau_{\rm{large}}^{\rm{pitch}}$ is an adjustable parameter.
$P_{\rm{large}}^{\rm{pitch}}$ is a negative constant as punishment.}

% (3) Speed reward: Encourages the aircraft to increase its speed, since a higher velocity provides greater escape potential in short-distance stage, while constraining the aircraft’s speed within a  safe operating range to avoid excessive or insufficient velocity.

\textcolor{black}{
$r_{\rm{large}}^{\rm{vel}}$ and $c_{\rm{large}}^{\rm{vel}}$ are the velocity reward and the velocity constraint, respectively.
$r_{\rm{large}}^{\rm{vel}}$ and $c_{\rm{large}}^{\rm{vel}}$ can be expressed as
\begin{equation}
    r_{\rm{large}}^{\rm{vel}} = \tanh(\frac{||\dot{\bm{x}}_{a}|| - \dot{x}_{a}^{\rm{ref}}}{\tau_{\rm{large}}^{\rm{vel}}})
\end{equation}
\begin{equation}
    c_{\rm{large}}^{\rm{vel}} =
    \begin{cases}
        P_{\rm{large}}^{\rm{vel}}, & \text{if } ||\dot{\bm{x}}_{a}|| > \dot{x}_{a}^{\rm{max}} \text{ or } ||\dot{\bm{x}}_{a}|| < \dot{x}_{a}^{\rm{min}} \\
        0,   & \text{otherwise}
    \end{cases}
\end{equation}
where
% where $v$ denotes the velocity of the aircraft. 
% $\dot{x}_{a}^{ref}$ denotes the target velocity. 
$\tau_{\rm{large}}^{\rm{vel}}$ is an adjustable parameter.
% $\dot{x}_{a}^{\rm{max}}$ and $\dot{x}_{a}^{\rm{min}}$ denotes the upper and lower boundaries, respectively.
$P_{\rm{large}}^{\rm{vel}}$ is a negative constant as punishment.}

% The overall form of the reward function is as follows:

%     \begin{equation}
%     \label{eq:long distance reward} 
%     \begin{aligned}   
%     r_{away} =& - \omega^l_\Psi \cdot (\Psi - \Psi^*) \\
%                 &+ \omega^l_v \cdot tanh(\frac{v - v^*}{\tau_4})  \\
%                 &+ \omega^l_\theta \cdot e^{-\displaystyle\frac{|\theta|}{\tau_5}}   
%                 + \omega^l_\phi \cdot e^{-\displaystyle\frac{|\phi|}{\tau_6}} \\
%                 & + f_{roll} + f_{v} + f_{speed}
%     \end{aligned}
% \end{equation}

% $\Psi$ denotes the angle between the aircraft’s velocity vector and the line-of-sight (LOS) vector from the obstacle to the aircraft, with $\Psi^*$ representing the desired angle. (zero angle) $v$ is the aircraft’s speed, while $v^*$ denotes the reference speed. $\theta$ and $\phi$ denote the pitch and roll angles, and $f_{roll}$ and $f_{v}$ represent penalty terms when the roll angle or velocity exceeds the range.

% \subsection{? Behavior Cloning}

% Introduce bc. 

% \hl{
% Loss function of BC

% how to handle data around the threshold

% }

%%%%%%%%%%%%%%%%%%%%%%%%%%%%%%%%%%%%%%%%%%

% \noindent \rule{1\linewidth}{0.5mm}

\section{Experiments \label{sec:experiments}}

\textcolor{black}{
To evaluate the effectiveness of the developed evasion strategy in addressing a missile, close-to-real experiments
% on the training and testing of the proposed evasion strategy 
are conducted. 
The experiments include the training and analysis of the three policies of the proposed evasion strategy, the training of a baseline RL-based evasion strategy, comparison of the proposed evasion strategy and existing evasion strategies, and tests of the proposed evasion strategy in addressing different navigation laws.
}

\textcolor{black}{
A personal computer with a 3.7 GHz CPU, 32 GB memory, and an RTX 3080 Ti GPU is used to conduct the experiments.
This study runs the simulator at a frequency of 200 Hz to improve simulation accuracy.
The aircraft is simulated based on the JSBSim simulator with the F-16 aircraft model \cite{Berndt2004JSBSim:C++}.
The JSBSim simulator is a flight simulator that covers the aircraft's physical characteristics, aerodynamic model, flight control systems, and propulsion, and has been used in several academic studies \cite{Zhou2025EnhancingStrategy,Scukins2024DeepEvasion,Pope2023HierarchicalTrials}.
To unleash maximum maneuver capability without being limited by a flight controller, the control inputs are designed to act on the elevator, aileron, rudder, and throttle of the aircraft.}

\textcolor{black}{
This study simulates the movement of a missile in three-dimensional space. 
A horizontal navigation law and a vertical navigation law (e.g., a PN law) are designed for the missile inspired by \cite{Tian2023MissileAircraft}.
The PN law can be expressed as
\begin{equation}
    \label{eq: 3dpn}
        a_m = N \cdot V_c \cdot \dot{\lambda}
\end{equation}
where $a_m$ is a missile commanded acceleration. 
$N$ is a navigation coefficient. 
$V_{c}$ is the closing velocity between the missile and the aircraft, 
$\dot{\lambda}$ is the LOS rate.
The yaw and pitch of the missile are updated according to the horizontal navigation law and a vertical navigation law in discrete form, inspired by \cite{Moran2005ThreeNavigation}.
The change of the yaw and pitch of the missile will be truncated proportionally, if the required overload is larger than the defined maximum overload of the missile.
Then, the position of the missile is updated based on the yaw, pitch, and velocity of the missile in discrete form, according to \cite{Yang2020NondominatedDogfight}.
}

\begin{figure}[!htb]
\centering
\includegraphics[scale=0.5]{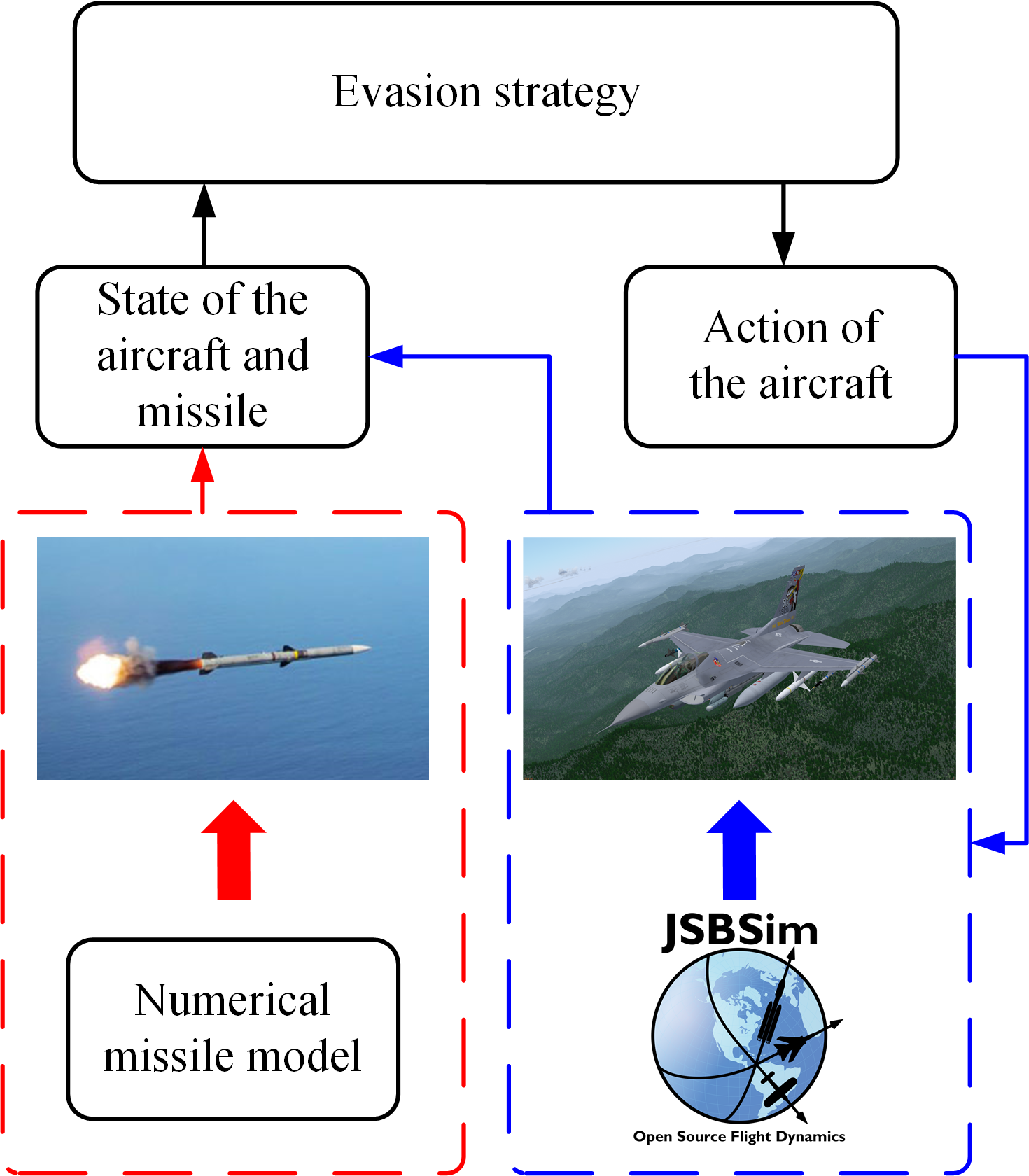}
\caption{\textcolor{black}{Framework of simulation based on the JSBSim simulator.}}
\label{fig: JSBSim flight simulator}
\end{figure}

% \subsection{Implementation of a missile}\label{sec:missile}

\subsection{Experiment setups}\label{sec:experiment setups}

% \hl{The reference frame used in the experiment is the North–East–Down (NED) coordinate system.}  

% \hl{background}

% \textcolor{black}{To unleash maximum maneuver capability without being limited by a flight controller, the control inputs are designed act on the elevator, aileron, rudder, and throttle of the aircraft.}

\textcolor{black}{
To simulate real-world scenarios of an aircraft evading a missile at a short distance, parameters are set for the scenarios in training and testing according to real parameters or references.
The initial velocity of the aircraft ranges from 280 m/s to 470 m/s. \cite{Geuzaine2003AeroelasticConditions,Hu2021ApplicationCombat}.
The initial altitude of the aircraft is set to range from 3000 m to 9000 m, inspired by the operational ceiling \cite{Kong2023HierarchicalCombat}.
The minimum altitude of the aircraft is set to 1000 m.
The initial heading of the aircraft is set between 0 deg and 360 deg.}

\textcolor{black}{A missile is set to approach the aircraft from an arbitrary horizontal direction. 
% In this study, the direction of the missile from the aircraft is define in a horizontal plane first from the negative direction of the y-axis (directly behind the aircraft). 
% The direction angle is positive if the missile is on the right side and vice verse.
The initial azimuth of the missile ranges from -180 deg to 180 deg randomly. 
% When the missile is directly behind the aircraft, the angle is defined as 0 deg.
It should be emphasized that the azimuth of the missile with respect to the aircraft is defined based on the opposite direction of the heading of the aircraft.
The initial elevation of the missile ranges from -15 deg to 15 deg in the initial state.
The initial distance between the aircraft and the missile is set within the range of 5–15 km to simulate a short distance scenario, inspired by \cite{Yan2022AGuidance,Alqudsi2018AAlgorithms,Yang2020TacticalCombat,Neele2005Two-colorSensors}.
The velocity of the missile ranges from 800 m/s and 1400 m/s according to \cite{Zhao2025DeepLimit,Moran2005ThreeNavigation}. 
The velocity of the missile is set to be constant, aiming to achieve an evasion strategy that can address a challenging missile with the dual pulse technique.}
% 第一篇中将被拦截方（也就是导弹）的速度设置在1200m/s
% 第二篇中将障碍物速度设置在1000m/s
\textcolor{black}{
The maximum overload of the missile ranges from 40 g to 50 g according to \cite{Yuan2023HierarchicalLearning,Wang2024Linear-quadraticEngagement,Li2022HierarchicalAvoidance}.
Unless otherwise specified, the missile tracks the aircraft following the PN law \cite{Moran2005ThreeNavigation}, with the navigation coefficient randomly ranging from 3 and 5 according to} \cite{Alqudsi2018AAlgorithms}.
\textcolor{black}{The maximum duration of a scenario is set to 25 s based on the operation time of a missile} \cite{Yang2020NondominatedDogfight}.
\textcolor{black}{
The lethal zone of the missile is assumed to be a spherical region with a radius of 10 m \cite{Yan2022AGuidance,Yan2024AGame}.}

% \noindent \rule{1\linewidth}{0.5mm}

% 这两篇文章说明了导弹的杀伤半径，分别设置成了7m和10m。

% 第一篇有提到他们的飞机速度设置在0.7马赫到1.4马赫之间，约为240-470；
% 第二篇将飞机速度设置在280，但其飞机的速度范围是200-400

% Existing literature \cite{Two-colour infrared missile warning sensors} shows that 
% \hl{Infrared (IR) seekers' maximum detection distance can reach up to approximately 30 km, but in different conditions, like HTPB/AP-Al are employed or if plume temperature decreases, the detection threshold may drop to around 15 km.} 
% Other studies also set the initial aircraft-obstacle distance to around 10 km. 
% 这三篇文章分别将障碍物距离设置在了9-11km、10-12km、10km

% 文章中说明经典导引系数在2-6之间，最常见的是3-5之间

% 文献说明导弹的最大工作时间是30秒，文章是这样设置的：文章说导弹的是指是基于“通用短程空空导弹”，其中导弹发动机工作时间是5.2s，最大制导飞行时间（t max，指导弹从发射到失去制导能力、触发自毁程序的最长时间）设置为30秒

% \noindent \rule{1\linewidth}{0.5mm}

% TODO: %这里仅仅写了障碍物的10m爆炸范围，是否需要补充其他的终止条件？

% The lethal radius is defined as $d_{max}$. 
% When the distance between the UAV and the obstacle satisfies $d \leq d_{max}$, the obstacle is assumed to destroy the UAV \cite{Yan2022AGuidance} \cite{Yan2024AGame}, and the simulation is immediately terminated.

% \hl{Elevation, Azimuth}

\textcolor{black}{
Table \ref{tab: initial conditions} lists the conditions of the scenarios discussed in the experiments.}

\begin{table}[!t]
    \caption{\textcolor{black}{Conditions of the scenarios investigated in the experiments} \label{tab: initial conditions}}
    \renewcommand{\arraystretch}{1.2}
    \centering
    \begin{tabular}{|c|c|}
    \hline
    \textbf{Parameter}	& \textbf{Value}\\
    \hline
    Initial altitude of the aircraft		    & 3000 m to 9000 m \\
    \hline
    Initial velocity of the aircraft		& 280 m/s to 470 m/s \\
    % \hline
    % aircraft longitude	    & $1.442031^\circ$\\
    % \hline
    % aircraft latitude		    & $43.607181^\circ$\\
    % \hline
    % aircraft roll		        & $0 (deg)$\\
    % \hline
    % aircraft pitch		        & $2.4 (deg)$\\
    \hline
    Initial heading of the aircraft 		    & 0 deg to 360 deg \\
    \hline
    Initial azimuth of the missile		& -180 deg to 180 deg \\
    \hline
    Initial elevation of the missile		& -15 deg to 15 deg \\
    \hline
    Initial distance		    & 5000 m to 15000 m \\
    \hline
    Constant velocity of the missile		& 800 m/s to 1400 m/s \\
    \hline
    Maximum overload of the missile		    & 40 g to 50 g \\
    \hline
    Lethal radius of the missile		    & 10 m \\
    \hline
    Maximum duration		    & 25 s \\
    \hline
    \end{tabular}
\end{table}

%  and \hl{the aircraft's heading randomly assigned between $0^{\circ}$ and $360^{\circ}$.}and the vertical angle with respect to the airplane was uniformly distributed between $-$15° and +15°

\subsection{Implementation of the multi-stage RL-based strategy}

\begin{figure}[!htb]
    \centering
    \includegraphics[scale=0.5]{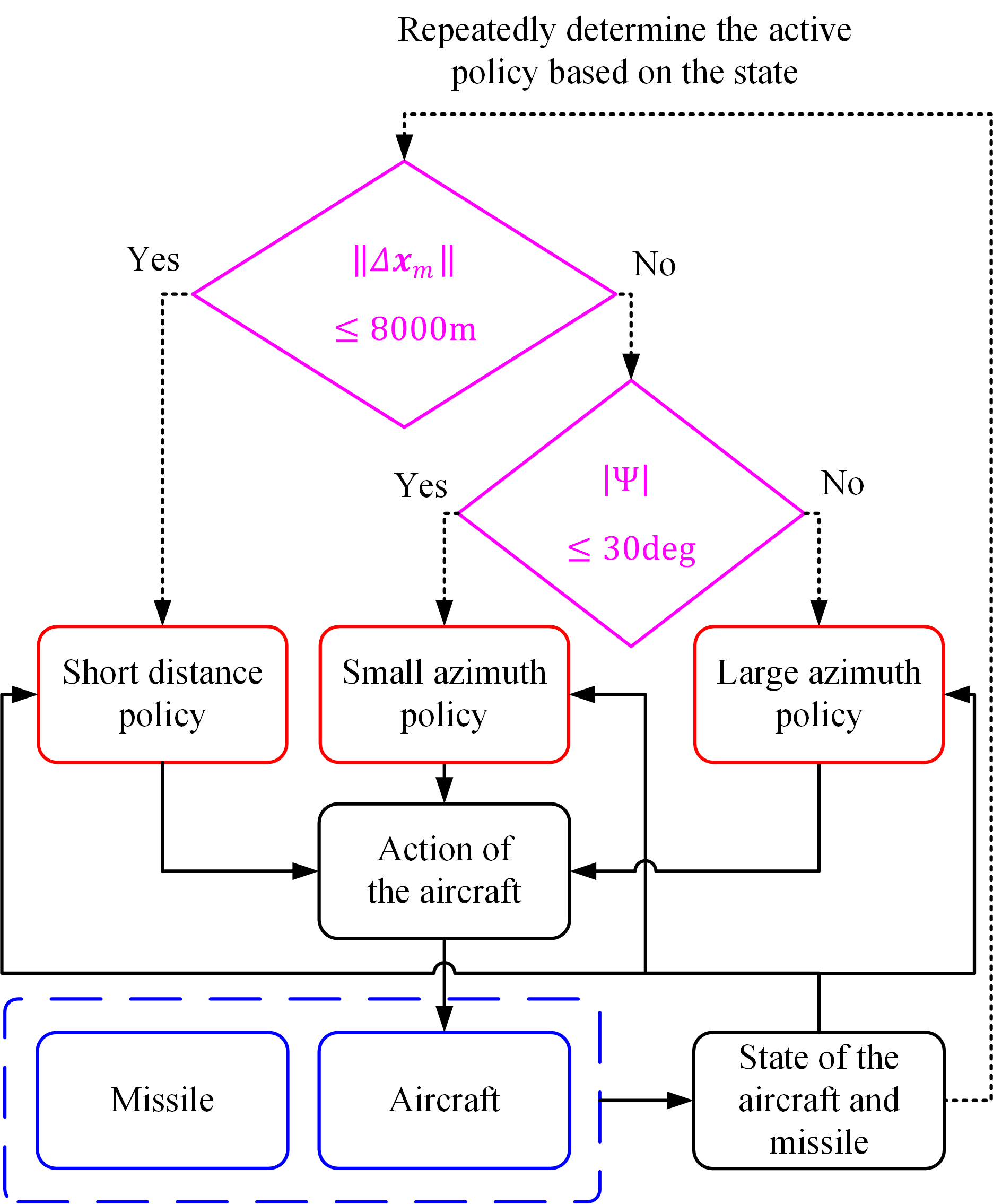}
    \caption{\textcolor{black}{Flow diagram of the obtained multi-stage RL-based evasion strategy.}}
    \label{fig: flow diagram experiment}
\end{figure}

\textcolor{black}{
This study applies the multi-stage RL-based evasion strategy to address a general scenario of an aircraft evading a missile, specified in Section \ref{sec:experiment setups}.
The implementation of the multi-stage RL-based evasion strategy is demonstrated in this section.
According to the multi-stage RL-based evasion strategy, this study first learns a short distance policy to address a missile that approaches the aircraft to a certain short distance.
The training of short distance policy is specified in Section \ref{sec:short distance policy}.
Then, the performance of the short distance policy is investigated thoroughly in Section \ref{sec:threshold}.
Based on the performance, this study determines that the short distance policy is activated if the distance between the aircraft and the missile is less than 8000 m.
The conditions that are beneficial for the short distance policy are determined and set as the objectives of the training of a small azimuth policy and a large azimuth policy.
The beneficial conditions include a high velocity and a small roll of the aircraft.
Next, this study set a threshold of 30 deg for the magnitude of the azimuth, and the training of a small azimuth policy and a large azimuth policy is specified in Section \ref{sec:small azimuth policy} and Section \ref{sec:large azimuth policy}, respectively.
The multi-stage RL-based evasion strategy implemented in the experiments can be expressed as Fig. \ref{fig: flow diagram experiment}.
}

\textcolor{black}{
The Proximal Policy Optimization (PPO) algorithm is used to train the three policies.
A policy based on the PPO algorithm has an actor network and a critic network. 
In this study, both the actor network and critic network have two hidden layers with 256 units.
Tanh is used as the activation function for all hidden layers and the output layer. 
The Adam optimizer is used for training.}

\textcolor{black}{
According to the problem statement and experiment setting, 
the state space of a policy is set to 
$\bm{s} = [\dot{x}_a, \dot{y}_a, \dot{z}_a, \theta, \phi, \psi, \Delta x_m, \Delta y_m, \Delta z_m, \Delta \dot{x}_m, \Delta \dot{y}_m, \Delta \dot{z}_m]$, 
where $[\dot{x}_a, \dot{y}_a, \dot{z}_a]$ reflects the velocity of the aircraft. 
$[\theta, \phi, \psi]$ denotes the roll, pitch, and heading of the aircraft. 
$[\Delta x_m, \Delta y_m, \Delta z_m]$ and $[\Delta \dot{x}_m, \Delta \dot{y}_m, \Delta \dot{z}_m]$ are the position and velocity of the missile with respect to the aircraft, respectively.
% Table \ref{tab:observation space range} lists the upper and lower bounds of the state space.
}
\textcolor{black}{
According to the experiment setting, the action space of a policy is set to $\bm{a}= [\delta_{e}, \delta_{a}, \delta_{r}, \delta_{t}]$. 
$[\delta_{e}, \delta_{a}, \delta_{r}, \delta_{t}]$ denotes normalized control inputs on the elevator, aileron, rudder, and throttle, respectively.
Table \ref{tab: state space and action space} lists the upper and lower bounds of the state space and action space.}
\textcolor{black}{The other hyperparameters used in this study are listed in Table \ref{tab: hyperparameters}.}

\begin{table}[!t]
    \caption{\textcolor{black}{Feasible range of the state space and action space}}
    \label{tab: state space and action space}
    \renewcommand{\arraystretch}{1.4}
    \centering
    \begin{tabular}{|c|c|c|}
    \hline
    \textbf{Parameter}	& \textbf{\makecell{Lower \\ bound}}  & \textbf{\makecell{Upper \\ bound}}\\
    \hline
    \makecell{$x$ element of the velocity of \\  the aircraft $\dot{x}_a$} & $-470$ m/s       & $470$ m/s\\
    \hline
    \makecell{$y$ element of the velocity of \\  the aircraft $\dot{y}_a$}		& $-470$ m/s       & $470$ m/s\\
    \hline
    \makecell{$z$ element of the velocity of \\  the aircraft $\dot{z}_a$}		& $-470$ m/s       & $470$ m/s\\
    \hline
    Roll of the aircraft $\phi$    & $-180$ deg   & $180$ deg \\
    \hline
    Pitch of the aircraft $\theta$      & $-90$ deg    & $90$ deg \\
    \hline
    Heading of the aircraft $\psi$      & $0$ deg      & $360$ deg \\
    \hline
    \makecell{$x$ element of the position of \\ the missile with respect to \\ the aircraft $\Delta x_m$}		& $-15000$ m    & $15000$ m\\
    \hline
    \makecell{$y$ element of the position of \\ the missile with respect to \\ the aircraft $\Delta y_m$}		& $-15000$ m    & $15000$ m\\
    \hline
    \makecell{$z$ element of the position of \\ the missile with respect to \\ the aircraft $\Delta z_m$}		& $-15000$ m    & $15000$ m\\
    \hline
    \makecell{$x$ element of the velocity of \\ the missile with respect to \\ the aircraft $\Delta \dot{x}_m$}	 & $-1870$ m/s   & $1870$ m/s\\
    \hline
    \makecell{$y$ element of the velocity of \\ the missile with respect to \\ the aircraft $\Delta \dot{y}_m$}		& $-1870$ m/s   & $1870$ m/s\\
    \hline
    \makecell{$z$ element of the velocity of \\ the missile with respect to \\ the aircraft $\Delta \dot{z}_m$}		& $-1870$ m/s   & $1870$ m/s\\
    \hline
    \makecell{Control input on \\ elevator $\delta_{e}$}		& -1   & 1\\
    \hline
    \makecell{Control input on \\ aileron  $\delta_{a}$}		& -1   & 1\\
    \hline
    \makecell{Control input on \\ rudder   $\delta_{r}$}		& -1   & 1\\
    \hline
    \makecell{Control input on \\ throttle $\delta_{t}$}		& 0    & 1\\
    \hline
    \end{tabular}
\end{table}

\begin{table}[!t]
    \caption{\textcolor{black}{Hyperparameters used in the training of policies \label{tab: hyperparameters}}}
    \renewcommand{\arraystretch}{1.2}
    \centering
    \begin{tabular}{|c|c|}
    \hline
    \textbf{Hyperparameter}	& \textbf{Value}\\
    \hline
    Discount factor		    & $0.99$\\
    \hline
    Learning rate		    & $3 \times 10^{-4}$\\
    \hline
    Maximum number of steps		& $7500$\\
    \hline
    % Max train steps		& $3 \times 10^{5}$\\
    % \hline
    Batch size		        & $1024$\\
    \hline
    \end{tabular}
\end{table}

\subsection{Training and validation of a short distance policy}\label{sec:short distance policy}

\textcolor{black}{
This study applies the curriculum learning technique to make a short-distance policy more trainable.
Since the short-distance policy needs to perform agile and aggressive maneuvers to evade a missile, the training of the short-distance policy includes two phases - 1) training a steep turn policy and 2) training a short-distance policy based on the steep turn policy.}

% \hl{steep turn policy} \cite{Yang2018EvasiveCombat}

% : first, a traditional circling policy is realized using reinforcement learning; 
% then, 
% the model is employed as a pre-trained model, an additional reward involved line-of-sight (LOS) rate term is added to enhance the strategy.

\textcolor{black}{
In both phases, scenarios with the same setting listed in Table \ref{tab: initial conditions} are adopted.
The initial position of the aircraft is set to [0, 0, $z_a$] (unit: m), where $z_a$ represents an altitude randomly selected from 3000 m to 9000 m.
The initial heading of the aircraft is randomly set ranging from 0 deg to 360 deg.
Both the initial roll and pitch of the aircraft are set to 0 deg.
The initial velocity of the aircraft is randomly selected from 280 m/s to 470 m/s.}

\textcolor{black}{
According to preliminary tests, it is shown that aggress evasion maneuvers of the aircraft are less effective, if the distance between the aircraft and the missile is larger than 12000 m. 
Thus, this study trains a short distance policy focusing on distance ranging from 5000 m to 12000 m, and thus the missile is set to has a random distance of 5000 m to 12000 m from the aircraft.
The azimuth of the missile is random.
The elevation of the missile randomly ranges from -15 deg to 15 deg.
The velocity of the missile is a constant ranging from 800 m/s to 1400 m/s.
The maximum overload of the missile is randomly selected from 40 g to 50 g.
The missile follows the PN law with a randomly selected navigation coefficient ranging from 3 to 5.
}
\textcolor{black}{The maximum number of steps of an episode is 7500.
An episode will be terminated if the aircraft is caught by the missile.
}

% \textcolor{black}{The above-mentioned initial conditions of the aircraft and the missile are applied to both the training of the circling policy and the training the avoidance policy.}

% Since the short duration of the short-distance avoidance phase, the obstacle’s velocity is assumed to remain constant throughout the engagement.

\subsubsection{Steep turn policy}\label{sec:steep turn policy}

% In the circling phase, the aircraft is required to achieve its maximum overload to evade the obstacle effectively. 
% Since the turning capability is strongly correlated with the load factor, 

\textbf{Training.}
\textcolor{black}{
This study learns a steep turn policy through training the policy to make an aircraft keep performing a large roll.
The target angle is set to 85 deg.
% At this attitude, the aircraft can sustain a load factor of about 6-7 g.
% Based on the reward function mentioned in the previous section, the target roll is set to 85 deg,
% while 
The target pitch is set to 0 deg to avoid losing too much altitude. 
According to (\ref{eq: steep turn reward}), the reward function of the steep turn policy is defined as
\begin{equation}
    \label{eq: steep turn reward experiment}
    \begin{aligned}
    r_{\rm{turn}} =& 0.5  e^{-\displaystyle\frac{|\phi - {\rm{sgn}}((\Delta\bm{x}_m \times \dot{\bm{x}}_{a}) \cdot \bm{u}_{\rm{up}}) \cdot 85|}{0.2}}\\
            &+ 0.5  e^{-\displaystyle\frac{|\theta|}{0.2}}
    \end{aligned}
\end{equation}
% ${\rm{sgn}}()$ determines the left or right turn based on the initial position of the aircraft and the missile and the initial velocity of the aircraft, aiming to force the missile to perform more aggressive maneuvers.
}

% The circling direction coefficient $d$ is determined based on the initial relative position of the obstacle, as shown in \ref{eq:circling direction coefficient}.
    
% \begin{equation}
%     \label{eq:circling direction coefficient}
%     % d = \text{sign}\left( (\dot{\bm{x}}_{a} \times \Delta\bm{x}_m) \cdot \bm{u}_{\rm{up}} \right)
%     d = 
%     \begin{cases} 
%     1 & \text{if } (\mathbf{\dot{\bm{x}}_{a}} \times \mathbf{\Delta\bm{x}_m}) \cdot \mathbf{\bm{u}_{\rm{up}}} > 0, \\
%     -1 & \text{else}.
%     \end{cases}
% \end{equation}

% This $d$ aims to control the turning direction of the aircraft.
% When $d=1$, the aircraft performs a clockwise circling maneuver, $d=1$, it performs a counterclockwise maneuver.
% This design increases the difficulty for the obstacle to maintain effective pursuit of the aircraft.

% \hl{
% Based on the above-mentioned training set-ups, the number of training steps is $4 \times 10^7$.
% Fig. \ref{fig:reward circling} shows the reward converges after 2000 episodes.}

\begin{figure}[!htb]
    \centering
    \includegraphics[scale=0.4, trim=0cm 0cm 0cm 0cm,clip]{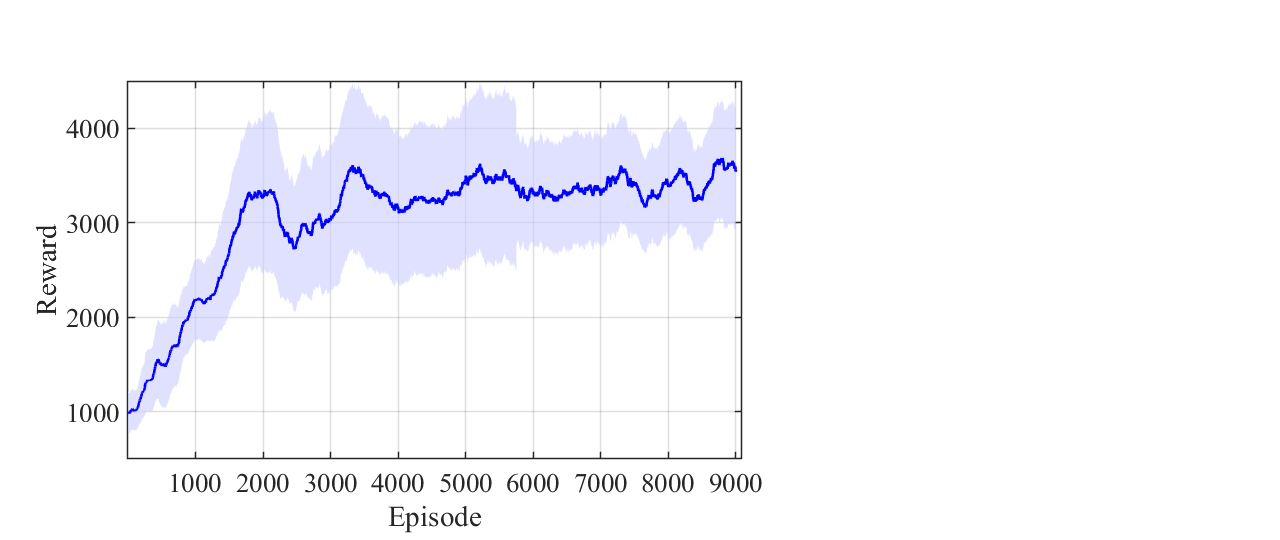}
    \caption{\textcolor{black}{Accumulated reward achieved in the training of a steep turn policy.}}
    \label{fig: turn policy reward}
\end{figure}

\textcolor{black}{
The learning process of the steep turn policy can be reflected by the achieved reward presented in Fig. \ref{fig: turn policy reward}. 
The learning process converges in 3000 episodes.
}

% \noindent \rule{1\linewidth}{0.5mm}

\textbf{Validation.}
\textcolor{black}{
To validate the performance of the learned steep turn policy, a test is performed.
The test is set to have a maximum number of 5000 steps (i.e., 25 s) and the missile is set to 21000 m from the aircraft, so as to enable the aircraft to finish a circle trajectory.
If the position of the missile $\Delta\bm{x}_m$ is beyond the state space, the position will be truncated.}

\textcolor{black}{
In the initial of the test, the position of the aircraft is set to [0, 0, 5097] (unit: m).
The heading of the aircraft is set to 0 deg.}
\textcolor{black}{
Both the roll and pitch of the aircraft are set to 0 deg.
The initial velocity of the aircraft is set to 280 m/s. 
The aircraft follows the learned steep turn policy.}
\textcolor{black}{
The initial missile is 21000 m from the aircraft.
The initial azimuth of the missile is 0 deg (i.e., behind the aircraft).
The initial elevation of the missile is -6.27 deg.
The velocity of the missile is constantly 800 m/s.
The maximum overload of the missile is randomly set to 42.97 g,
The missile follows the PN law with a randomly selected navigation coefficient of 3.82.}

\begin{figure}[!htb]
    \centering
    \includegraphics[scale=0.4, trim=13cm 0cm 0cm 0cm,clip]{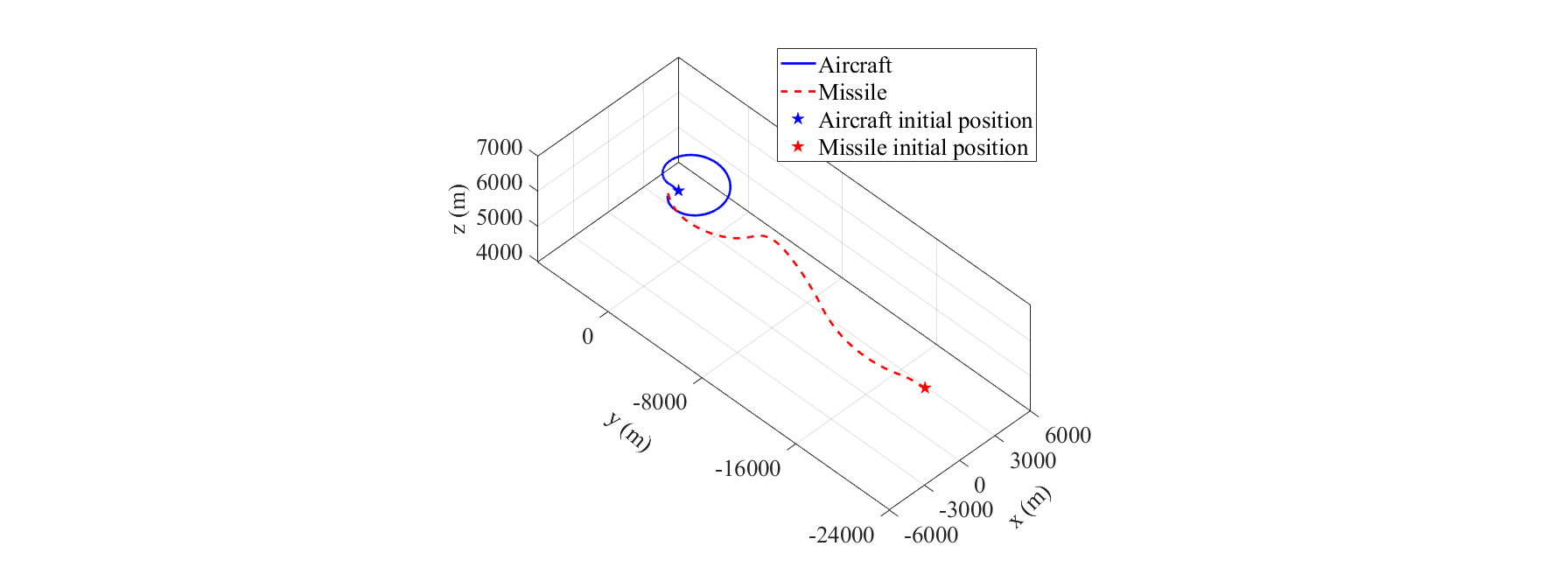}
    \caption{\textcolor{black}{Trajectory of the aircraft and the missile in the test of the steep turn policy.}}
    \label{fig: turn trajectory}
\end{figure}

\begin{figure}[!htb]
    \centering
    \includegraphics[scale=0.4]{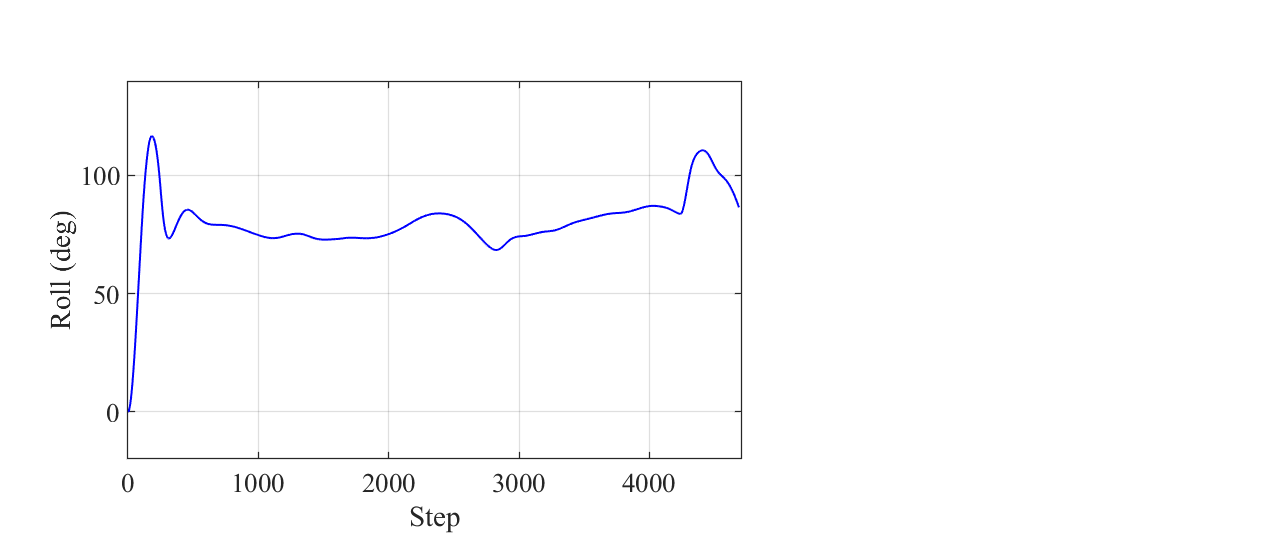}
    \caption{\textcolor{black}{Roll of the aircraft in the test of the steep turn policy.}}
    \label{fig: turn roll}
\end{figure}

\begin{figure}[!htb]
    \centering
    \includegraphics[scale=0.4]{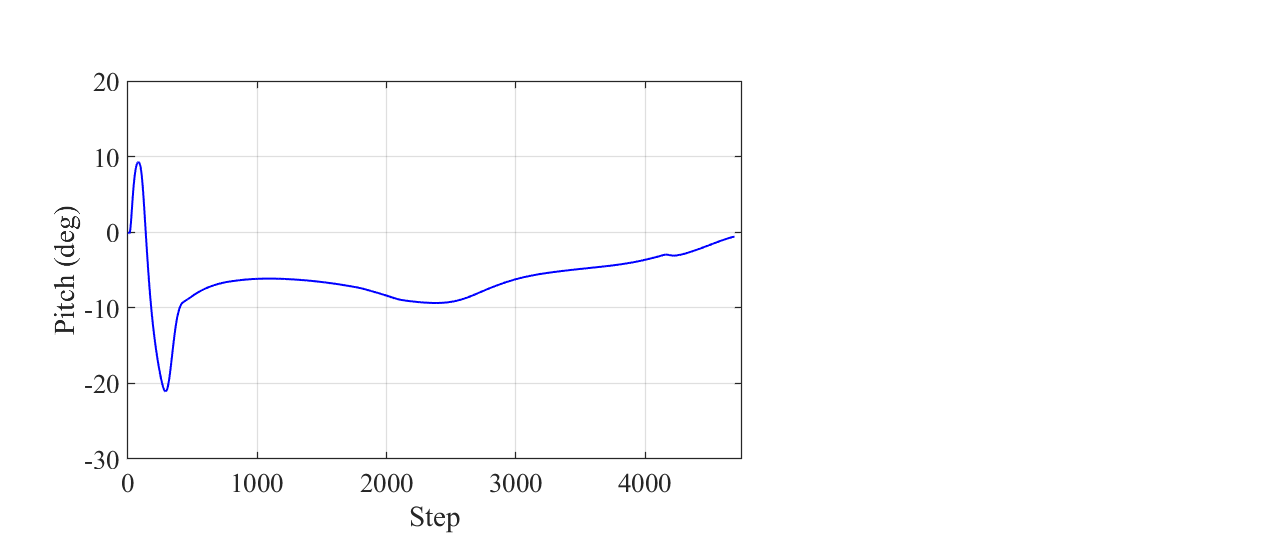}
    \caption{\textcolor{black}{Pitch of the aircraft in the test of the steep turn policy.}}
    \label{fig: turn pitch}
\end{figure}

\textcolor{black}{
% The results of the test are presented in Fig. \ref{}
Fig. \ref{fig: turn trajectory} shows the trajectory of the aircraft and the missile.
Fig. \ref{fig: turn roll} and Fig. \ref{fig: turn pitch} show the roll and pitch angle of the aircraft.
% The result is meet our expectation.
It is shown that the roll of the aircraft is around 85 deg and the pitch of the aircraft is about 0 deg.
The learned steep turn policy can guide the aircraft to perform a circle trajectory.
The effectiveness of the steep turn policy is verified.}

% Fig. \ref{fig:circling trajectory} shows the trajectory. 

% \subsection{? Improved Circling Strategy}\label{sec:threshold}

\subsubsection{Short distance policy}

% In the basic circling strategy, the aircraft performs sustained a high overload, approaching its maneuvering limits, but it is insufficient to guarantee successful evasion against an aggressive obstacle. The aggressive obstacle is guided by the 3-dimensional proportional navigation law(3DPN), and it can chase the aircraft continuously. 

% This implies that increasing the LOS rate will directly increase the obstacle’s required acceleration. 
% If this acceleration demand exceeds the obstacle’s maximum maneuvering capability, it will directly lead to degraded tracking performance and increased miss distance, ultimately raising the escape probability of the aircraft.
% The commanded lateral acceleration of the obstacle is directly 

\textbf{Training.}
\textcolor{black}{
A PN law determines a reference lateral acceleration proportional to LOS rate to chase a target \cite{Tyan2015AnalysisTarget}.
The PN law inspires this study to design a reward term based on the LOS rate for the short distance policy.
% In addition to sustaining high roll-induced load factors, the aircraft is guided to maximize the LOS rate.
% By adding a new reward function$r_{\dot{\lambda}}$, the aircraft increases the likelihood of forcing the obstacle to exceed its acceleration limits, thus improving the probability of successful evasion. 
Based on the reward function of the steep turn policy \ref{eq: steep turn reward experiment}, the reward function of the short distance policy is defined as}

\textcolor{black}{
\begin{equation}
    \label{eq: short distance reward experiment}
    \begin{aligned}
    r_{\rm{short}} = & 0.5 e^{-\displaystyle\frac{|\phi- {\rm{sgn}}((\Delta\bm{x}_m \times \dot{\bm{x}}_{a}) \cdot \bm{u}_{\rm{up}}) \cdot 85|}{0.2}} \\
            &+ 0.5 e^{-\displaystyle{\frac{|\theta|}{0.2}}}\\
            &+ 0.6 \tanh({\frac{{\rm{sgn}}((\Delta\bm{x}_m \times \dot{\bm{x}}_{a}) \cdot \bm{u}_{\rm{up}}) \cdot \dot{\lambda}^s}{0.1}})
    \end{aligned}
\end{equation}
}

\textcolor{black}{
% where $\lambda$ is LOS.
% $\widetilde{\lambda}$ is a smoothed LOS rate. 
Since the simulation frequency is 200 Hz, leading to a fluctuating LOS rate, this study smooths $\dot{\lambda}$ by
\begin{equation}
    \label{eq: los rate smooth}
    \dot{\lambda}^{s}(t) = 0.25 \dot{\lambda}^{s}_{\rm{raw}}(t) + 0.75 \dot{\lambda}^{s}(t-1)
\end{equation}
where $\dot{\lambda}^{s}_{\rm{raw}}(t)$ is the raw value of the signed LOS rate at step $t$.} 
% where the roll and pitch are two attitude angles of an aircraft, and 
% $\lambda$ refers to the line-of-sight angular rate of an obstacle relative to aircraft.

% \hl{
% Based on the above-mentioned training set-ups, The maximum number of steps in a training episode is 7500. And the circling model is used as pre-train model.
% Fig. \ref{fig:improved reward} shows the reward converges after 4000 episodes.}

\begin{figure}[!htb]
    \centering
    \includegraphics[scale=0.4]{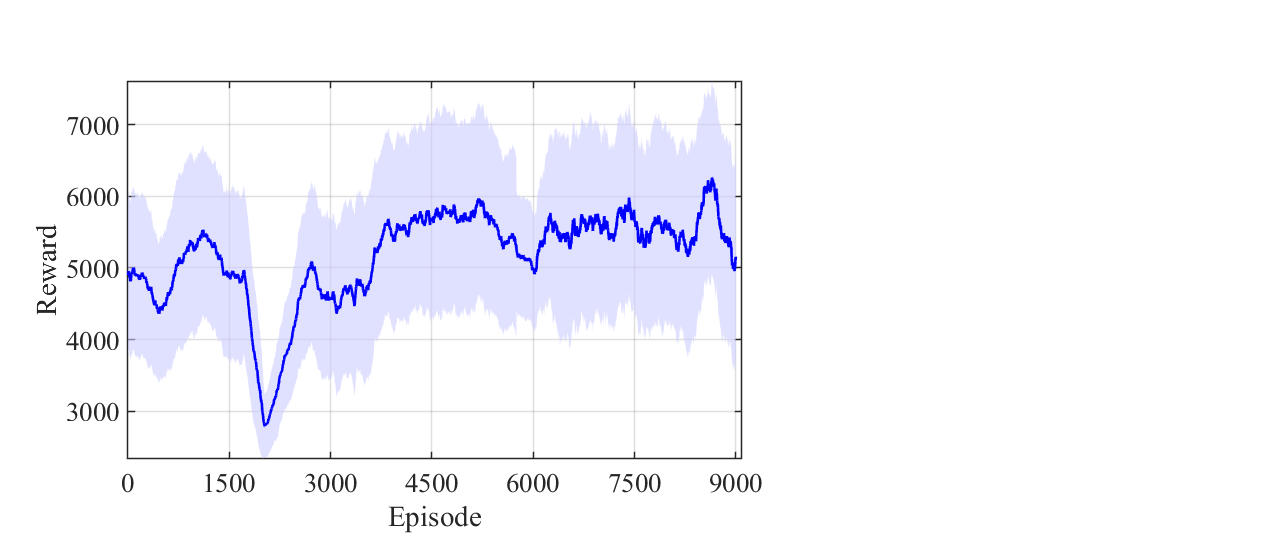}
    \caption{\textcolor{black}{Accumulated reward achieved in the training of the short distance policy.}}
    \label{fig: short policy reward}
\end{figure}

\textcolor{black}{
The learning process of the short distance policy can be reflected by the achieved reward presented in Fig. \ref{fig: short policy reward}. 
The learning process converges in 4000 episodes.
}

% \noindent \rule{1\linewidth}{0.5mm}

\textbf{Validation.}
\textcolor{black}{
To evaluate the performance of the short distance policy, a test is performed. 
The test is set to have a maximum number of 5000 steps (i.e., 25 s).
In the initial of the test, the position of the aircraft is set to [0, 0, 4820] (unit: m). 
The heading of the aircraft is set to 0 deg. 
Both the roll and pitch of the aircraft are set to 0 deg. 
The velocity of the aircraft is set to 280 m/s. 
The aircraft follows the short distance policy.
The missile is 6000 m from the aircraft. 
The azimuth of the missile is 30 deg. 
The elevation of the missile is -3.41 deg. 
The velocity of the missile is 800 m/s. 
The maximum overload of the missile is randomly set to 46.92 g. 
The missile follows the PN law with a randomly selected navigation coefficient of 4.17.
}

\begin{figure}[!htb]
    \centering
    \includegraphics[scale=0.4, trim=10cm 0cm 0cm 0cm,clip]{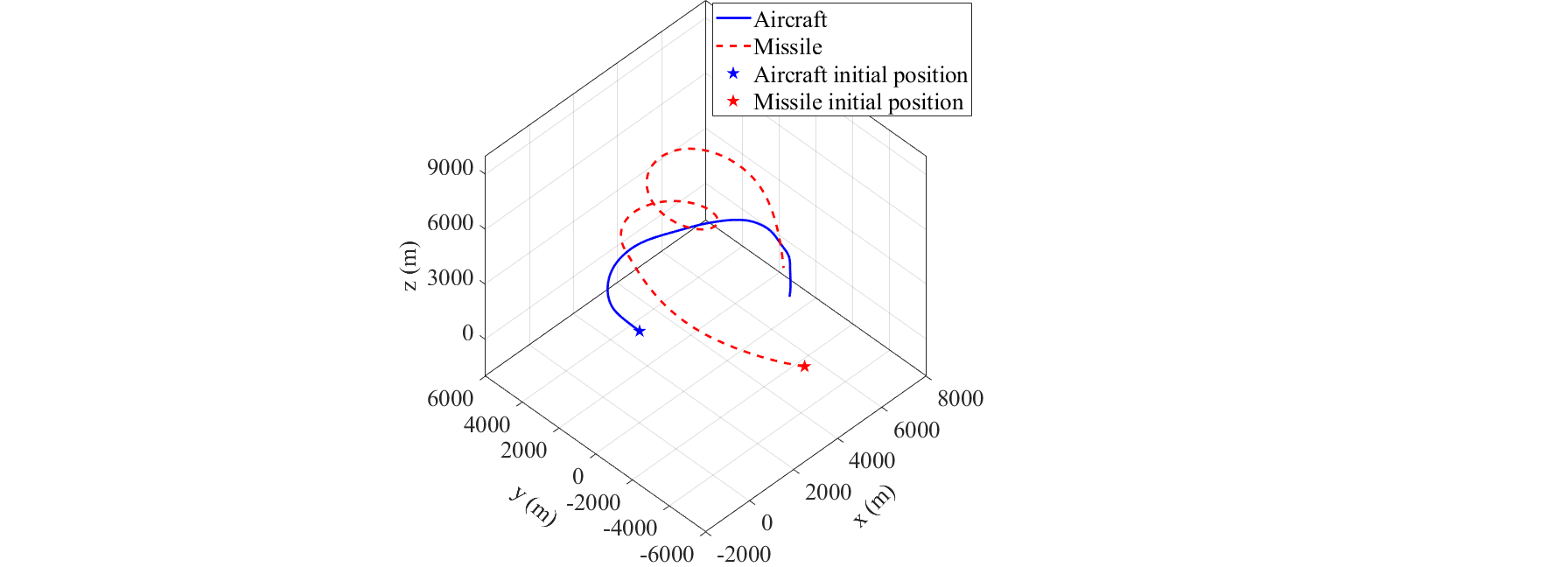}
    \caption{\textcolor{black}{Trajectory of the aircraft and the missile in the test of the short distance policy.}}
    \label{fig: short trajectory}
\end{figure}

\begin{figure}[!htb]
    \centering
    \includegraphics[scale=0.4, trim=0cm 0cm 0cm 0cm,clip]{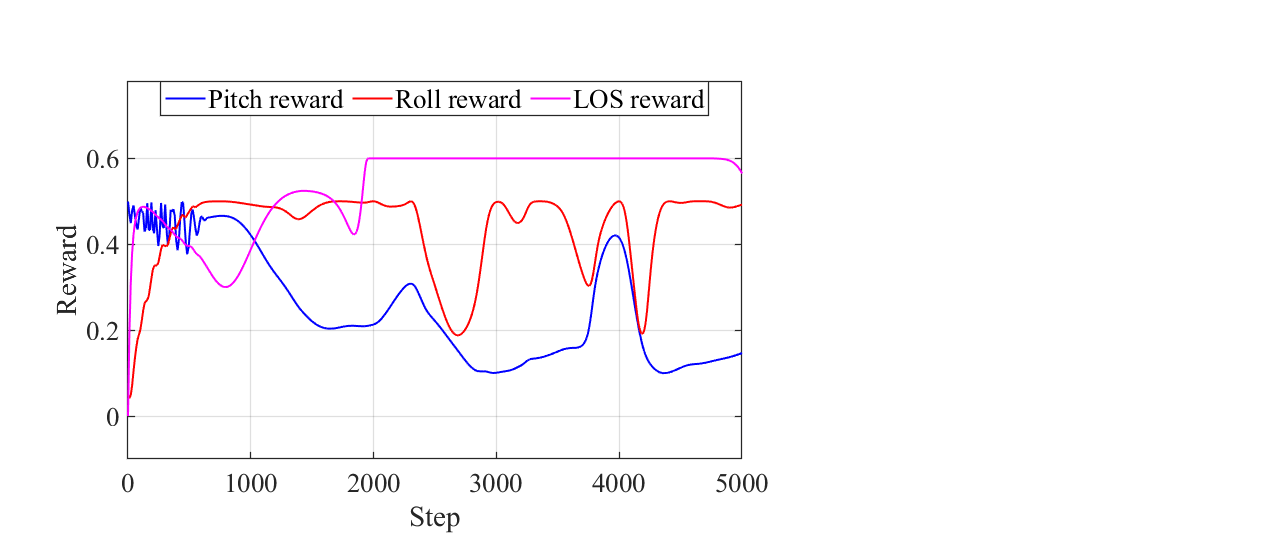}
    \caption{\textcolor{black}{Itemized rewards per step calculated based on the state of the aircraft and the missile in the test of the short distance policy.}}
    \label{fig: short item reward}
\end{figure}

\begin{figure}[!htb]
    \centering
    \includegraphics[scale=0.4]{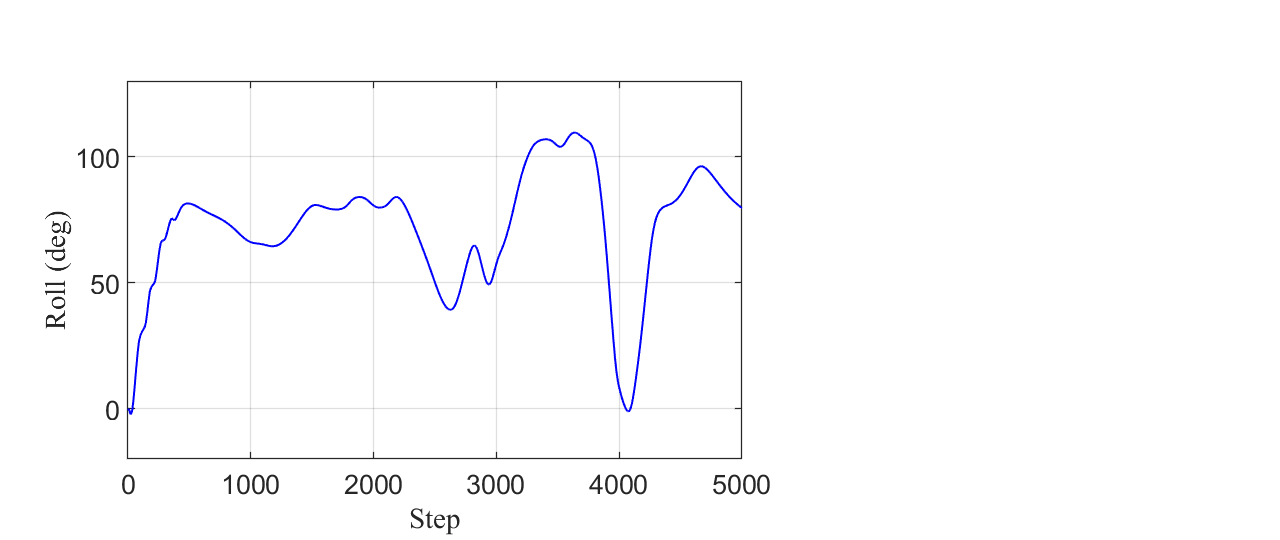}
    \caption{\textcolor{black}{Roll of the aircraft in the test of the short distance policy.}}
    \label{fig: short roll}
\end{figure}

\begin{figure}[!htb]
    \centering
    \includegraphics[scale=0.4]{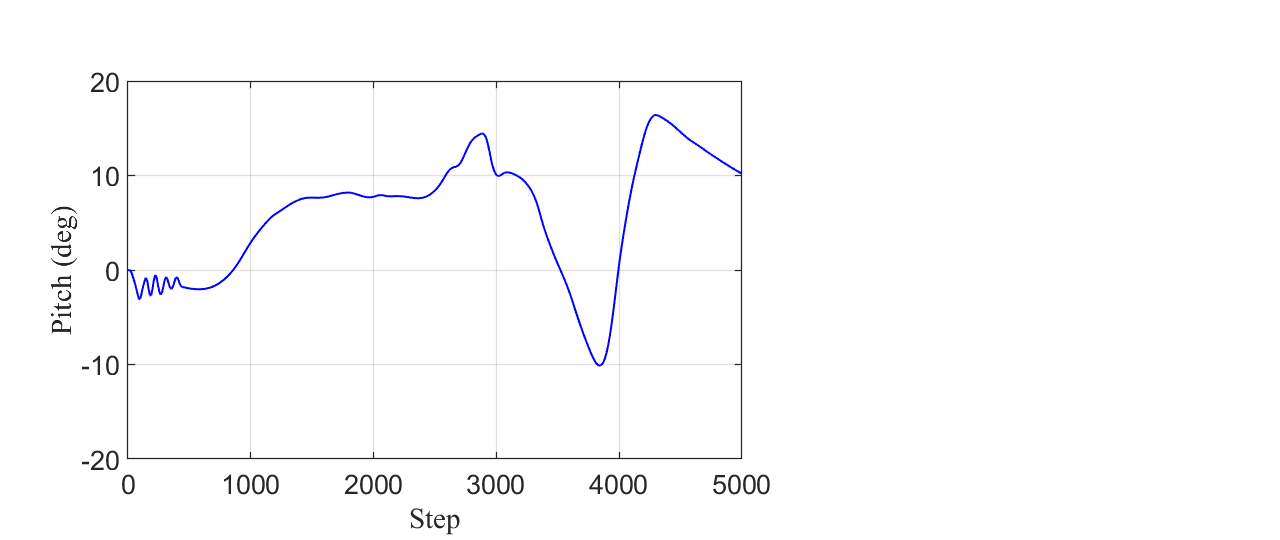}
    \caption{\textcolor{black}{Pitch of the aircraft in the test of the short distance policy.}}
    \label{fig: short pitch}
    \end{figure}

\begin{figure}[!htb]
    \centering
    \includegraphics[scale=0.4]{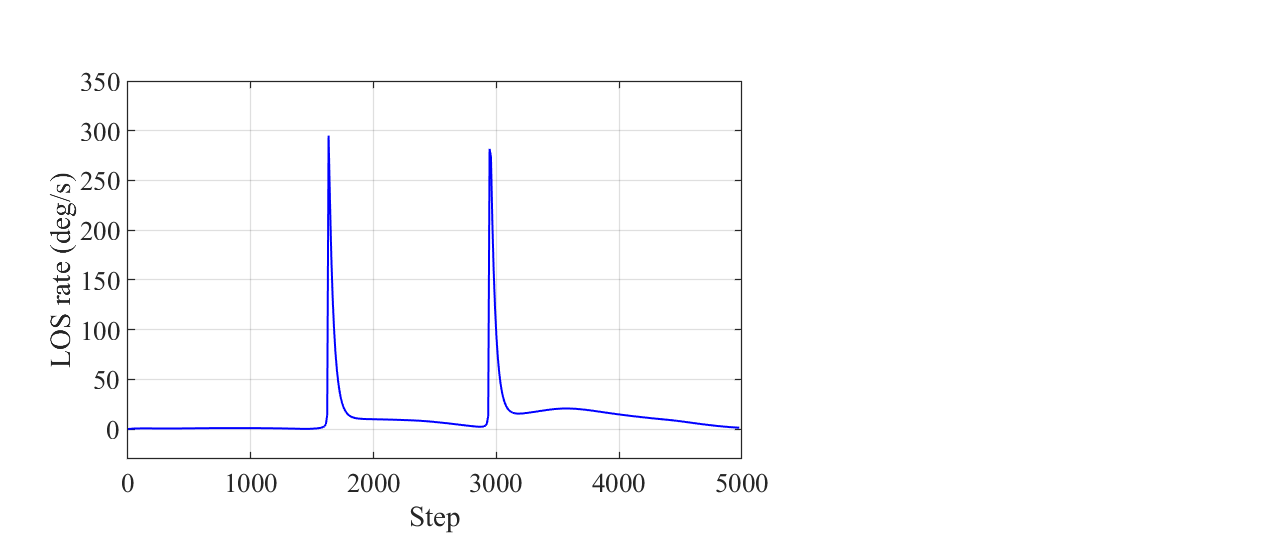}
    \caption{\textcolor{black}{LOS rate in the test of the short distance policy}}
    \label{fig: short LOS rate}
    \end{figure}

\begin{figure}[!htb]
    \centering
    \includegraphics[scale=0.4]{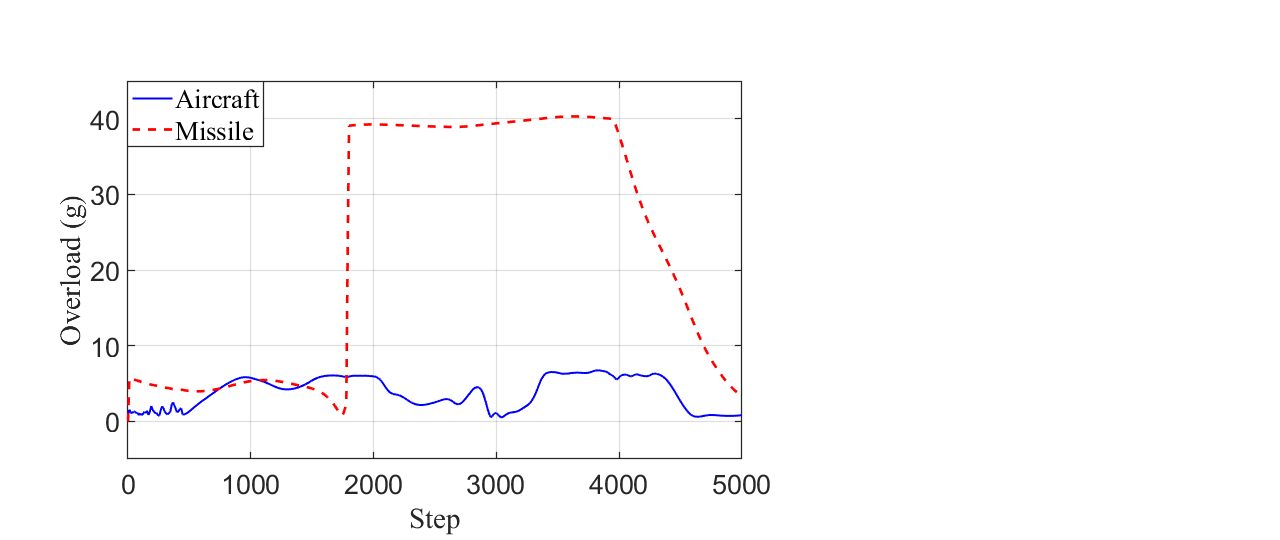}
    \caption{\textcolor{black}{Overload of the aircraft and the missile in the test of the short distance policy.}}
    \label{fig: short overload}
    \end{figure}

\begin{figure}[!htb]
    \centering
    \includegraphics[scale=0.4]{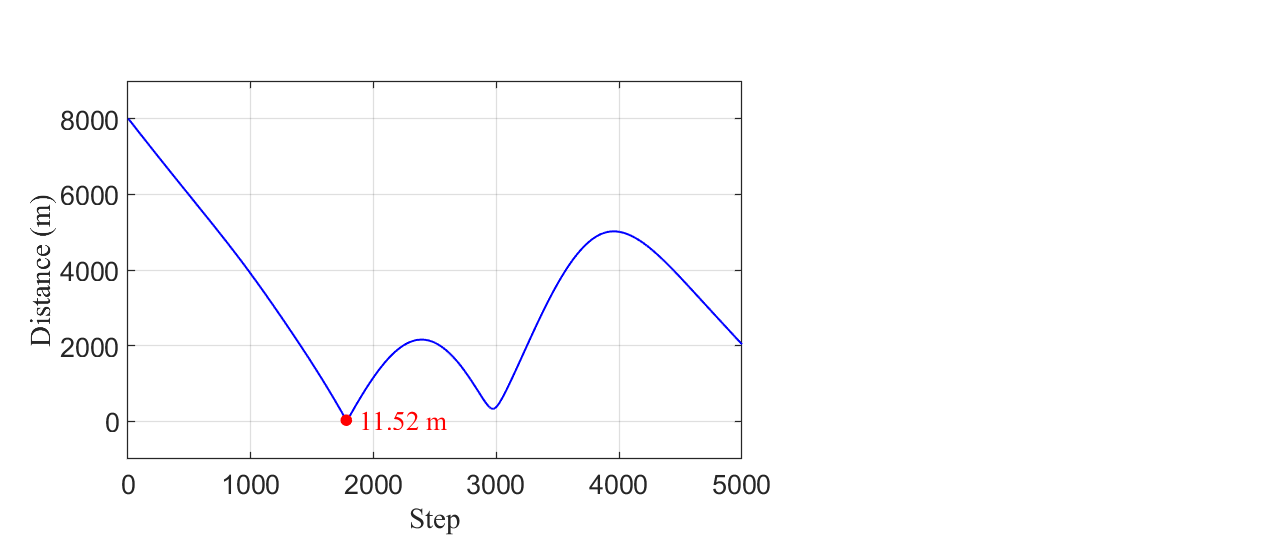}
    \caption{\textcolor{black}{Distance between the aircraft and the missile in the test of the short distance policy.}}
    \label{fig: short distance}
    \end{figure}

\textcolor{black}{
Fig. \ref{fig: short trajectory} shows the trajectory of the aircraft and the missile.
Fig. \ref{fig: short item reward} shows rewards corresponding to the roll and pitch of the aircraft and LOS rate. 
It is shown that the short distance policy can balance the objectives in controlling the roll and pitch, and increasing the LOS rate, since the magnitude of the rewards is comparable.
Fig. \ref{fig: short roll}, Fig. \ref{fig: short pitch}, and Fig. \ref{fig: short LOS rate} show the roll and pitch of the aircraft and LOS rate. 
The overload of the aircraft and the missile is shown in Fig. \ref{fig: short overload}. 
When the missile is close to the aircraft, the overload of the missile increases sharply.
Fig. \ref{fig: short distance} shows the distance between the aircraft and the missile.
According to the figures, one can see that the short distance policy can maintain the roll and pitch under control (Fig. \ref{fig: short roll} and Fig. \ref{fig: short pitch}) and considerably increase the LOS rate when the missile is close to the aircraft (Fig. \ref{fig: short LOS rate}) and maximize the overload of the missile (Fig. \ref{fig: short overload}).
As a result, the short distance policy enable the aircraft to successfully avoid the missile (Fig. \ref{fig: short trajectory}).}

\subsection{Determination of the conditions of policy switching}\label{sec:threshold}

\textcolor{black}{
According to the proposed multi-stage RL-based policy, one should investigate the short distance policy and determine the beneficial conditions for switching to the short distance policy.
To investigate the short-distance policy, a statistical analysis is conducted through dividing the initial conditions of the aircraft and the missile into intervals and then performing tests.
The maximum number of steps of a test is 5000 (i.e., 25 s).
% under different initial conditions. 
% The test settings were defined as follows:
(1) The initial velocity of the aircraft is divided into intervals ranging from 280 m/s to 470 m/s, with a step of 40 m/s.
(2) The initial velocity of the missile is divided into intervals ranging from 800 m/s and 1400 m/s, with a step of 100 m/s.
(3) The initial distance between the aircraft and the missile is divided into intervals ranging from 5000 m to 12000 m, with a step of 1000 m.
(4) The initial azimuth of the missile is divided into intervals ranging from -180 deg to 180 deg, with a step of 30 deg.} 
\textcolor{black}{
The initial roll and pitch of the aircraft are set to 0 deg. 
% O Initial conditions other than the aforementioned 
The other initial conditions are randomly selected, according to Table \ref{tab: initial conditions}. 
% Specifically, the initial altitude of the aircraft is randomly selected from 3000 m to 9000 m. 
% The heading of the aircraft is random.
% The angle between the vector from the missile to the aircraft and a horizontal plane ranges from -15 deg to 15 deg in the initial state.
% The maximum overload of the missile is randomly selected from 40 g to 50 g, 
The missile follows the PN law with a randomly selected navigation coefficient ranging from 3 to 5.
}

\begin{figure}[!htb]
    \centering
    \includegraphics[scale=0.4, trim=0cm 0cm 0cm 0cm,clip]{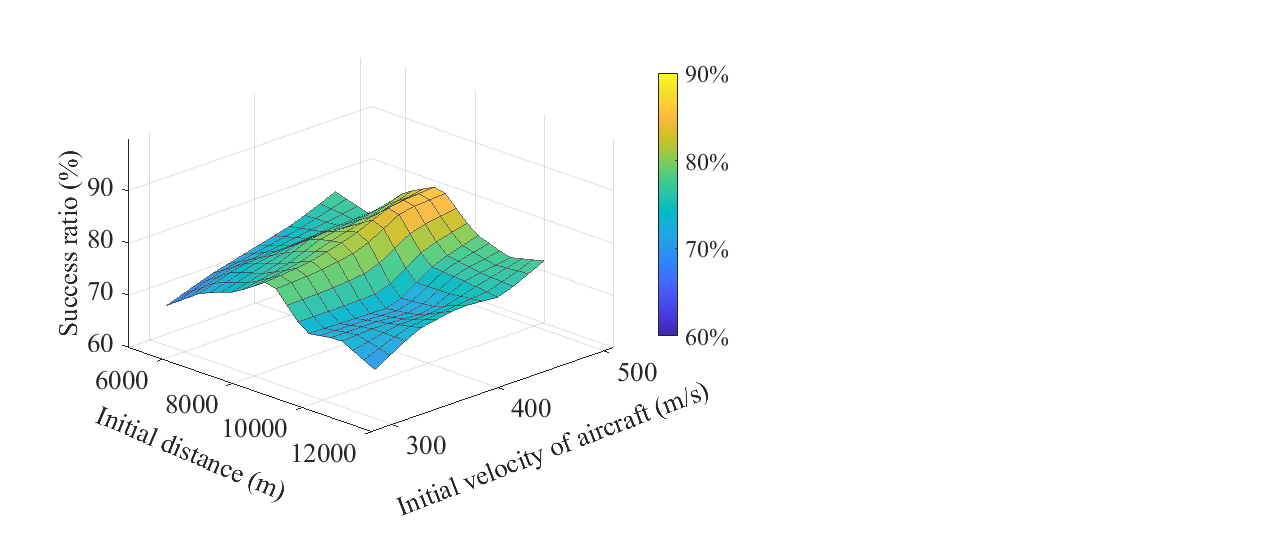}
    \caption{\textcolor{black}{Success ratio of the short distance policy with different initial conditions.}}
    \label{fig: short success ratio}
\end{figure}

\textcolor{black}{
% By combining these parameter groups, multiple experiments were constructed. 
For an interval determined by the initial velocities of the aircraft and the missile, the initial distance between the aircraft and the missile, and the initial azimuth of the missile, the other initial conditions are randomly selected and 40 tests have been conducted.
The success ratio achieved by the short-distance policy is calculated to investigate the performance, as shown in Fig. \ref{fig: short success ratio}.}

% shows the success ratio in different initial velocity of the aircraft and initial distance. 

\textcolor{black}{
The experimental results indicate that the best success ratio is achieved if the initial distance is around 8000 m to 9000 m.
Moreover, the increase of the velocity of the aircraft increases the success ratio.
According to the experimental results, 
this study involves the distance of 8000 m as one of the conditions of switching to the short distance policy and set accelerating the aircraft as an objective for the small azimuth policy and the large azimuth policy.}

% Moreover, the initial distance between the aircraft and the obstacle also has a influence on the success ratio.
% When the distance is relatively short (around 5000 m), the success ratio is low. 
% As the distance increases, the success ratio gradually improves, reaching its peak between 8000 m and 9000 m, and then declines as the distance continues to increase.

% \textbf{ Roll angle experiment.}

\begin{figure}[!htb]
    \centering
    \includegraphics[scale=0.4]{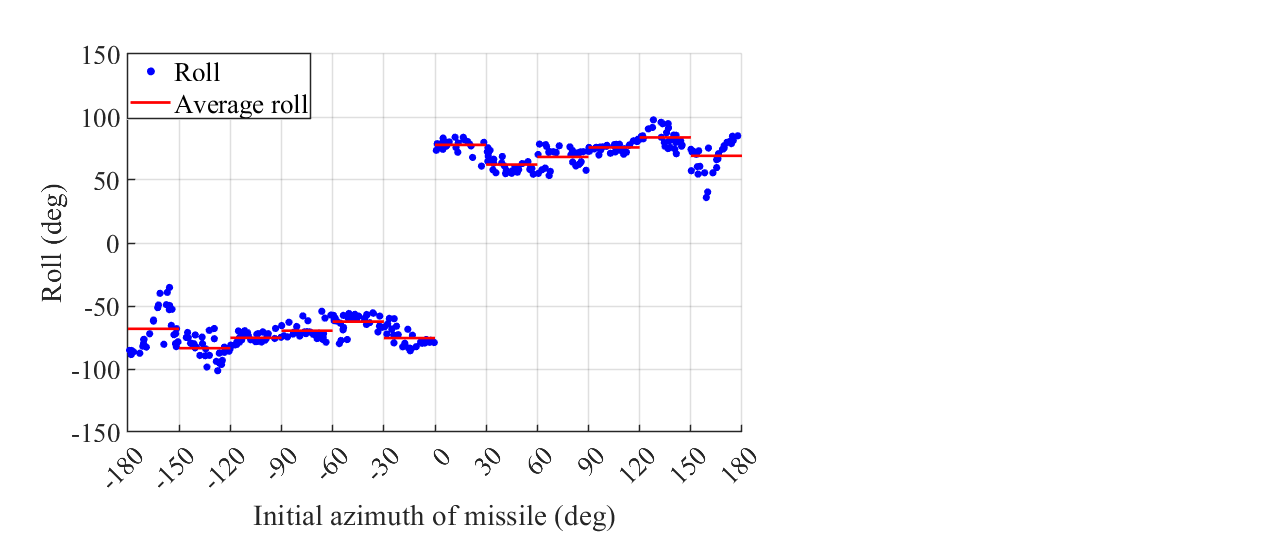}
    \caption{\textcolor{black}{
    Roll of the aircraft at a distance of 8000 m following the short distance policy to address a missile initialized at a distance of 12000 coming from different directions.}}
    \label{fig: short roll form 12 to 8}
\end{figure}

\textcolor{black}{
It should be noticed that the performance of the short-distance policy in addressing a long initial distance scenario is not as good as the performance in addressing an 8000 m initial distance scenario.
This result is not reasonable, since if an aircraft with a long initial distance from a missile maintains level flight and then switch to the short-distance policy at 8000 m distance, the aircraft can achieve the same performance in addressing an 8000 m initial distance scenario.}
\textcolor{black}{
To investigate the unreasonable result,
another statistical analysis is conducted to disclose the difference between the state of an aircraft with an 8000 m initial distance and the state of an aircraft when the distance decrease from 12000 m to 8000 m following the short-distance policy.}
\textcolor{black}{
The statistical analysis divides the initial conditions of the aircraft and the missile into intervals and initializes the other initial conditions in the same way as the statistical analysis corresponding to Fig. \ref{fig: short success ratio}, except for the 12000 m initial distance.  
The maximum number of steps of a test is 5000 (i.e., 25 s).} 
\textcolor{black}{
% By combining these parameter groups, multiple experiments were constructed.
For an interval determined by the initial velocities of the aircraft and the missile,
%initial distance between the aircraft and the missile, 
and the initial azimuth of the missile, 5 tests 
% with other initial conditions that are randomly selected 
have been conducted.}
\textcolor{black}{It is shown that if an aircraft starts to follow the short distance policy at an initial distance of 12000 m, the roll of the aircraft is approximately 85 deg or -85 deg when the distance is decreased by the missile to 8000 m, as shown in Fig. \ref{fig: short roll form 12 to 8}.
The major difference between the state of an aircraft with an 8000 m initial distance and the state of an aircraft when the distance decrease from 12000 m to 8000 m is the roll of the aircraft.
}

% The maximum number of steps of a test is 5000. 
% (1) The initial velocity of the aircraft is divided into intervals ranging from 280 m/s to 470 m/s, with a step of 40 m/s. 
% (2) The initial velocity of the missile is divided into intervals ranging from 800 m/s and 1400 m/s, with a step of 100 m/s. 
% \hl{(3) The initial distance is set to 12000 m. }
% (4) The initial azimuth of the missile is divided into intervals ranging from -180 deg to 180 deg, with a step of 30 deg.

% The initial roll and pitch of the aircraft are set to 0 deg. 
% Initial conditions other than the aforementioned initial conditions are randomly selected, according to Table \ref{tab:initial conditions of circling}. 
% The initial altitude of the aircraft is randomly selected from 3000 m to 9000 m. 
% The heading of the aircraft is random.
% The angle between the vector from the missile to the aircraft and a horizontal plane ranges from -15 deg to 15 deg in the initial state.
% The maximum overload of the missile is randomly selected from 40 g to 50 g, 
% The missile follows the 3DPN law with a randomly selected navigation coefficient ranging from 3 to 5.

% \textbf{Relationship with Roll and success ratio}

\begin{figure}[!htb]
    \centering
    \includegraphics[scale=0.4]{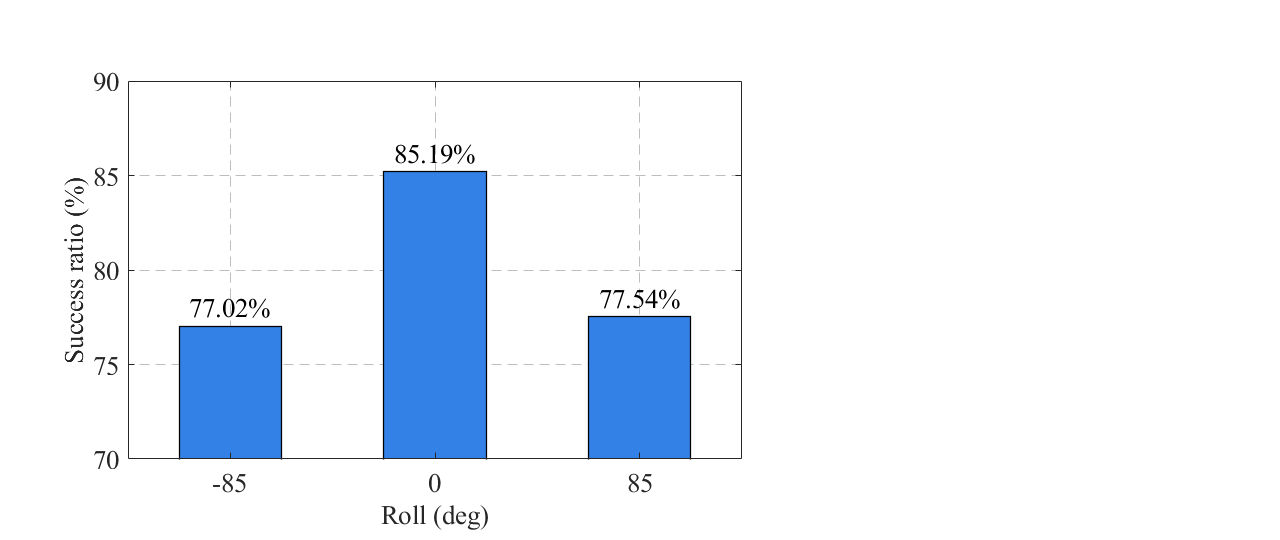}
    \caption{\textcolor{black}{Success ratio of the short distance policy with the initial distance of 8000 m and different initial roll conditions.}}
    \label{fig: short roll success ratio}
\end{figure}

\textcolor{black}{
To evaluate the influence of the initial roll of the aircraft on the success ratio,
a statistical analysis is conducted to disclose the success ratio of an aircraft with an 8000 m initial distance but with different roll angles.}
\textcolor{black}{
The statistical analysis divides the initial conditions of the aircraft and the missile into intervals and initializes the other initial conditions in the same way as the statistical analysis corresponding to Fig. \ref{fig: short success ratio}, except for the 8000 m initial distance and different initial roll angles of -85 deg, 0 deg, and 85 deg.  
The maximum number of steps of a test is 5000 (i.e., 25 s).} 
\textcolor{black}{
% By combining these parameter groups, multiple experiments were constructed. 
For an interval determined by the initial velocities of the aircraft and the missile, the initial azimuth of the missile, and the roll of the aircraft, 5 tests 
% with other initial conditions that are randomly selected 
have been conducted.}
\textcolor{black}{
The result of the influence of the initial roll of the aircraft on the success ratio is shown in Fig. \ref{fig: short roll success ratio}.
The result indicates that for an aircraft with an 8000 m initial distance,
the 0 deg roll in the initial state can provide a considerably higher success ratio than a large roll (e.g., -85 deg or 85 deg) in the initial state.
According to the experimental result, 
this study set reducing the roll of the aircraft to about 0 deg when switching to the short distance policy as an objective for the small azimuth policy and the large azimuth policy.}

% The initial velocity of the aircraft is set from 280 m/s to 470 m/s.
% The initial velocity of the missile is set from 800 m/s and 1400 m/s.
% \hl{The initial distance is set to 8000 m.}
% The initial azimuth of the missile is set from -180 deg to 180 deg.

% \hl{The initial roll of the aircraft is set to -85 deg, 0 deg and 85 deg.}
% The initial pitch of the aircraft are set to 0 deg. 
% Initial conditions other than the aforementioned initial conditions are randomly selected, according to Table \ref{tab:initial conditions of circling}. 
% The initial altitude of the aircraft is randomly selected from 3000 m to 9000 m. 
% The heading of the aircraft is random.
% The angle between the vector from the missile to the aircraft and a horizontal plane ranges from -15 deg to 15 deg in the initial state.
% The maximum overload of the missile is randomly selected from 40 g to 50 g, 
% The missile follows the 3DPN law with a randomly selected navigation coefficient ranging from 3 to 5.

\textcolor{black}{
\textbf{Remark 1}. According to the analysis of the short distance policy in this section, 1) the condition of switching to the short distance policy is the distance of 8000 m and the objectives of the small azimuth policy and the large azimuth policy include 2) accelerating the aircraft and 3) reducing the roll of the aircraft to around 0 deg when switching to the short distance policy.
}

\subsection{Training and validation of a small azimuth policy}\label{sec:small azimuth policy}

% When the velocity–LOS angle between the aircraft and the obstacle is relatively small, the moving-away policy no longer focuses on  reducing this angle. 
% Instead, the main purpose is to keep the angle within an acceptable range, preventing it from growing rapidly into an unfavorable condition.

% \textbf{Training.}
\textcolor{black}{
According to the proposed multi-stage RL-based evasion strategy and the conditions and objectives determined based on the analysis of the short distance policy in Section \ref{sec:threshold}, a small azimuth policy is trained in this section.
The small azimuth policy will be activated if the distance between the aircraft and the missile is larger than 8000 m and the magnitude of the azimuth of the missile is smaller than 30 deg. 
The objectives of small azimuth policy include
(1) maintaining the magnitude of azimuth within 30 deg,
(2) keeping the pitch and roll of the aircraft under control to avoid critical loss of altitude,
(3) reducing the magnitude of the azimuth of the missile,
and (4) accelerating the aircraft.}

\textbf{Training.}
\textcolor{black}{
According to the multi-stage RL-based evasion strategy and the objectives, the reward function of the small azimuth policy is defined as
\begin{equation}
    \label{eq: small azimuth reward experiment}
    \begin{aligned}
     r_{\rm{small}} =& r_{\rm{small}}^{\rm{roll}} + r_{\rm{small}}^{\rm{pitch}} + r_{\rm{small}}^{\rm{azimuth}} + r_{\rm{small}}^{\rm{vel}} +  \\
                & c_{\rm{small}}^{\rm{roll}} + c_{\rm{small}}^{\rm{pitch}} + c_{\rm{small}}^{\rm{azimuth}} + c_{\rm{small}}^{\rm{vel}}
    \end{aligned} 
\end{equation}
where the reward terms are defined as
\begin{equation}
    r_{\rm{small}}^{\rm{roll}} = 0.5 e^{-\displaystyle \frac{|\phi|}{0.2}}
\end{equation}
\begin{equation}
    r_{\rm{small}}^{\rm{pitch}} = 0.5 e^{-\displaystyle \frac{|\theta|}{0.2}}
\end{equation}
\begin{equation}
    r_{\rm{small}}^{\rm{azimuth}} =  1.0 e^{-\displaystyle\frac{|\Psi|}{0.2}}
\end{equation}
\begin{equation}
    r_{\rm{small}}^{\rm{vel}} =  0.2  \tanh(\frac{||\dot{\bm{x}}_{a}||-350}{80})
\end{equation}
\begin{equation}
    c_{\rm{small}}^{\rm{roll}} =
    \begin{cases}
        -20, & \text{if } |\phi| > 135 \text{ deg}\\
        0,   & \text{otherwise}
    \end{cases}
\end{equation}
\begin{equation}
    c_{\rm{small}}^{\rm{pitch}} =
    \begin{cases}
        -20, & \text{if } |\theta| > 22.5 \text{ deg}\\
        0,   & \text{otherwise}
    \end{cases}
\end{equation}
\begin{equation}
    c_{\rm{small}}^{\rm{azimuth}} =
    \begin{cases}
        -20, & \text{if } |\Psi| > 30 \text{ deg}\\
        0,   & \text{otherwise}
    \end{cases}
\end{equation}
\begin{equation}
    c_{\rm{small}}^{\rm{vel}} =
    \begin{cases}
        -20, & \text{if } ||\dot{\bm{x}}_{a}|| < 240 \text{ m/s}\text{ or } ||\dot{\bm{x}}_{a}|| > 510\text{ m/s} \\
        0,   & \text{otherwise}
    \end{cases}
\end{equation}
}

% \hl{training setting?}

\textcolor{black}{
To train a small azimuth policy, 
% the initial position of the aircraft is set to [0, 0, $z_a$](unit: m), where $z_a$ is an altitude randomly selected from 3000 m and 9000 m.
% The initial heading of the aircraft is randomly set ranging from 0 deg to 360 deg. 
% The initial roll and pitch are Both  set to 0 deg.
the initial velocity of the aircraft is divided into five intervals (i.e., 280 m/s to 320 m/s, 320 m/s to 360 m/s, 360 m/s to 400 m/s, 400 m/s to 440 m/s, and 440 m/s to 470 m/s) and the ratio of the possibility of selecting from the five intervals are 16, 8, 4, 2, and 1, respectively. 
If an interval is selected, then the initial velocity of the aircraft will be uniformly distributed in the interval.
The training process initializes the other initial conditions of the aircraft in the same way as Section \ref{sec:short distance policy}.
}
\textcolor{black}{
The initial distance between the aircraft and the missile is divided into five intervals (i.e., 5000 m to 7000 m, 7000 m to 9000 m, 9000 m to 11000 m, 11000 m to 13000 m, and 13000 m to 15000 m) and the ratio of the possibility of selecting from the five intervals are 1, 2, 4, 8, and 16, respectively.}
\textcolor{black}{
The training process initializes the other initial conditions of the missile in the same way as Section \ref{sec:short distance policy}.
% The initial azimuth of the missile is ranging from -30 deg to 30 deg. 
% The angle between the vector from the missile to the aircraft and a horizontal plane is randomly selected from -15 deg to 15 deg. 
% The initial velocity of the missile is ranging from 800 m/s to 1400 m/s. 
% The maximum overload of the missile ranges from 40 g to 50 g. 
% The missile follows the PN law with a randomly selected navigation coefficient ranging from 3 to 5.
}
\textcolor{black}{
The maximum number of steps of an episode is 7500. 
An episode will be terminated if the distance between the aircraft and the missile is less than 5000 m.}

% \hl{
% Training results.
% Fig. \ref{fig: small reward} shows the episode-reward in training. 
% The reward converge after 6000 episode. 
% }

\begin{figure}[!htb]
    \centering
    \includegraphics[scale=0.4]{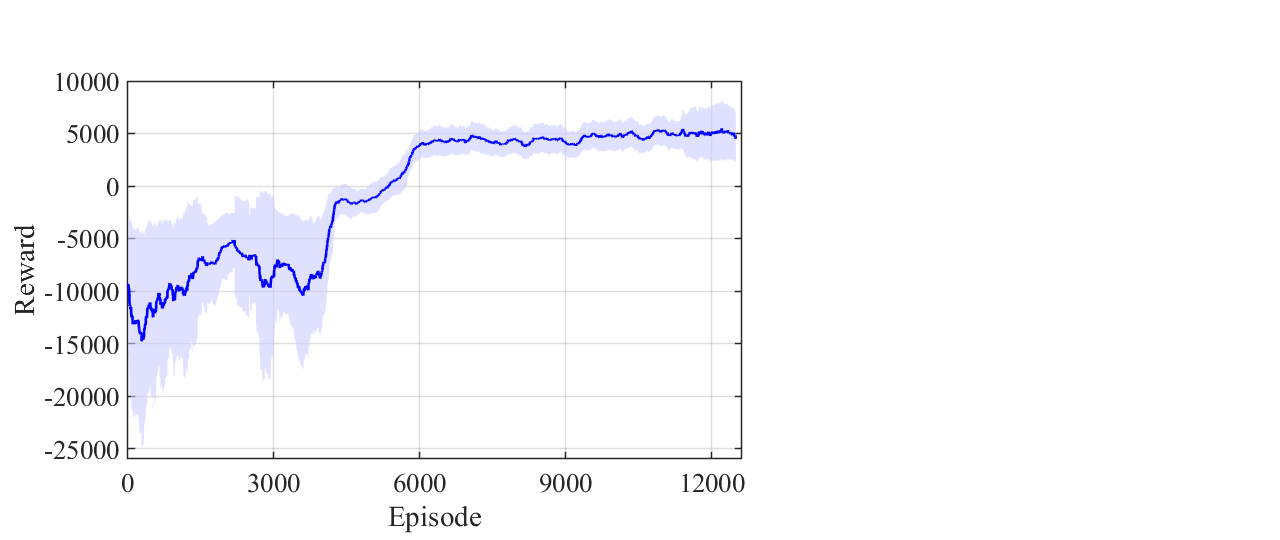}
    \caption{\textcolor{black}{Accumulated reward achieved in the training of the small azimuth policy.}}
    \label{fig: small policy reward}
\end{figure}

\textcolor{black}{
The learning process of the small azimuth policy can be reflected by the achieved reward presented in Fig. \ref{fig: small policy reward}. 
The learning process converges in 6000 episodes.
}

% \begin{figure}[!htb]
%     \centering
%     \includegraphics[scale=1]{experiment_result/temporary_result/small_trajectory.png}
%     \caption{Trajectory of the aircraft and missle.}
%     \label{fig:reward circling}
%     \end{figure}

% \noindent \rule{1\linewidth}{0.5mm}

\textbf{Validation.}
\textcolor{black}{
To validate the small azimuth policy, a statistical analysis is conducted through dividing the initial conditions of the aircraft and the missile into intervals and then performing tests. 
The maximum number of steps of a test is 5000 (i.e., 25 s).
(1) The initial velocity of the aircraft is divided into intervals ranging from 280 m/s to 470 m/s, with a step of 40 m/s.
(2) The initial velocity of the missile is divided into intervals ranging from 800 m/s and 1400 m/s, with a step of 100 m/s.
(3) The initial distance is set at 15000 m.
(4) The initial azimuth of the missile is divided into intervals ranging from -30 deg to 30 deg, with a step of 12 deg.
If an interval is selected, then the initial velocity of the aircraft, the initial velocity of the missile, and the initial azimuth of the missile will be uniformly distributed in the interval.
The validation tests initialize the other initial conditions of the aircraft and the missile in the same way as Section \ref{sec:short distance policy}.
}
% \textcolor{black}{
% The initial roll and pitch of the aircraft are set to 0 deg. 
% Initial conditions other than the aforementioned initial conditions are randomly selected, according to Table \ref{tab: initial conditions}. 
% The initial altitude of the aircraft is randomly selected from 3000 m to 9000 m. 
% The heading of the aircraft is random.
% The angle between the vector from the missile to the aircraft and a horizontal plane ranges from -15 deg to 15 deg in the initial state.
% The maximum overload of the missile is randomly selected from 40 g to 50 g, 
% The missile follows the PN law with a randomly selected navigation coefficient ranging from 3 to 5.
% }

\begin{figure}[!htb]
    \centering
    \includegraphics[scale=0.4]{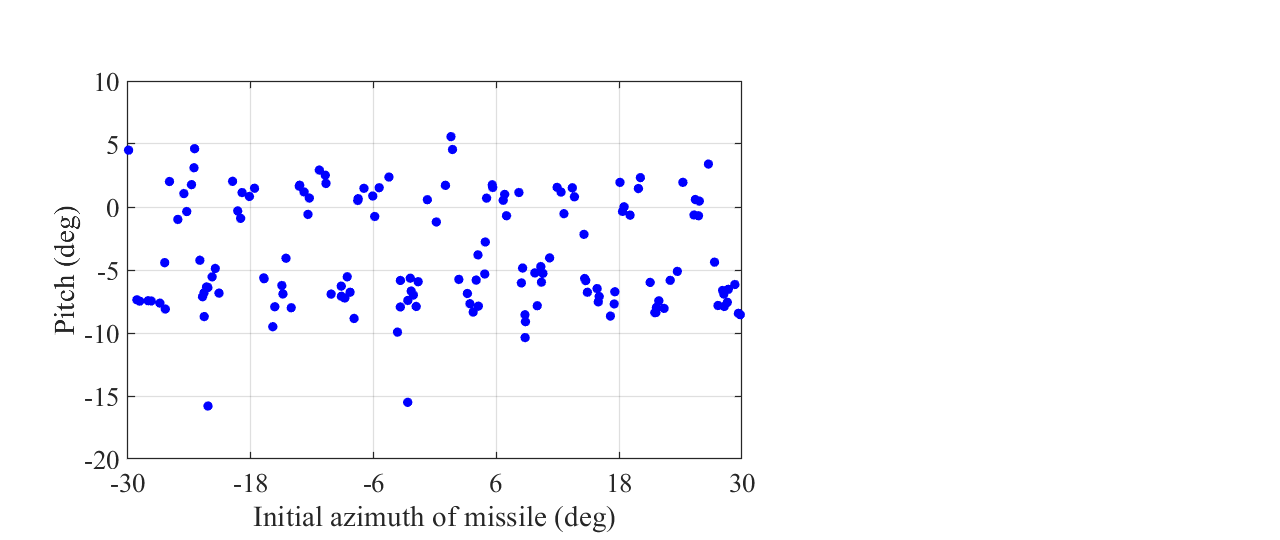}
    \caption{\textcolor{black}{Average pitch of the aircraft in a test following the small azimuth policy with different initial azimuth conditions.}}
    \label{fig: small pitch}
    \end{figure}

\begin{figure}[!htb]
    \centering
    \includegraphics[scale=0.4]{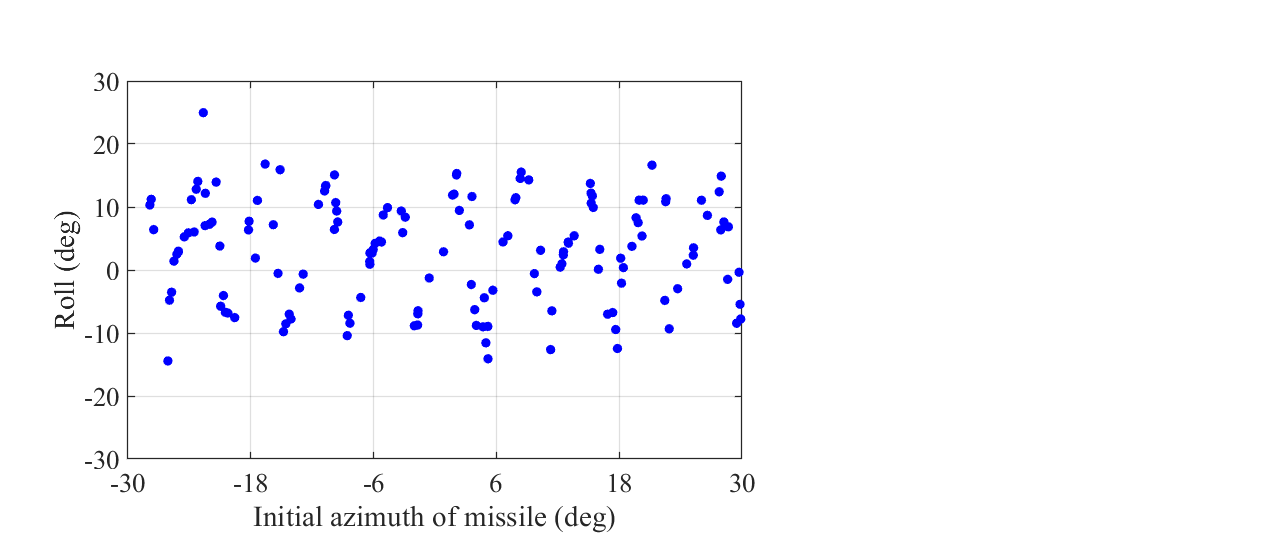}
    \caption{\textcolor{black}{Average roll of the aircraft in a test following the small azimuth policy with different initial azimuth conditions.}}
    \label{fig: small roll}
    \end{figure}

\begin{figure}[!htb]
    \centering
    \includegraphics[scale=0.4]{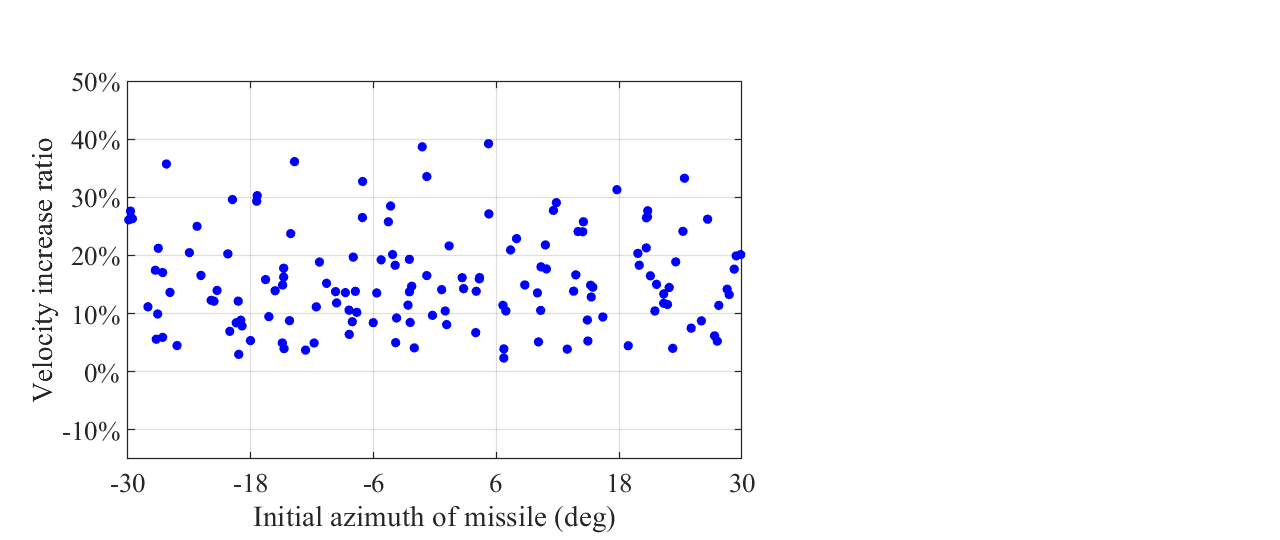}
    \caption{\textcolor{black}{Velocity increase rate of the final velocity to the initial velocity achieved by the small azimuth policy with different initial azimuth conditions.}}
    \label{fig: small velocity}
    \end{figure}

\textcolor{black}{
% By combining these parameter groups, multiple experiments were constructed. 
For an interval determined by initial velocities of the aircraft and the missile, and the initial azimuth of the missile, 
the other initial conditions are randomly selected and one test have been conducted.
% one test with other initial conditions that are randomly selected have been conducted.
}
\textcolor{black}{
The number of out-of-bounds occurrences for roll, pitch, velocity, and azimuth is listed in Table \ref{tab: small out of bound}.
% shows the Out-of-Bounds Rate of the four critical flight parameters, including roll, pitch, speed, and horizontal projection angle. 
The number of all out-of-bounds occurrences is zero, indicating that the constraint terms can effectively maintain the roll, pitch, velocity, and azimuth within corresponding defined ranges.
For every interval, one representative test is randomly selected from the 30 tests.
% The Fig. \ref{fig:small average pitch}, Fig. \ref{fig:small average roll.}, Fig. \ref{fig: small final angle},  and Fig. \ref{fig:small average roll.} are based on one representative data out of the 30 tested data.
Fig. \ref{fig: small pitch} and Fig. \ref{fig: small roll} show the average pitch and average roll of the aircraft in one episode according to the representative tests. 
It is shown that both average pitch and average roll are around 0 deg, indicating that the aircraft doesn't continuously perform aggress pitch or roll to climb, descend, or turn.
Fig. \ref{fig: small roll} shows the azimuth of the missile at the last step. 
With different initial conditions, the azimuth ranges from -3.5 deg to 3.5 deg, suggesting that the small azimuth policy has learned to reduce the azimuth of the missile.
Fig. \ref{fig: small velocity} shows the velocity increase ratio, indicating that the small azimuth policy can 
accelerate the aircraft.}

\textcolor{black}{
The results indicate that the small azimuth policy can maintain the azimuth within 30 deg (Table \ref{tab: small out of bound}) and keep the pitch and roll of the aircraft under control (Fig. \ref{fig: small pitch} and Fig. \ref{fig: small roll}).
The small azimuth policy can reduce the magnitude of azimuth to around 3 deg or less (Fig. \ref{fig: small roll}).
The small azimuth policy can accelerate the aircraft also (Fig. \ref{fig: small velocity}). 
It is validated that the small azimuth policy can achieve the set objectives.}

\begin{table}[!t]    
    \centering  
    \caption{\textcolor{black}{Number of out-of-bounds occurrences for roll, pitch, velocity, and azimuth}}
    \label{tab: small out of bound}
    \begin{tabular}{|c|c|c|c|c|} 
    \hline 
    \textbf{Variable} & \textbf{Roll} & \textbf{Pitch} & \textbf{Velocity} &  \textbf{Azimuth}\\ 
    \hline
    \textbf{\makecell{Number of \\ out-of-bounds}} & 0 & 0 & 0 & 0 \\  
    \hline
    \end{tabular}
\end{table}

    \begin{figure}[!htb]
    \centering
    \includegraphics[scale=0.4]{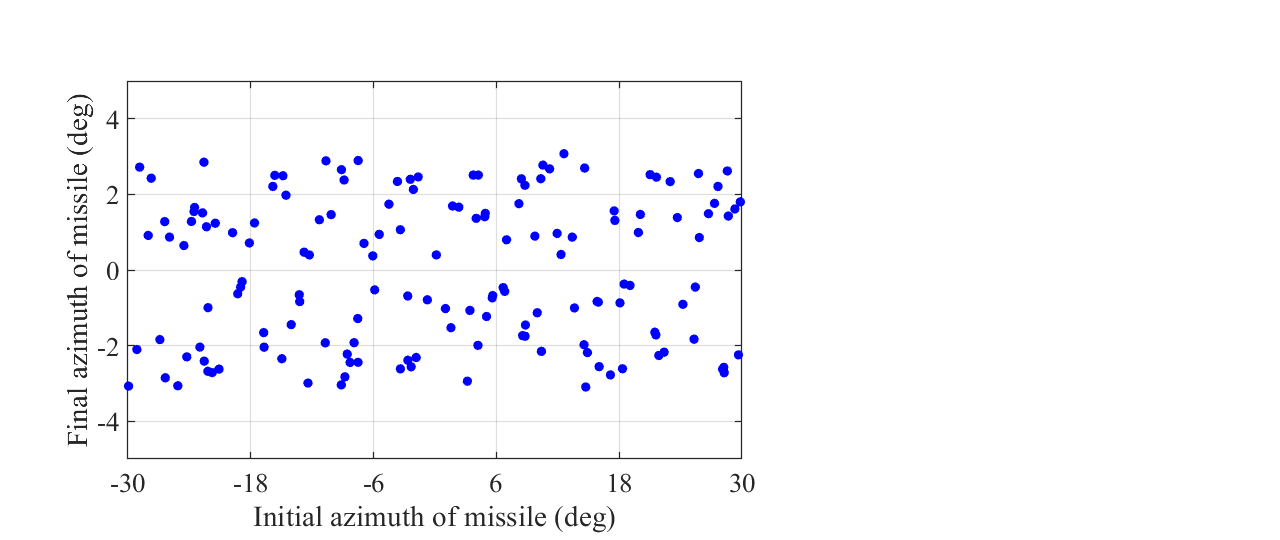}
    \caption{\textcolor{black}{Final azimuth of a test achieved by the small azimuth policy with different initial azimuth conditions.}}
    \label{fig: small azimuth}
    \end{figure}

% enables the aircraft to maintain a stable attitude, with the roll angle is constrained within $[-30 deg, 30 deg]$, 
% and pitch angle is constrained within $[-15 deg, 15 deg]$.

% figure 3: aircraft's velocity. 

% The result shows that the aircraft will accelerate without exceeding the speed range.

\subsection{Training and validation of a large azimuth policy}\label{sec:large azimuth policy}

% The avoidance policy test results show that higher aircraft speed at the short distance and switching at an appropriate distance lead to a higher escape success ratio. Thus, we designed the long-distance moving-away policy. 
% Before reaching the switching threshold, this policy aims to separate from the obstacle and to establish favorable initial conditions for the subsequent short-distance avoidance policy.

% When the angle between the aircraft’s velocity direction and the obstacle’s line-of-sight (LOS) is large, the moving-away policy focuses on three objectives:

\textcolor{black}{
According to the proposed multi-stage RL-based evasion strategy and the conditions and objectives determined based on the analysis of the short distance policy in Section \ref{sec:threshold}, a large azimuth policy is trained in this section.
The large azimuth policy will be activated if the distance between the aircraft and the missile is larger than 8000 m and the magnitude of the azimuth of the missile is larger than 30 deg. 
The objectives of small azimuth policy include
(1) reducing the azimuth of the missile,
% velocity–LOS angle: Align the aircraft’s velocity vector with the LOS vector, thereby delaying potential collision.
(2) reducing the magnitude of roll of the aircraft if the distance between the aircraft and the obstacle less than 8500 m to achieve a roll around 0 deg when switching to the short distance policy,
(3) keeping the pitch of the aircraft under control, 
% Maintain attitude stability: Constrain pitch deviations to preserve steady flight.
and (4) accelerating the aircraft.}

\textbf{Training.}
\textcolor{black}{
According to the multi-stage RL-based evasion strategy and the objectives, the reward function of the large azimuth policy is defined as
\begin{equation}
    \label{eq:large azimuth policy reward}
    \begin{aligned}
    r_{\rm{large}} = & r_{\rm{large}}^{\rm{roll}} + r_{\rm{large}}^{\rm{pitch}} + r_{\rm{large}}^{\rm{vel}} +  \\
                & c_{\rm{large}}^{\rm{roll}} + c_{\rm{large}}^{\rm{pitch}} + c_{\rm{large}}^{\rm{vel}}
    \end{aligned} 
\end{equation}
It should be noted that the definition of $r_{\rm{large}}^{\rm{roll}}$ depends on the distance between the aircraft and the missile. 
$r_{\rm{large}}^{\rm{roll}}$ is defined as 
\begin{equation}
    r_{\rm{large}}^{\rm{roll}} =
    \begin{cases}
        r_{\rm{large}}^{\rm{far}}, & || \Delta\bm{x}_m || > 8500 \text{ m} \\[6pt]
        r_{\rm{large}}^{\rm{close}}, & || \Delta\bm{x}_m || \le 8500 \text{ m}
    \end{cases}
    \label{eq: large azimuth roll reward experiment}
\end{equation}
If the distance between the aircraft and the missile is larger than the threshold (i.e., $ || \Delta\bm{x}_m || > 8500$),
the reward function is
\begin{equation}
    r_{\rm{large}}^{\rm{far}} =
    0.5  e^{\displaystyle - \frac{|\phi - {\rm{sgn}}((\Delta\bm{x}_m \times \dot{\bm{x}}_{a}) \cdot \bm{u}_{\rm{up}}) \cdot 85|}{0.2}}
    \label{eq: large azimuth roll far reward experiment}
\end{equation}
% where $\phi$ is the roll of the aircraft.
If the distance between the aircraft and the missile is smaller than the threshold (i.e., $ || \Delta\bm{x}_m || \le 8500$), the reward function is
\begin{equation}
    r_{\rm{large}}^{\rm{close}} =
    \begin{cases}
        -20, & |\phi| > 30 \text{ deg} \\[6pt]
        0.5, & |\phi| \le 30 \text{ deg}
    \end{cases}
    \label{eq: large azimuth roll close reward experiment}
\end{equation}
The other reward terms are defined as
% \begin{equation}
%     r_{\rm{large}}^{\rm{roll}} = 0.5 \times e^{-\displaystyle \frac{|\phi-{\rm{sgn}}((\Delta\bm{x}_m \times \dot{\bm{x}}_{a}) \cdot \bm{u}_{\rm{up}}) \cdot 85|}{0.2}}
% \end{equation}
\begin{equation}
    r_{\rm{large}}^{\rm{pitch}} = 0.5  e^{-\displaystyle \frac{|\theta|}{0.2}}
\end{equation}
\begin{equation}
    r_{\rm{large}}^{\rm{vel}} = 0.3 \tanh(\frac{||\dot{\bm{x}}_{a}||-350}{60})
\end{equation}
\begin{equation}
    c_{\rm{large}}^{\rm{roll}} =
    \begin{cases}
        -20, & \text{if } |\phi| > 135 \text{ deg} \\
        0,   & \text{otherwise}
    \end{cases}
\end{equation}
\begin{equation}
    c_{\rm{large}}^{\rm{pitch}} =
    \begin{cases}
        -20, & \text{if } |\theta| > 22.5 \text{ deg}\\
        0,   & \text{otherwise}
    \end{cases}
\end{equation}
\begin{equation}
    c_{\rm{large}}^{\rm{vel}} =
    \begin{cases}
        -20, & \text{if } ||\dot{\bm{x}}_{a}|| < 240 \text{ m/s}\text{ or } ||\dot{\bm{x}}_{a}|| >510 \text{ m/s}\\
        0,   & \text{otherwise}
    \end{cases}
\end{equation}
% where $\phi$ is the roll of the aircraft.
}

% \hl{training setting?}

\textcolor{black}{
To train a small azimuth policy,
% The initial position of the aircraft is set to [0, 0, $z_a$](unit: m), where $z_a$ is an altitude randomly selected from 3000 m and 9000 m.
% The initial heading of the aircraft is randomly set ranging from 0 deg to 360 deg.
% Both the initial roll and pitch are set to 0 deg.
% The initial velocity of the aircraft ranges from 280 m/s to 470 m/s. 
% It should be noticed that since this section aims to obtain a truing policy that tends to accelerate the aircraft, this study utilizes a velocity initialization method to address the issue that too much collected data for training is of high velocity.
the initial velocity of the aircraft is divided into five intervals (i.e., 280 m/s to 320 m/s, 320 m/s to 360 m/s, 360 m/s to 400 m/s, 400 m/s to 440 m/s, and 440 m/s to 470 m/s) and the ratio of the possibility of selecting from the five intervals are 16, 8, 4, 2, and 1, respectively. 
% If an interval is selected, then the initial velocity is uniformly distributed in the interval.
% in order to make a uniform distribution of velocity ranges.
If an interval is selected, then the initial velocity of the aircraft will be uniformly distributed in the interval.
The training process initializes the other initial conditions of the aircraft in the same way as Section \ref{sec:short distance policy}.
}
\textcolor{black}{
% The missile is 5–15 km from the aircraft. 
The initial distance between the aircraft and the missile is divided into five intervals (i.e., 5000 m to 7000 m, 7000 m to 9000 m, 9000 m to 11000 m, 11000 m to 13000 m, and 13000 m to 15000 m) and the ratio of the possibility of selecting from the five intervals are  1, 2, 4, 8, and 16, respectively.}
\textcolor{black}{
% The missile is initially generated with the angle between the line-of-sight (LOS) of obstacle and the aircraft’s velocity vector is set between [30, 180] (unit: deg) and [-180, -30] (unit: deg). 
The initial azimuth of the missile is divided into ten intervals (i.e., 30 deg to 60 deg, 60 deg to 90 deg, 90 deg to 120 deg, 120 deg to 150 deg, 150 deg to 180 deg, -180 deg to -150 deg, -150 deg to -120 deg, -120 deg to -90 deg, -90 deg to -60 deg, and -60 deg to -30 deg) and the ratio of the possibility of selecting from the ten intervals are 1, 2, 4, 8, 16, 16, 8, 4, 2, and 1, respectively.
The training process initializes the other initial conditions of the missile in the same way as Section \ref{sec:short distance policy}.
}
% \textcolor{black}{
% The angle between the vector from the missile to the aircraft and a horizontal plane is randomly selected from -15 deg to 15 deg. 
% The initial velocity of the missile is ranging from 800 m/s to 1400 m/s. 
% The maximum overload of the missile ranges from 40 g to 50 g.
% The missile  follows the PN law with a randomly selected navigation coefficient ranging from 3 to 5.
% }
\textcolor{black}{
The maximum number of steps of an episode is 7500.
An episode will be terminated if the distance between the aircraft and the missile is less than 5000 m or the azimuth of the missile is large -15 deg and smaller than 15 deg.
}

% Since this study focuses on scenarios where the missile is at a long distance and large relative angle, two additional termination conditions are defined:
% (1) 
% (2) when the horizontal angle between the obstacle and the aircraft is smaller than $\frac{\pi}{12}$.

\begin{figure}[!htb]
    \centering
    \includegraphics[scale=0.4]{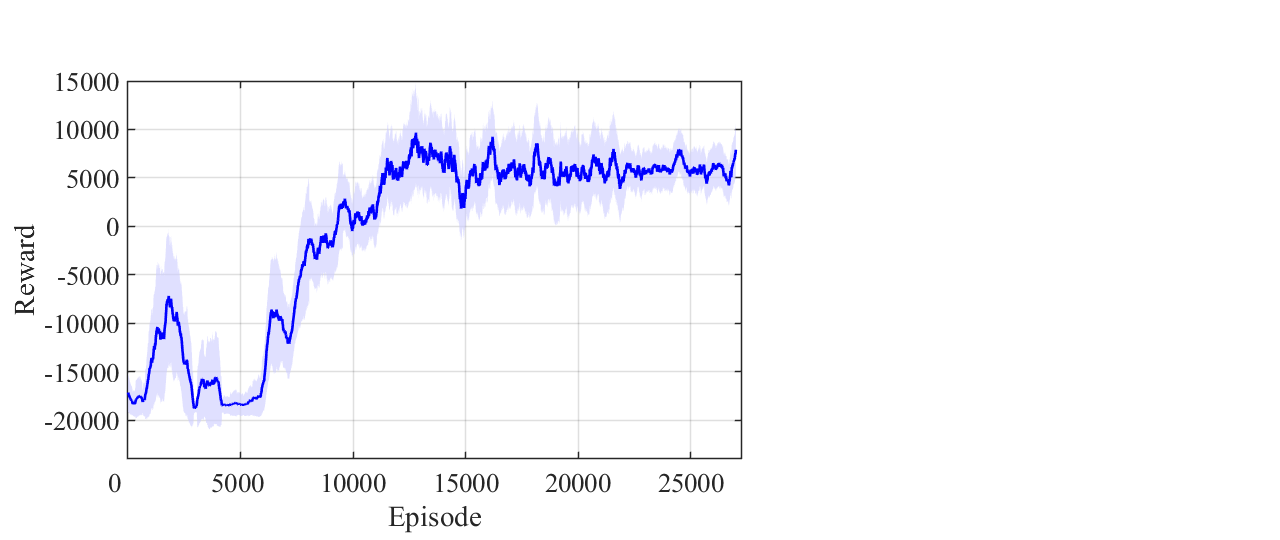}
    \caption{\textcolor{black}{Accumulated reward achieved in the training of the large azimuth policy.}}
    \label{fig: large policy reward}
\end{figure}

\textcolor{black}{
The learning process of the large azimuth policy can be reflected by the achieved reward presented in Fig. \ref{fig: large policy reward}. 
The learning process converges in 14000 episodes.
}

% \noindent \rule{1\linewidth}{0.5mm}

\textbf{Validation.}
\textcolor{black}{To evaluate the large azimuth policy, a statistical analysis is conducted based on different initial conditions of the aircraft and the missile. 
The maximum number of steps of a test is 5000 (i.e., 25 s).
(1) The initial velocity of the aircraft is set to 280 m/s.
(2) The initial velocity of the missile is set to 800 m/s.
(3) The initial distance is set at 15000 m.
(4) The initial azimuth of the missile is set to -30 deg, -60 deg, -90 deg, -120 deg, -150 deg, 30 deg, 60 deg, 90 deg, 120 deg, 150 deg, and 180 deg, respectively.
The validation tests initialize the other initial conditions of the aircraft and the missile in the same way as Section \ref{sec:short distance policy}.
}
% \textcolor{black}{
% The initial roll and pitch of the aircraft are set to 0 deg. 
% Initial conditions other than the aforementioned initial conditions are randomly selected, according to Table \ref{tab: initial conditions}. 
% The initial altitude of the aircraft is randomly selected from 3000 m to 9000 m. 
% The heading of the aircraft is random.
% The angle between the vector from the missile to the aircraft and a horizontal plane ranges from -15 deg to 15 deg in the initial state.
% The maximum overload of the missile is randomly selected from 40 g to 50 g, 
% The missile follows the PN law with a randomly selected navigation coefficient ranging from 3 to 5.}
% By combining these parameter groups, multiple experiments were constructed. 

% \begin{figure}[!htb]
    % \centering
    % \includegraphics[scale=1]{experiment_result/temporary_result/big_trajectory.png}
    % \caption{\hl{Trajectory of the aircraft and the missile.}}
    % \label{fig:reward circling}
% \end{figure}

\begin{figure}[!htb]
    \centering
    \includegraphics[scale=0.4]{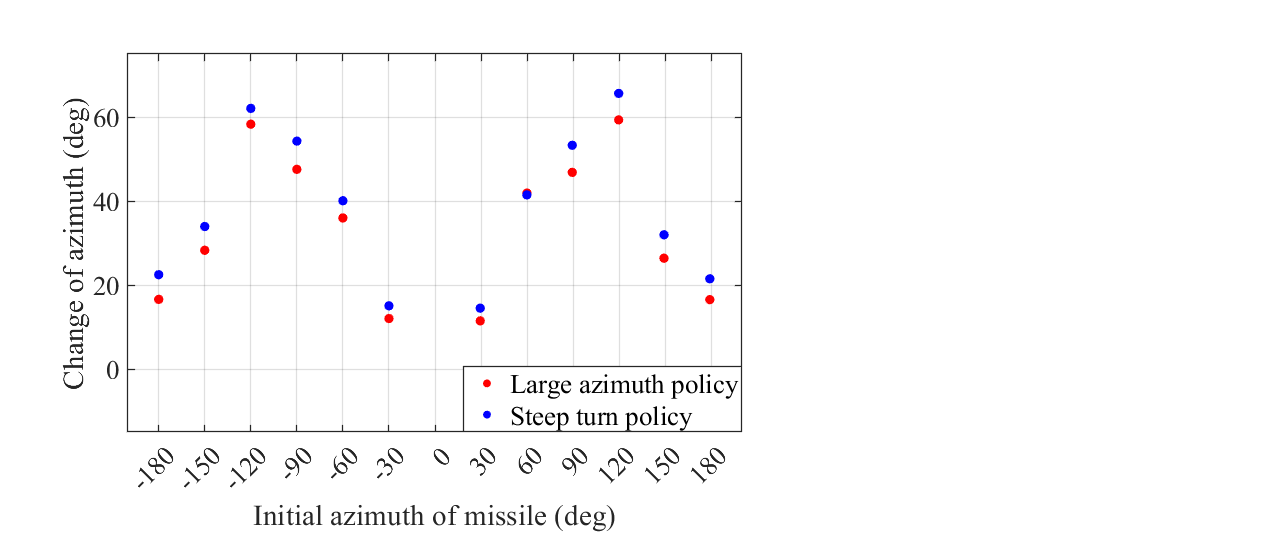}
    \caption{\textcolor{black}{Change of azimuth achieved by the large azimuth policy and the steep turn policy with different initial azimuth conditions.}}
    \label{fig: large azimuth change}
\end{figure}

\begin{figure}[!htb]
    \centering
    \includegraphics[scale=0.4]{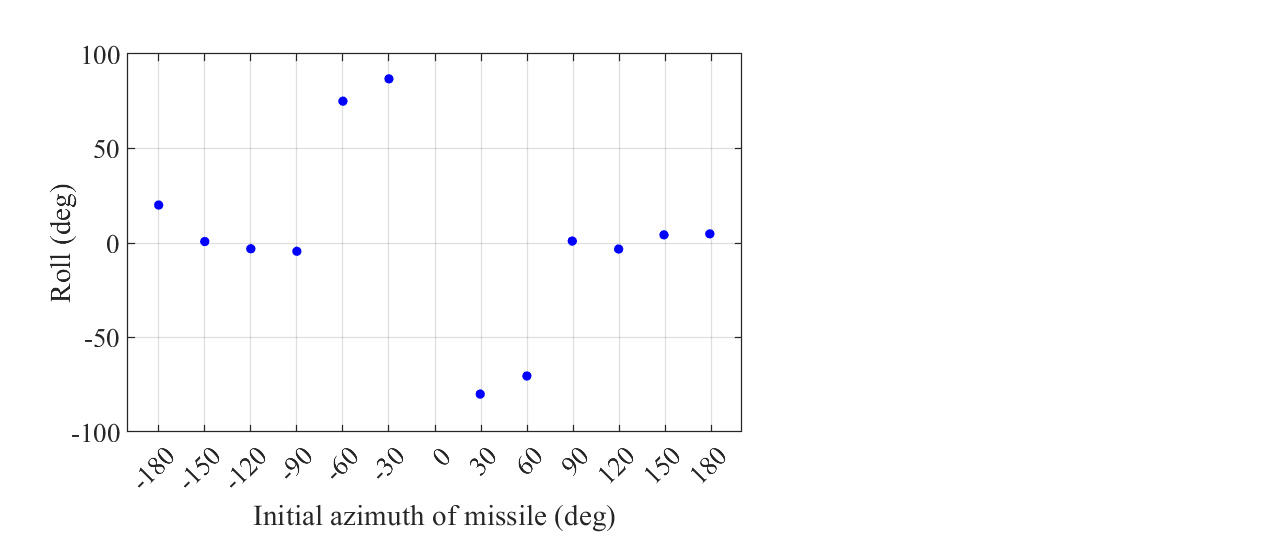}
    \caption{\textcolor{black}{Final roll of the aircraft following the large azimuth policy with different initial azimuth conditions.}}
    \label{fig: large roll}
\end{figure}

\begin{figure}[!htb]
    \centering
    \includegraphics[scale=0.4]{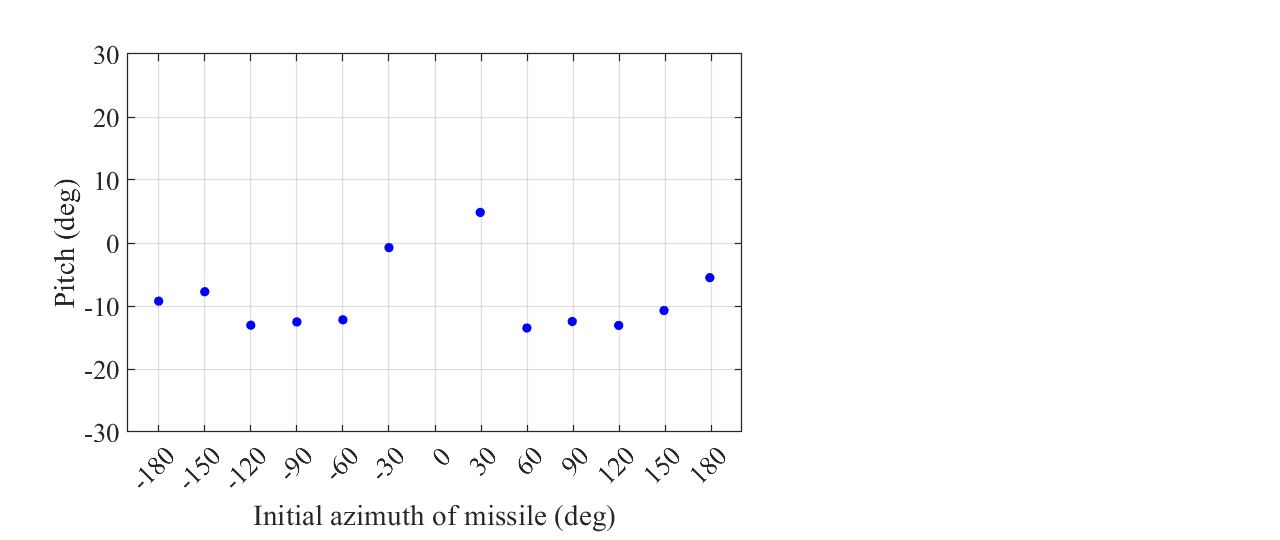}
    \caption{\textcolor{black}{Average pitch of the aircraft following the large azimuth policy with different initial azimuth conditions.}}
    \label{fig: large pitch}
\end{figure}

\begin{figure}[!htb]
    \centering
    \includegraphics[scale=0.4]{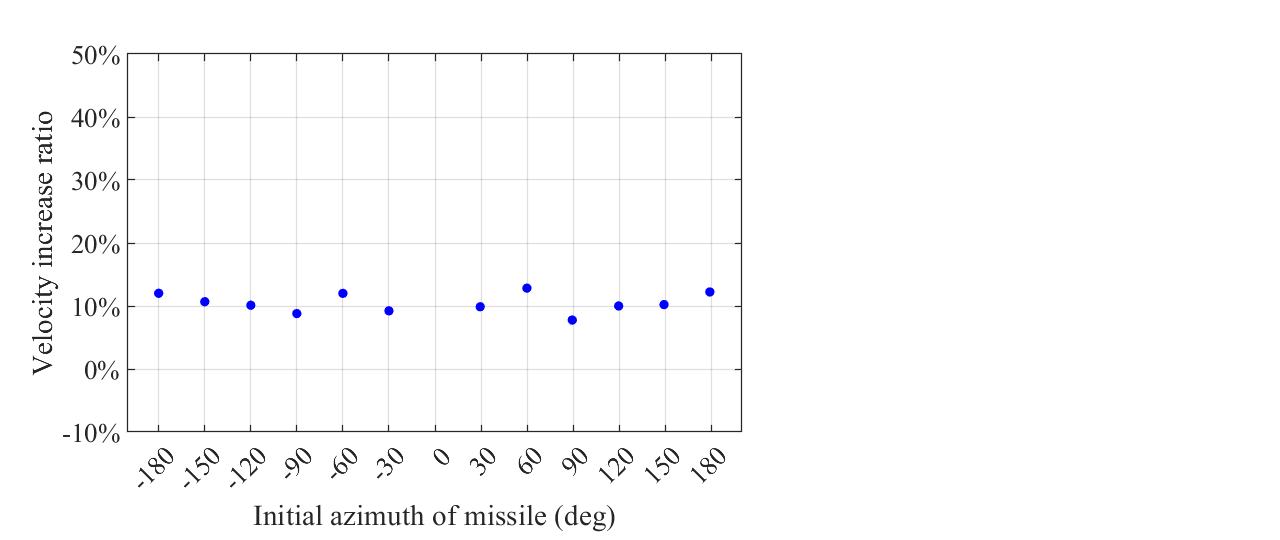}
    \caption{\textcolor{black}{Velocity increase rate of the final velocity to the initial velocity achieved by the large azimuth policy with different initial azimuth conditions.}}
    \label{fig: high velocity}
\end{figure}

\textcolor{black}{For an interval determined by initial velocities of the aircraft and the missile, and the initial azimuth of the missile, 
% one test with other initial conditions that are randomly selected have been conducted.
the other initial conditions are randomly selected and one test have been conducted.}
\textcolor{black}{
The number of out-of-bounds occurrences for roll, pitch, and velocity is listed in Table \ref{tab: large out of bound}. 
The number of all out-of-bounds occurrences is zero, indicating that the constraint terms can effectively maintain the roll, pitch, and velocity within corresponding defined ranges.
Fig. \ref{fig: large azimuth change} shows the change of azimuth achieved by the large azimuth policy and the steep turn policy with the same set of different initial directions. The large azimuth policy shows a turning capability comparable to the steep turn policy.
Fig. \ref{fig: large roll} presents the roll of the aircraft at the last step of one episode. 
If the magnitude of the initial azimuth is small (e.g., 30 deg and 60 deg), the large azimuth policy can turn the aircraft to achieve the termination condition - the magnitude of azimuth smaller than 15 deg, and the aircraft can terminate an episode by a large roll.
% completes the turn before reaching the distance threshold (d=8500 m), resulting in a direction smaller than 15 deg. 
If the magnitude of the initial azimuth is large, 
the large azimuth policy cannot finish a turn, and the large azimuth policy can reduce the roll of the aircraft to around 0 deg to achieve the beneficial roll condition of the short distance policy.
% the turn is not completed before the distance decreases to 8500 m, so after 8500 m the aircraft prefer a stable attitude, so the roll is close to 0 deg.
Fig. \ref{fig: large pitch} shows the average pitch of the aircraft in one episode.
It is shown that the average pitch is around 0 deg, indicating that the aircraft doesn't continuously perform aggress pitch to climb or descend.
Fig. \ref{fig: high velocity} shows the velocity increase ratio, indicating that the large azimuth policy can accelerate the aircraft by about 10 percent.
}

\textcolor{black}{
The results indicate that the large azimuth policy can reduce the azimuth and has a performance comparable to the steep turn policy (Fig. \ref{fig: large azimuth change}).
The large azimuth policy can reduce the magnitude of roll to around 0 deg if the distance reaches 8000 m (Fig. \ref{fig: large roll}).
The large azimuth policy can keep the pitch the aircraft under control (Fig. \ref{fig: large pitch}) and accelerate the aircraft (Fig. \ref{fig: high velocity}).
It is validated that the large azimuth policy can achieve the set objectives.}

\begin{table}[!t]    
    \centering  
    \caption{\textcolor{black}{Number of out-of-bounds occurrences for roll, pitch, and velocity}}
    \label{tab: large out of bound}
    \begin{tabular}{|c|c|c|c|} 
    \hline 
    \textbf{Variable} & \textbf{Roll} & \textbf{Pitch} & \textbf{Velocity} \\ 
    \hline
    \textbf{\makecell{Number of \\ out-of-bounds}} & 0 & 0 & 0 \\  
    \hline
    \end{tabular}
\end{table}

% the obstacle is from different direction, the speed and airspeed is random. 
% check if the policy helps to adjust the airplane and obstacle  to meet our expectation.

\subsection{Implementation of a baseline RL-based strategy}\label{sec:baseline}

% In , 
% \hl{the state space consists of five variables: 
% the relative distance between the missile and the aircraft, the line-of-sight (LOS) azimuth angle, the LOS rate, the flight path angles of both the aircraft and the obstacle.
% As for the action space, Yan's RL-based strategy uses the aircraft’s overload (acceleration) as the control variable.
% However, this state space is insufficient to fully describe the problem in three-dimensional space.

% Our proposed Multi-stage RL-based strategy state space covers the aforementioned five variables (either directly or through indirect calculation), 

% \hl{The state space of Yan's RL-based strategy is set to $\bm{s} = [v_x, v_y, v_z, \theta, \phi, \psi, \Delta x, \Delta y, \Delta z, \Delta v_x, \Delta v_y, \Delta v_z]$, where $ v_x, v_y, v_z$ is aircraft's velocity components in x, y and z axis, $\theta, \phi, \psi$ is roll, pitch and heading of the aircraft, $\Delta x, \Delta y, \Delta z, \Delta v_x, \Delta v_y, \Delta v_z $ is the relative position and relative velocity between the aircraft and the obstacle.

% The action space is set to $\bm{a}= [\delta_{e}, \delta_{a}, \delta_{r}, \delta_{t}]$, consisting of normalized control inputs on the elevator, aileron, rudder, and throttle.}

% \textbf{Initial state:} 

% In \cite{Yan2024AGame}, 
% \hl{both the aircraft and the obstacle start from fixed initial states.

\textcolor{black}{
To evaluate the performance of the proposed multi-stage RL-based evasion strategy, this study compares the proposed multi-stage RL-based evasion strategy to an RL-based evasion strategy proposed in \cite{Yan2024AGame} and a steep turn strategy that is a classical evasion maneuver performed by human pilots \cite{Yang2018EvasiveCombat}.
Although scholars have proposed other RL-based evasion strategies, such as \cite{Cook2025MissileLearning,Zhang2025AdaptiveGame}, this study believe that the RL-based evasion strategy proposed in \cite{Yan2024AGame} is an appropriate baseline. 
Other RL-based evasion strategies may have one or more issues of lack of implementation details, considering issues additional to evasion (e.g., decoy), simple reward function, and obsolescence.}

\textbf{Training.}
\textcolor{black}{
In this section, an RL-based evasion policy is trained according to \cite{Yan2024AGame}.
To ensure a fair comparison, for the training of a policy according to \cite{Yan2024AGame}, the aircraft and the missile are initialized in the same way as the training of the short distance policy except that the missile is at a distance ranging from 5000 m to 15000 m from the aircraft.
The policy uses the same state space and action space as the proposed multi-stage RL-based evasion strategy also.
}

% \hl{
% The initial position of the aircraft is set to [0, 0, $z$] (unit: m), where $z$ represent an altitude randomly selected from 3000 m to 9000 m.
% The initial heading of the aircraft is randomly set ranging from 0 deg to 360 deg.
% Both the initial roll and pitch are set to 0 deg.
% The initial velocity of the aircraft is randomly selected from 280 m/s to 470 m/s.
% }

% The direction of the missile from the aircraft in a horizontal plane is random.
% The angle between the vector from the missile to the aircraft and a horizontal plane ranges from -15 deg to 15 deg.
% The velocity of the missile is a constant ranging from 800 m/s to 1400 m/s.
% The maximum overload of the missile is randomly selected from 40 g to 50 g.
% The missile follows the 3DPN law with a randomly selected navigation coefficient ranging from 3 to 5.

% \textbf{Terminal conditions:}The termination condition of the baseline stops the simulation immediately once the aircraft and the obstacle meet for the first time. In our training period, this terminal condition is not applied. Instead, the minimum distance between the aircraft and the obstacle $d_{min}$ is used in reward function.

\textcolor{black}{
Inspired by \cite{Yan2024AGame}, the reward function of the baseline RL-based policy is defined as
\begin{equation}\label{eq:baseline17 total reward}
    r_{\rm{baseline}} = r_{\rm{baseline}}^{\rm{distance}} + r_{\rm{baseline}}^{\rm{LOS}} + r_{\rm{baseline}}^{\rm{overload}} + r_{\rm{baseline}}^{\rm{pitch}} + r_{\rm{baseline}}^{\rm{roll}}
\end{equation}
where $r_{\rm{baseline}}^{\rm{distance}}$ is a terminal reward related to distance and is defined as
\begin{equation}
    \label{eq: baseline terminal reward} 
        r_{\rm{baseline}}^{\rm{distance}} =
        \begin{cases}
        -200 (\min_{1}^{T}(|| \Delta\bm{x}_m(t) ||) - 10)^2, \\ 
            \quad \quad \quad \quad \quad \quad \quad\text{if} , || \Delta\bm{x}_m (T) || < 10 \, \text{m} \\
        400 (\min_{1}^{T}(|| \Delta\bm{x}_m(t) ||)- 10)+4000, \\
            \quad \quad \quad \quad \quad \quad \quad\text{if} \, || \Delta\bm{x}_m (T) || \geq 10 \, \text{m}
    \end{cases}
\end{equation}
where $T$ represents the number of steps of an episode.
The LOS reward is defined as
\begin{equation}
    \label{eq: baseline LOS reward}
    r_{\rm{baseline}}^{\rm{LOS}} = 2.4 \ln (|{\dot{\lambda}|)}
\end{equation}
Based on \cite{Yan2024AGame}, the absolute LOS rate $|\dot{\lambda}|$ is calculated by
\begin{equation}
    \label{eq: baseline LOS rate}
    |\dot{\lambda}| = \frac{||\bm{u}_{m}(t) - \bm{u}_{m}(t-1)||}{\Delta t}
\end{equation}
% We define the unit vector in horizontal $\bm{u}_{\rm{up}}=\frac{1}{\rho}[\Delta x, \Delta z]$, where $\rho =\sqrt{{\Delta x}^2+{\Delta z}^2}$.
% The time derivative of u can be calculated by
% \begin{equation}\label{eq: baseline LOS rate}
%     \dot{\bm{u}_{\rm{up}}} = \frac{d \bm{u}_{\rm{up}}}{dt} = \frac{1}{\rho}[\Delta \dot{x}, \Delta \dot{z}] + \frac{\dot{\rho}}{\rho^2}[\Delta x, \Delta z]
% \end{equation}
% \begin{equation}\label{eq: baseline LOS rate}
%     ||\dot{\bm{u}_{\rm{up}}}|| = \frac{|\Delta x \Delta \dot{z} - \Delta z \Delta \dot{x}|}{{\Delta x}^2+{\Delta z}^2} = |\dot{\lambda}|
% \end{equation}
% If $\Delta t$ is small enough,then
% \begin{equation}\label{eq: baseline LOS rate}
%     \frac{||\bm{u}_{\rm{up}}(t) - \bm{u}_{\rm{up}}(t-1)||}{\Delta t} = ||\dot{\bm{u}_{\rm{up}}}|| = |\dot{\lambda}|
% \end{equation}
The overload reward is defined as
\begin{equation}
    \label{eq: baseline overload reward}
    r_{\rm{baseline}}^{\rm{overload}} = -0.01 n_m^2
\end{equation}
where $n_m$ is the overload of the missile.
It should be noted that the above-mentioned reward terms (\ref{eq: baseline terminal reward}), (\ref{eq: baseline LOS reward}), and (\ref{eq: baseline overload reward}) are proposed in \cite{Yan2024AGame}. 
To adapt the strategy to a three-dimensional scenario, 
the pitch reward and roll constraint of the short distance policy are applied to the baseline RL-based policy.
The pitch reward aims to avoid severe loss of altitude, which is not a problem in the two-dimensional scenario discussed in \cite{Yan2024AGame}.
The roll constraint prevents the aircraft rollover.
The pitch reward and roll constraint are defined as
\begin{equation}\label{eq:baseline17 LOS reward}
    r_{\rm{baseline}}^{\rm{pitch}} = 0.5 e^{-\displaystyle\frac{|\theta|}{0.2}}
\end{equation}
\begin{equation}
    r_{\rm{baseline}}^{\rm{roll}} =
    \begin{cases}
        -20, & \text{if } |\phi| > 135 \text{ deg} \\
        0,   & \text{otherwise}
    \end{cases}
\end{equation}
}

% \hl{training results
% Fig. \ref{fig:reward baseline 17} shows the episode-reward in training. The reward converge after 4500 episode.}

\textcolor{black}{
The maximum number of steps of an episode is 7500.
% An episode will be terminated if the altitude of the aircraft is less than 1000 m. 
If the distance between the aircraft and the missile is less than 10 m, the episode will not be terminated, aiming to record the minimum distance for achieved reward calculation.
% or when the number of simulation steps exceeds the maximum limit of 7500.
}

\begin{figure}[!htb]
    \centering
    \includegraphics[scale=0.4]{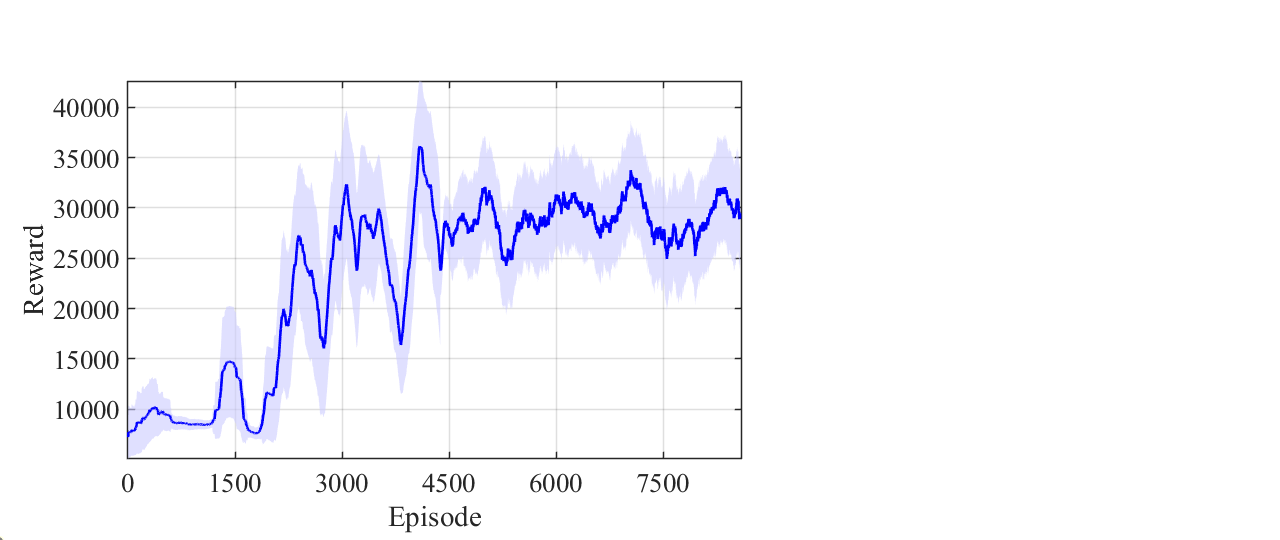}
    \caption{\textcolor{black}{Accumulated reward achieved by the training of the baseline RL-based evasion policy.}}
    \label{fig: baseline policy reward}
\end{figure}

\textcolor{black}{
The learning process of the baseline RL-based policy can be reflected by the achieved reward presented in Fig. \ref{fig: baseline policy reward}. 
The learning process converges in 4500 episodes.
}

% \noindent \rule{1\linewidth}{0.5mm}

\textbf{Validation.}
\textcolor{black}{
To evaluate the performance of the baseline RL-based policy, a test is performed. 
The test has a maximum number of 5000 steps (i.e., 25 s).
In the initial of the test, the position of the aircraft is set to [0, 0, 3970] (unit: m). 
The heading of the aircraft is set to 0 deg. 
Both the roll and pitch of the aircraft are set to 0 deg. 
The velocity of the aircraft is set to 280 m/s. 
The aircraft follows the baseline RL-based policy.
The missile is 6000 m from the aircraft. 
The azimuth of the missile is -30 deg in a horizontal plane. 
The elevation of the missile is 10.72 deg. 
The velocity of the missile is 800 m/s. 
The maximum overload of the missile is randomly set to 40.53 g. 
The missile follows the PN law with a randomly selected navigation coefficient of 3.50.
}

\begin{figure}[!htb]
    \centering
    \includegraphics[scale=0.4, trim=12cm 0cm 0cm 0cm,clip]{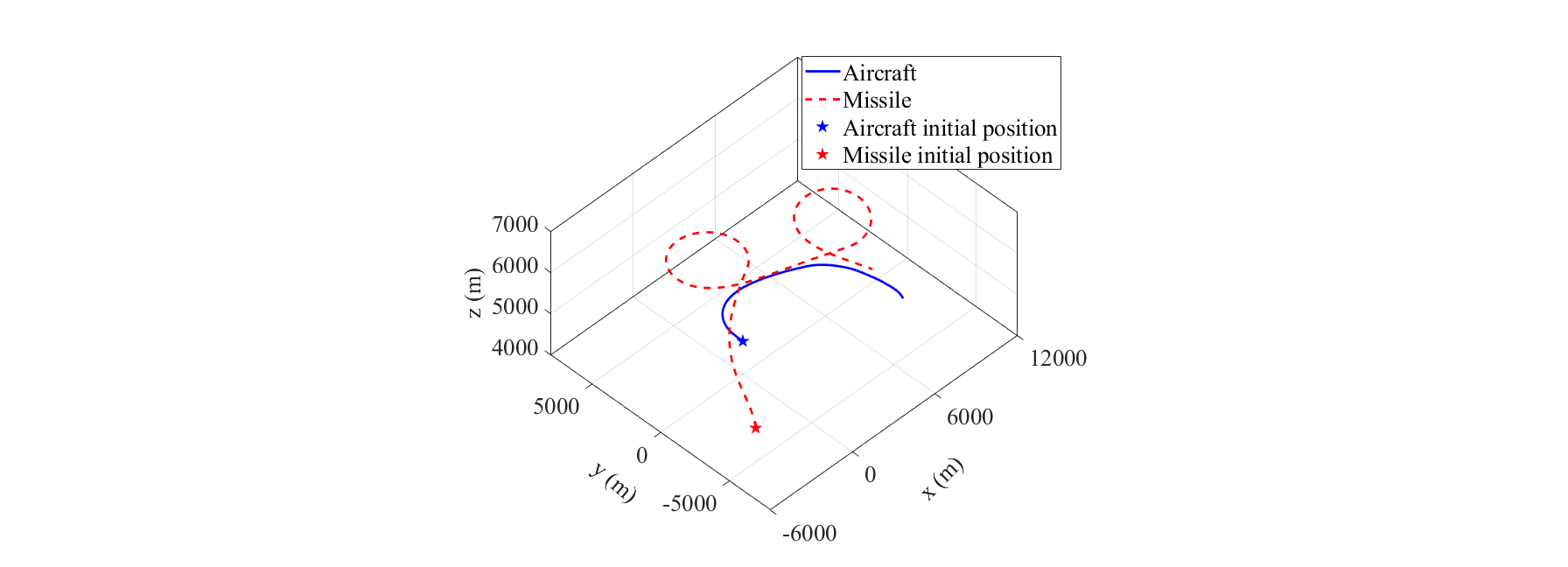}
    \caption{\textcolor{black}{Trajectory of the aircraft and the missile in the test of the baseline RL-based evasion policy.}}
    \label{fig: baseline trajectory}
\end{figure}

\begin{figure}[!htb]
    \centering
    \includegraphics[scale=0.4, trim=0cm 0cm 0cm 0cm,clip]{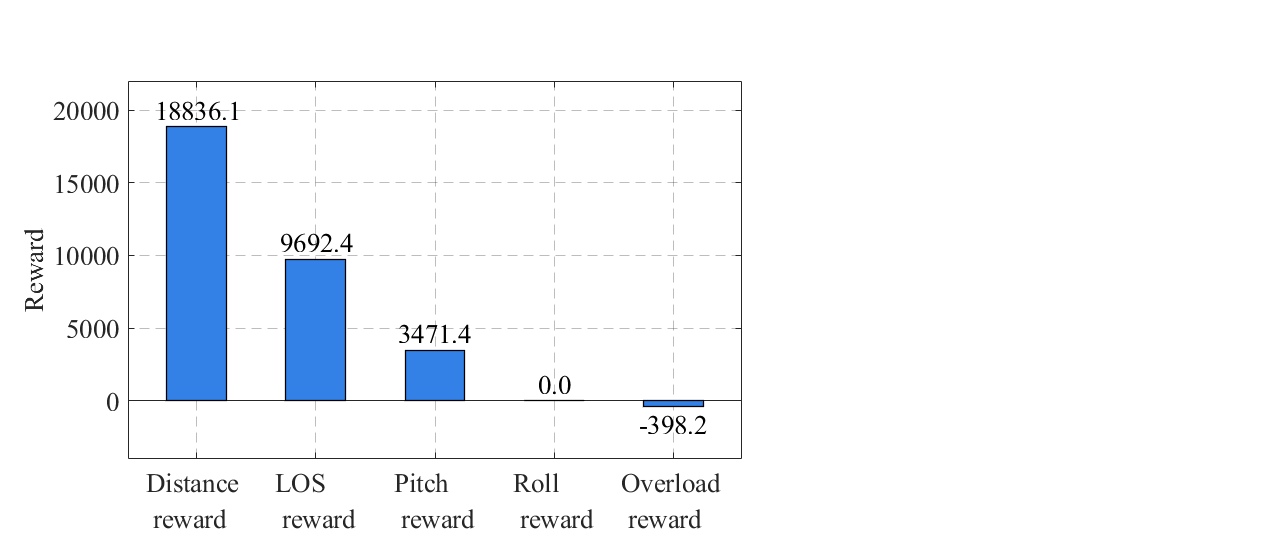}
    \caption{\textcolor{black}{Itemized accumulated rewards of an episode calculated based on the state of the aircraft and the missile in the test of the baseline RL-based evasion policy.}}
    \label{fig: baseline item reward}
\end{figure}

\begin{figure}[!htb]
    \centering
    \includegraphics[scale=0.4]{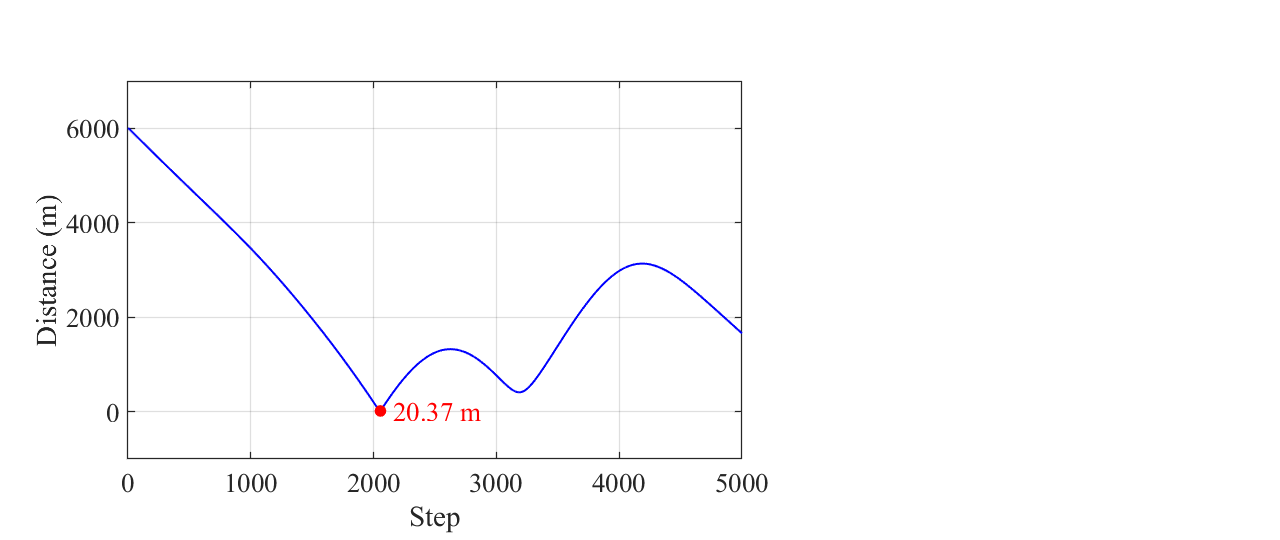}
    \caption{\textcolor{black}{Distance between the aircraft and the missile in the test of the baseline RL-based evasion policy.}}
    \label{fig: baseline distance}
\end{figure}

\begin{figure}[!htb]
    \centering
    \includegraphics[scale=0.4]{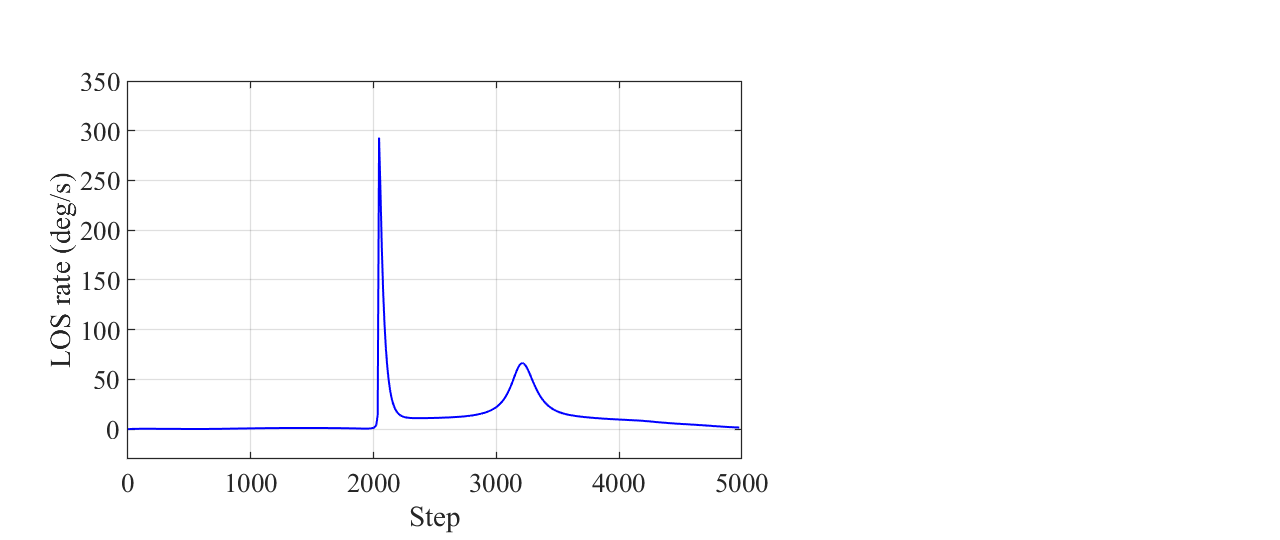}
    \caption{\textcolor{black}{LOS rate in the test of the baseline RL-based evasion policy.}}
    \label{fig: baseline LOS rate}
\end{figure}

\begin{figure}[!htb]
    \centering
    \includegraphics[scale=0.4]{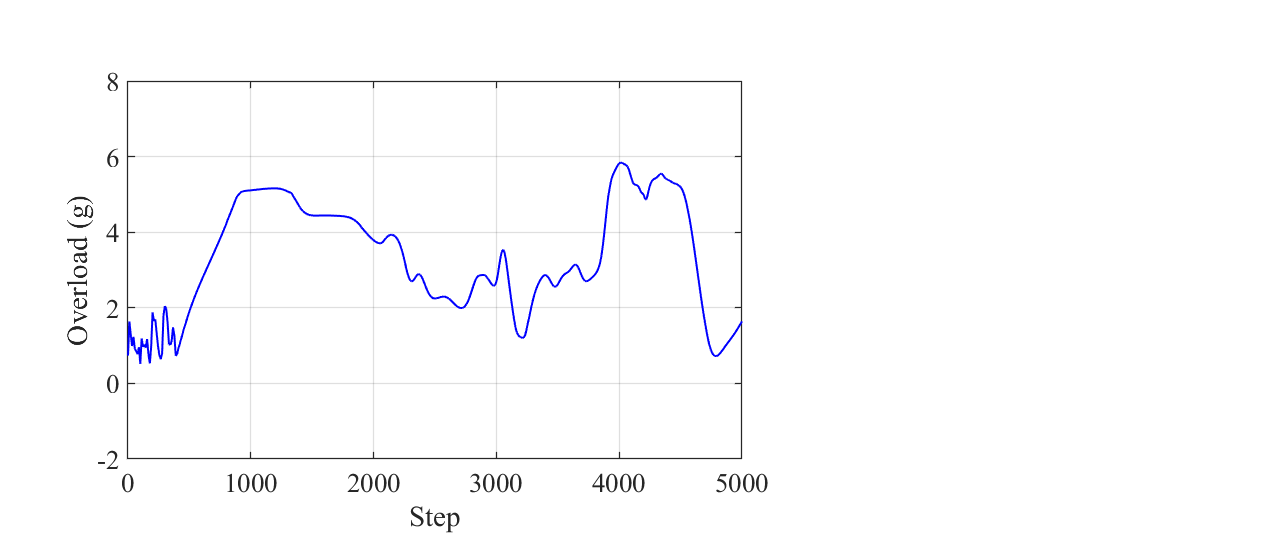}
    \caption{\textcolor{black}{Overload of the aircraft in the test of the baseline RL-based evasion policy.}}
    \label{fig: baseline overload}
\end{figure}

\begin{figure}[!htb]
    \centering
    \includegraphics[scale=0.4, trim=0cm 0cm 0cm 0cm,clip]{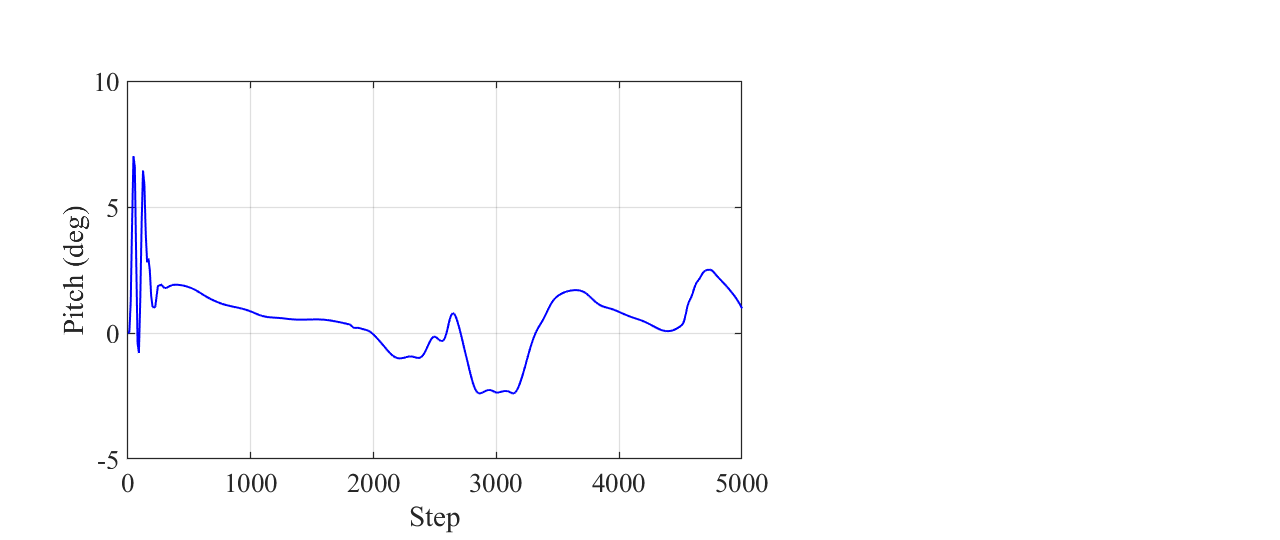}
    \caption{\textcolor{black}{Pitch of the aircraft in the test of the baseline RL-based evasion policy.}}
    \label{fig: baseline pitch}
\end{figure}

\begin{figure}[!htb]
    \centering
    \includegraphics[scale=0.4, trim=0cm 0cm 0cm 0cm,clip]{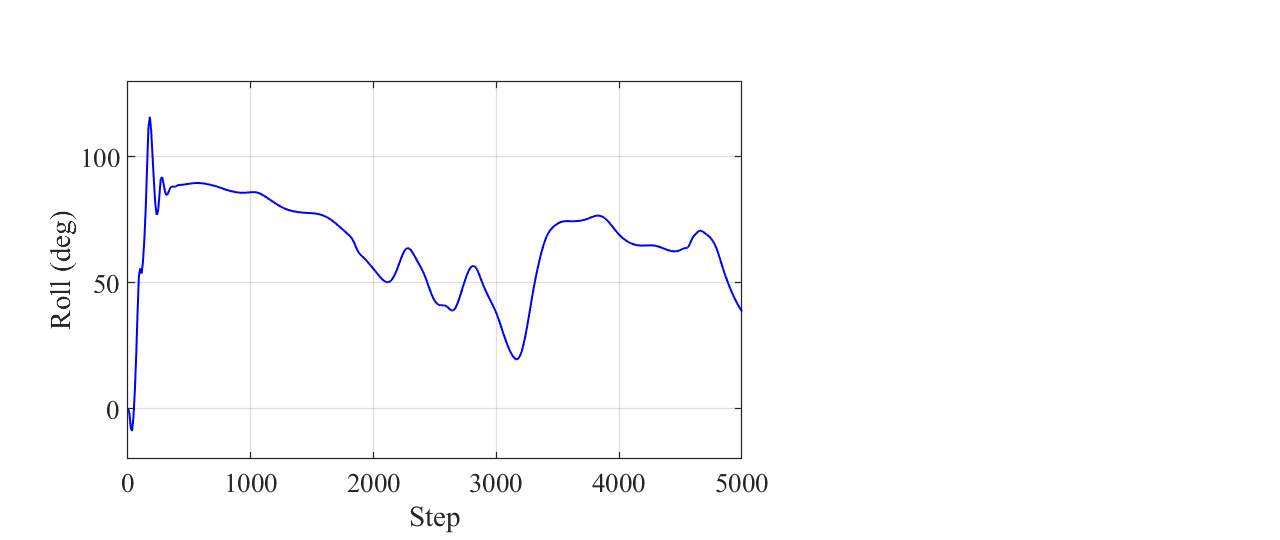}
    \caption{\textcolor{black}{Roll of the aircraft in the test of the baseline RL-based evasion policy.}}
    \label{fig: baseline roll}
\end{figure}

\textcolor{black}{
Fig. \ref{fig: baseline trajectory} shows the trajectory of the aircraft and the missile.
Fig. \ref{fig: baseline item reward} shows the itemized accumulated rewards in one episode corresponding to distance, LOS rate, pitch, roll, and overload. 
It is shown that the baseline RL-based policy can balance the objectives in addressing the distance, LOS rate, and pitch,
avoiding since the magnitude of the corresponding rewards is comparable, in avoiding large overload and roll loss of control.
Fig. \ref{fig: baseline distance} shows the distance between the aircraft and the missile and the minimum distance used to calculate the distance reward is 20.37 m.
Fig. \ref{fig: baseline LOS rate}, Fig. \ref{fig: baseline overload}, Fig. \ref{fig: baseline pitch}, and Fig. \ref{fig: baseline roll} show the LOS rate, overload of the aircraft, pitch, and roll. 
One can see that the baseline RL-based policy can maintain the roll and pitch under control (Fig. \ref{fig: baseline roll} and Fig. \ref{fig: baseline pitch}) and considerably increase the LOS rate when the missile is close to the aircraft (Fig. \ref{fig: baseline LOS rate}).
The overload of the aircraft is less than 6 g (Fig. \ref{fig: baseline overload}), resulting from the overload reward that tends to attenuate the overload of the aircraft.
It is shown that the baseline RL-based policy has learned to guide the aircraft to successfully avoid the missile (Fig. \ref{fig: baseline trajectory})}

% \hl{Since the reward function penalizes large load factors, the maximum overload of the aircraft in this baseline method is limited to about 6 g, which is lower than that achieved by our proposed strategy.}

\subsection{Comparison of the multi-stage RL-based strategy and baselines strategies}\label{sec:comparosion}

\textcolor{black}{
This section compares the proposed multi-stage RL-based evasion strategy to an RL-based evasion strategy proposed in \cite{Yan2024AGame} and a steep turn strategy that is a classical evasion maneuver performed by human pilots \cite{Yang2018EvasiveCombat}, to evaluate the performance of the proposed strategy.
The baseline RL-based evasion strategy is achieved in Section \ref{sec:baseline}.
The steep turn strategy is based on the steep turn policy is achieved in Section \ref{sec:short distance policy}.
}

% To evaluate the performance of the proposed evasion strategy, the proposed evasion strategy is compared with the traditional circling strategy and Existing RL-method named LSRC-TD3 \cite{Yan2024AGame}. 

% The traditional circling strategy adopts the circling policy trained in Section C, 
% while the LSRC-TD3 strategy is trained following the settings described in Section G. 
% \hl{test}

\textcolor{black}{
To investigate the three strategies, a statistical analysis is conducted through dividing the initial conditions of the aircraft and the missile into intervals and then performing tests. 
The maximum number of steps of a test is 5000 (i.e., 25 s).
(1) The initial velocity of the aircraft is divided into intervals ranging from 280 m/s to 470 m/s, with a step of 40 m/s. 
(2) The initial velocity of the missile is divided into intervals ranging from 800 m/s and 1400 m/s, with a step of 100 m/s.
(3) The initial distance between the aircraft and the missile is divided into intervals ranging from 5000 m to 15000 m, with a step of 1000 m.
(4) The initial azimuth of the missile is divided into intervals ranging from -180 deg to 180 deg, with a step of 30 deg.
The tests initialize the other initial conditions in the same way as the statistical analysis corresponding to Fig. \ref{fig: short success ratio}.  
}

% \textcolor{black}{
% The initial roll and pitch of the aircraft are set to 0 deg. 
% Initial conditions other than the aforementioned initial conditions are randomly selected, according to Table \ref{tab: initial conditions}. 
% The initial altitude of the aircraft is randomly selected from 3000 m to 9000 m. 
% The heading of the aircraft is random.
% The elevation of the missile ranges from -15 deg to 15 deg in the initial state. 
% The maximum overload of the missile is randomly selected from 40 g to 50 g. 
% The missile follows the PN law with a randomly selected navigation coefficient ranging from 3 to 5.
% }

% By combining these parameter groups, multiple experiments were conducted for the three methods. 

\begin{figure}[!htb]
    \centering
    \includegraphics[scale=0.4, trim = 0cm 0cm 0cm 0cm, clip]{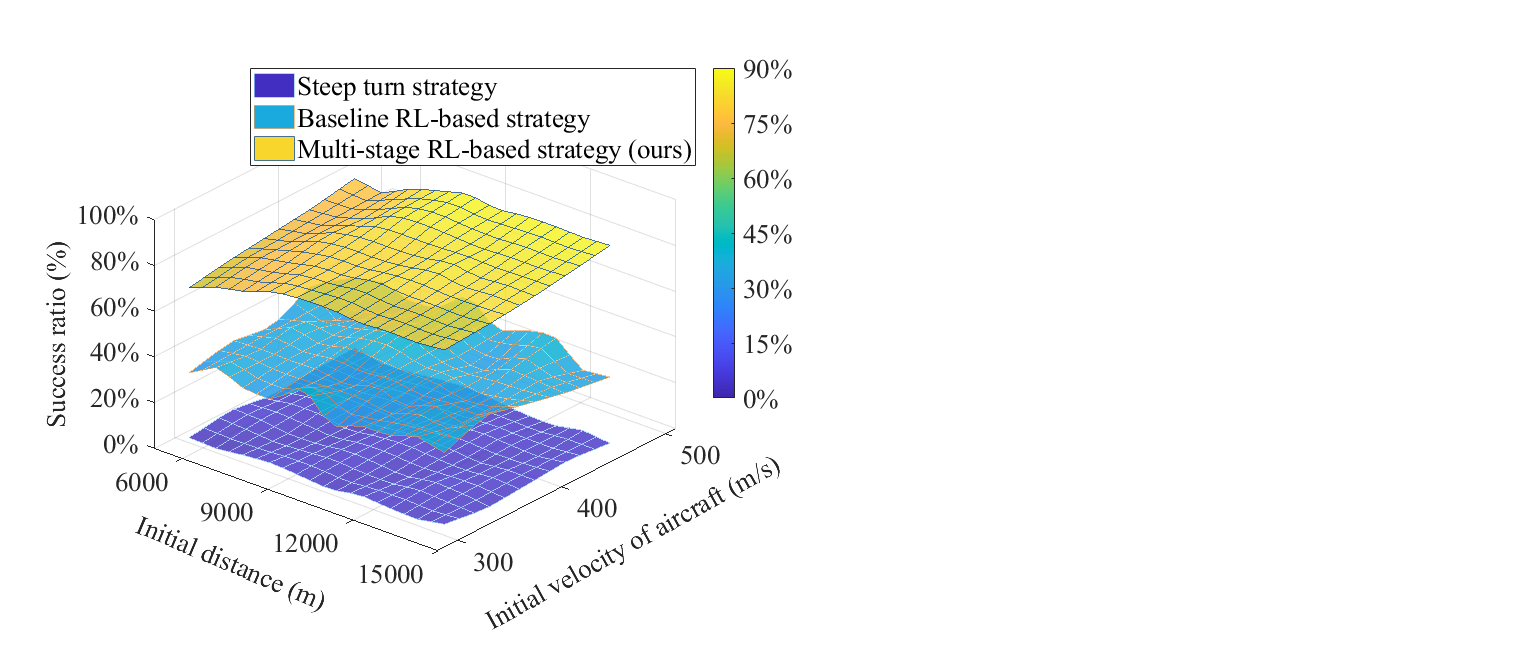}
    \caption{\textcolor{black}{Success ratio of the multi-stage RL-based evasion strategy, baseline RL-based evasion strategy, and baseline steep turn evasion strategy.}}
    \label{fig: comparison success ratio}
\end{figure}

\begin{figure}[!htb]
    \centering
    \includegraphics[scale=0.4, trim = 0cm 0cm 0cm 0cm, clip]{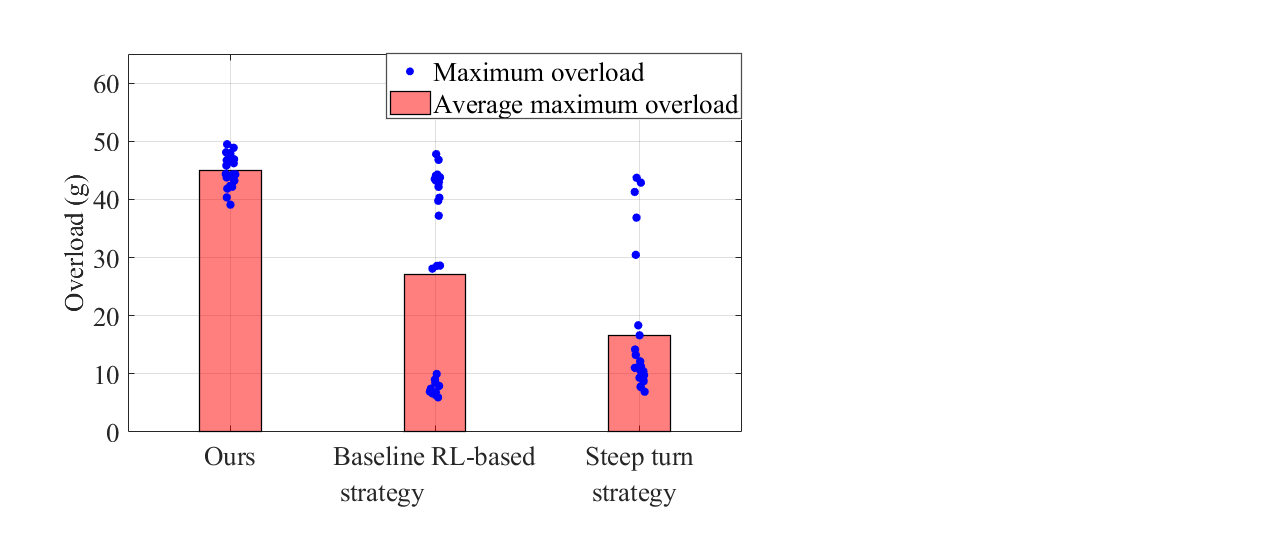}
    \caption{\textcolor{black}{Maximum overload performed by the missile to chase the aircraft with different strategies.}}
    \label{fig: comparison overload}
\end{figure}

\textcolor{black}{
For an interval determined by initial velocities of the aircraft and the missile, the initial distance between the aircraft and the missile, and the initial azimuth of the missile, 
% 20 tests with other initial conditions that are randomly selected have been conducted for all three strategies.
the other initial conditions are randomly selected and 20 tests have been conducted for all three strategies.
To ensure fairness in comparison, for the 20 tests, the aircraft addresses the same set of 20 scenarios determined according to the above-mentioned method, following the proposed strategy, traditional steep turn strategy, and baseline RL-based strategy, respectively.}
\textcolor{black}{
Fig. \ref{fig: comparison success ratio} presents the success ratio of the proposed multi-stage RL-based evasion strategy, baseline RL-based evasion strategy, and baseline steep turn evasion strategy. 
The proposed strategy achieves a success ratio of 80.89 percent in average, while the baseline RL-based strategy and the steep turn strategy achieve a success ratio of 34.14 percent and 3.34 percent in average, respectively, as shown in Table \ref{tab: comparison success ratio}.
It is validated that the proposed multi-stage RL-based evasion strategy can guide the F-16 aircraft to avoid, at 80.89 percent probability, a missile with a random azimuth, a velocity of 800 m/s to 1400 m/s, a maximum overload of 40 g to 50 g, and a detection distance of 5000 m to 15000 m.
If a missile is detected beyond 8000 m, the proposed strategy can achieve a success ratio of 85.06 percent in average.
For every interval, one representative test is randomly selected from the 20 tests.
Fig. \ref{fig: comparison overload} shows the maximum overload in one episode performed by the missile in chasing the aircraft with different evasion strategies, according to the representative tests.
The proposed strategy can always force the missile to perform almost the maximum overload and the average maximum overload is 44.92 g. 
The baseline RL-based evasion strategy and steep turn evasion strategy cannot always force the missile to perform the maximum overload and the average maximum overload is 27.46 g and 16.27 g, respectively.
The advantages of the proposed multi-stage RL-based evasion strategy in success ratio and in forcing the missile to perform maximum overload is validated.
}

\begin{table}[h]
    \centering
    \caption{\textcolor{black}{Success ratio of the multi-stage RL-based evasion strategy, baseline RL-based evasion strategy, and baseline steep turn evasion strategy}}
    \label{tab: comparison success ratio}
    \begin{tabular}{|c|c|c|c|}
    \hline
    \textbf{Strategy} & \makecell{Multi-stage \\ RL-based \\ strategy} & \makecell{Baseline \\ RL-based \\ strategy} & \makecell{Steep turn \\ strategy}   \\
    \hline
    \textbf{\makecell{Success \\ ratio}} &80.89 \% &34.14 \% & 3.34 \% \\
    \hline
    \end{tabular}
\end{table}

% \subsection{Further evaluation of the multi-stage RL-based strategy} \label{sec:threshold}

% \subsubsection{> Influence of the delay of obstacle}\label{sec:threshold}

% \checkmark Table:  if success ratio is increasing?

% TODO: Figure: why time delay is important? maybe delay will influence the obstacle overload? or LOS angle? find the key factors and draw the result. 

% The time delay is set from 0 to 0.05s.(add reference)

\subsection{Validation based on a missile with the augmented proportional navigation law} \label{sec:APN}

\textcolor{black}{
This study further evaluates the generalizability of the multi-stage RL-based strategy based on 
comparing the performance of the strategy in addressing a missile with the APN law to the performance of the strategy in addressing a missile with the PN law. 
It should be noted that the multi-stage RL-based strategy has not been trained based on a missile with the APN law.}

\textcolor{black}{
The APN law introduces an additional term that accounts for the acceleration of the aircraft to determine the missile command acceleration $a_m$
and can be expressed as \cite{Shtessel2007SmoothApplication} 
\begin{equation}
    \label{eq:APN}
    a_m = N \cdot V_c \cdot \dot{\lambda} + N' \cdot a_a^{\perp}
\end{equation}
where $a_a^{\perp}$ represents the perpendicular component of the aircraft’s acceleration relative to the line of sight.
$N'$ is an aircraft acceleration correction coefficient.
}
% Compared to PN guidance, APN considers the acceleration of the aircraft and therefore improves tracking performance.
\textcolor{black}{
% For a missile following the 3DPN law, the navigation coefficient is randomly selected ranging from 3 to 5.
In the comparison, for a missile following the PN law, the navigation coefficient $N$ is set randomly from 3 to 5.
For a missile following the APN law, the navigation coefficient $N$ is set randomly from 3 to 5 and the aircraft acceleration correction coefficient $N'$ is randomly set from $0.5N$ to $N$ according to} \cite{Alqudsi2018AAlgorithms}.
\textcolor{black}{
The aircraft and the missile are initialized in the same way as the comparison of different strategies in Section \ref{sec:comparosion}.
For an interval determined by the initial velocities of the aircraft and the missile, the initial distance between the aircraft and the missile, and the initial azimuth of the missile, 
% 20 tests with other initial conditions that are randomly selected have been conducted for a missile with two different navigation laws.
the other initial conditions are randomly selected and 20 tests have been conducted for a missile with two different navigation laws.
To ensure fairness in comparison, for the 20 tests, the aircraft addresses the same set of 20 scenarios determined according to the above-mentioned method.}

% \hl{
% % To evaluate the robustness of our proposed multi-stage RL-based strategy, 
% 20 independent simulations were performed under the same setup for both PN and APN-guided missile. }

% The maximum number of steps of a test is 5000.
% (1) The initial velocity of the aircraft is divided into intervals ranging from 280 m/s to 470 m/s, with a step of 40 m/s.
% (2) The initial velocity of the missile is divided into intervals ranging from 800 m/s and 1400 m/s, with a step of 100 m/s.
% \hl{(3) The initial distance between the aircraft and the missile is divided into intervals ranging from 5000 m to 15000 m, with a step of 1000 m.}
% (4) The initial azimuth of the missile is divided into intervals ranging from -180 deg to 180 deg, with a step of 30 deg.

% The initial roll and pitch of the aircraft are set to 0 deg. 
% Initial conditions other than the aforementioned initial conditions are randomly selected, according to Table \ref{tab:initial conditions of circling}. 
% Specifically, the initial altitude of the aircraft is randomly selected from 3000 m to 9000 m. 
% The heading of the aircraft is random.
% The angle between the vector from the missile to the aircraft and a horizontal plane ranges from -15 deg to 15 deg in the initial state.
% The maximum overload of the missile is randomly selected from 40 g to 50 g.
% The missile follows the 3DPN law with a randomly selected navigation coefficient ranging from 3 to 5.

\begin{figure}[!htb]
    \centering
    \includegraphics[scale=0.4]{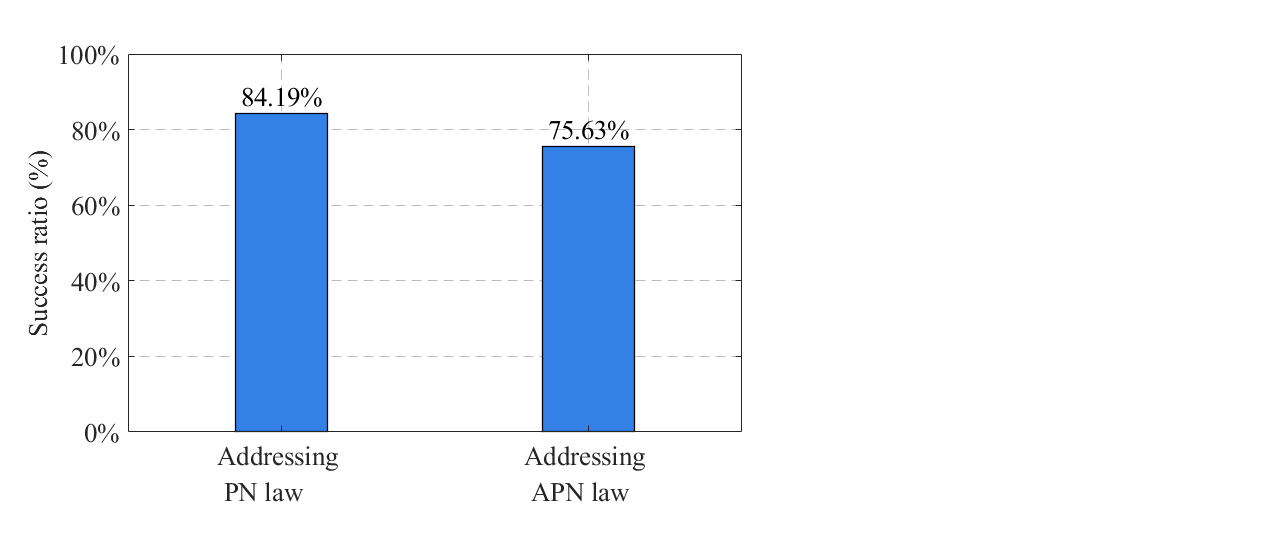}
    \caption{\textcolor{black}{Success ratio of the multi-stage RL-based evasion strategy in addressing a missile following different navigation laws.}}
    \label{fig: APN success ratio}
\end{figure}

\textcolor{black}{
The success ratio of the multi-stage RL-based evasion strategy in addressing a missile following the PN law and a missile following the APN law is presented in Fig. \ref{fig: APN success ratio}.
Although the success ratio in addressing a missile following the APN law is lower than that in addressing a missile following the PN law, the success ratio in the former case can still reach 75.63 percent.
The result is reasonable since APN law requires a higher initial acceleration command but the acceleration demand decreases at short distance, resulting in more overload available to fight against the aggressive evasion maneuver of the aircraft \cite{Liu2019NovelRendezvous}.
The experimental results show that the multi-stage RL-based strategy is with certain generalizability to address different missiles that have not been involved in training.
}

\section{Conclusion \label{sec:conclusion}}

\textcolor{black}{
This study developed a multi-stage RL-based strategy for aircraft to evade incoming missiles, particularly those detected at short range. The proposed strategy adopts a multi-stage architecture that learns a large-azimuth turning policy, a small-azimuth escape policy, and a short-range aggressive maneuver policy. At each stage, one of the three policies is activated based on distance and azimuth.}

\textcolor{black}{
The proposed multi-stage RL-based strategy, a baseline RL-based strategy, and a conventional steep-turn strategy were evaluated against missiles under various conditions in a high-fidelity simulation environment modeling an F-16 aircraft and a missile. Experimental results show that the proposed method achieves superior performance, enabling the F-16 aircraft to avoid missiles with a probability of 80.89 percent for velocities ranging from 800 m/s to 1400 m/s, maximum overloads from 40 g to 50 g, detection distances from 5000 m to 15000 m, and random azimuths. When the missile is detected beyond 8000 m, the success ratio increases to 85.06 percent. In comparison, the baseline RL-based and steep-turn strategies achieved success ratios of only 34.14 percent and 3.34 percent, respectively. The statistical analysis indicates that the machine-learning-based multi-stage strategy can significantly strengthen the final line of defense for aircraft in modern air combat.}

\bibliographystyle{IEEEtran}
\bibliography{references}

\end{document}